# Google's Chrome Antitrust Paradox

*Shaoor Munir*[*], *Konrad Kollnig*[**], *Anastasia Shuba*[***], *Zubair Shafiq*[*]

**ABSTRACT**

*This Article examines Google's dominance of the browser market, highlighting how Google's Chrome browser plays a critical role in reinforcing Google's dominance in other markets. While Google portrays Chrome as a neutral platform built on open-source technologies, this Article shows that Chrome is, in fact, instrumental in Google's strategy to reinforce its dominance in the online advertising, publishing, and browser markets. The examination of Google's strategic acquisitions, anticompetitive practices, and the implementation of so-called "privacy controls" underlines that Chrome is far from a neutral gateway to the web. Rather, it serves as a key tool for Google to maintain and extend its market power, often to the detriment of competition and innovation in the digital economy.*

*This Article illustrates how Chrome not only bolsters Google's position in online advertising and publishing through practices such as coercion and self-preferencing, but also leverages its advertising clout to engage in a "pay-to-play" paradigm—the cornerstone of Google's larger strategy of market control. It also outlines potential regulatory interventions and remedies by drawing on historical antitrust precedents. Lastly, this Article proposes a triad of solutions motivated by an analysis of Google's abuse of Chrome, including behavioral remedies targeting specific anticompetitive practices, structural remedies involving an internal separation of Google's divisions, and divestiture of Chrome from Google into an independent organization.*

*Despite Chrome's dominance and its critical role in Google's ecosystem, as well as its recent legal troubles with the Department of Justice, it so far has avoided significant antitrust action. A key reason for this inaction lies in the long-standing precedent supporting the hegemony of technology firms and the uncertainty surrounding*

[*]University of California, Davis

[**]Maastricht University

[***]Independent Researcher. Anastasia Shuba was employed by DuckDuckGo during the project, but this work was completed independently and does not necessarily reflect the company's position.

*Chrome's viability as a standalone entity. This Article attempts to address these issues to enable antitrust actions that are essential in remedying current market imbalances. Such actions are also critical to mitigate future threats to competition from an increasingly monopolistic technology landscape, thereby fostering a competitive digital environment that promotes innovation and protects consumer interests.*

**TABLE OF CONTENTS**

## I. INTRODUCTION

Google is one of the largest companies on the planet. Alphabet, its parent company, has a market capitalization of nearly $2.4 trillion.[1] In 2023, Alphabet generated more than $307 billion in revenue, 77 percent of which came from Google's online advertising services.[2] Key

1. *Market Capitalization of Alphabet (Google) (GOOG)*, COMPANIESMARKETCAP, https://companiesmarketcap.com/alphabet-google/marketcap/ [https://perma.cc/KLH6-YGMN] (last visited Jan. 30, 2025) ("As of January 2025 Alphabet (Google) has a market cap of $2.4 Trillion USD.").

2. Alphabet Inc., Annual Report (Form 10-K) 7 (Jan. 30, 2024), https://abc.xyz/assets/ff/7c/06d6f493f6462caf08e8502ffc33/596de1b094c32cf0592a08edfe84ae74.html [https://perma.cc/X6VN-GMQP] ("We have built world-class advertising technologies for advertisers, agencies, and publishers to power their digital marketing businesses. Our advertising solutions help millions of companies grow their businesses through our wide range of products across devices and formats, and we aim to ensure positive user experiences by serving the right ads at the right time and by building deep partnerships with brands and agencies. AI has been foundational to our advertising business for more than a decade. Products like Performance Max and Product Studio use the full power of our AI to help advertisers find untapped and incremental to Google's dominance in the market for online advertising has been a series of strategic acquisitions.[3] These strategic acquisitions expanded Google's portfolio of advertising services.[4] Most crucially, they consolidated its dominant position across the advertising stack. This stack includes the buy-side, also known as the demand-side platform (DSP); the sell-side, known as the supply-side platform (SSP); and the ad exchange, which connects the buy-side and the sell-side.[5] In 2003, Google acquired Applied Semantics, later rebranded as AdSense, which is now Google's SSP.[6] In 2005, Google acquired Urchin Software, later rebranded as Google Analytics, which is now Google's analytics service.[7] This was followed by the acquisition of DoubleClick in 2007, which is now Google's ad exchange (AdX) service.[8] The trend continued with the acquisition of AdMob in 2009, extending Google's presence in mobile advertising.[9] In 2010, Google acquired Invite Media, which was subsequently integrated into Google's ad exchange to offer open

---

conversion opportunities. Google Services generates revenues primarily by delivering both performance and brand advertising that appears on Google Search & other properties, YouTube, and Google Network partners' properties ('Google Network properties'). We continue to invest in both performance and brand advertising and seek to improve the measurability of advertising so advertisers understand the effectiveness of their campaigns.").

3. *See* Karina Montoya, *How Three Mergers Buttressed Google's Ad Tech Monopoly, Per DOJ*, TECHPOLICY.PRESS (Mar. 9, 2023), https://www.techpolicy.press/how-three-mergers-buttressed-googles-ad-tech-monopoly-per-doj/ [https://perma.cc/Z6YD-KX2C].

4. *See id.*

5. *See id.*

6. Dawn Kawamoto, *Google Buys Applied Semantics*, ZDNET (Apr. 24, 2003, 5:07 AM), https://www.zdnet.com/article/google-buys-applied-semantics/ [http://perma.cc/Y32F-PUUY] ("The company, through its AdSense product, will deliver text advertisements to Web pages based on keyword relevance to the page. This type of content targeting is an area Google entered last month . . . .").

7. Scott Crosby, *Urchin Software Corp.*, MEDIUM (Sept. 2, 2016), https://urchin.biz/urchin-software-corp-89a1f5292999 [https://perma.cc/52L8-REYS] ("In April 2005 the company was acquired by Google, and the Urchin product became 'Urchin from Google,' then later simply Google Analytics.").

8. Louise Story & Miguel Helft, *Google Buys DoubleClick for $3.1 Billion*, N.Y. TIMES (Apr. 14, 2007), https://www.nytimes.com/2007/04/14/technology/14DoubleClick.html [https://perma.cc/7J6D-L8VK] ("The sale offers Google access to DoubleClick's advertisement software and, more importantly, its relationships with Web publishers, advertisers and advertising agencies. . . . DoubleClick's exchange is different from the ad auctions that Google uses on its networks because the exchange is open to any Web publisher or ad network — not just the sites in Google's network.").

9. Susan Wojcicki & Vic Gundotra, *Investing in a Mobile Future with AdMob*, GOOGLE: OFF. BLOG (Nov. 9, 2009), https://googleblog.blogspot.com/2009/11/investing-in-mobile-future-with-admob.html [https://perma.cc/U22G-WHEU] ("[W]e are looking forward to having them join the Google team and work with us on the future of mobile advertising.").

bidding. Google also acquired its DSP service called DV360 in 2010.[10] In 2011, Google acquired AdMeld, an SSP which was subsequently integrated into Google's AdX service.[11] By systematically building dominance across every layer of the advertising stack—analytics, demand-side, supply-side, and AdX—Google established an unparalleled position in the online advertising ecosystem.[12] Google is now by far the most dominant player in the online advertising landscape—its AdX has greater than or equal to 50 percent market share.[13]

Google also conducted a series of strategic acquisitions to become dominant in the online publisher market, whose advertising space is sold exclusively by Google's advertising services.[14] Google acquired Outride in 2001 and Kaltix in 2003,[15] integrating the personalized search technology of both in Google Search.[16] In 2006, Google acquired YouTube, which is now one of the most popular social media sites in the United States.[17] This is despite the general perception that Google does not own a social media service after shutting down Google Plus in 2019.[18] In 2004, Google acquired Where 2, Keyhole, and ZipDash, which formed the basis of Google Maps and Google Earth.[19] Google also acquired Waze, a Google Maps competitor in 2013.[20] In 2004, Google acquired Picasa, a photo management and editing platform, which was later subsumed by Google Photos.[21] Google also acquired DocVerse as a precursor to launching Google Docs in 2010.[22]

---

10. *See* Neal Mohan, *Investing in Exchange Bidding*, GOOGLE: DOUBLECLICK ADVERTISER BLOG (June 3, 2010), https://doubleclick-advertisers.googleblog.com/2010/06/investing-in-exchange-bidding.html [https://perma.cc/BKV2-BZHW] ("[G]oing to continue to invest significantly in improving Invite Media's technology and products as a separate platform and, in time, make it work seamlessly with our DoubleClick for Advertisers (DFA) ad serving product."); Montoya, *supra* note 3 ("Google merged Invite Media with DV360 and gained access to a much larger network of advertisers.").

11. *The Best of Admeld, Now in DoubleClick Ad Exchange*, GOOGLE: ADMELD, https://admeld.com [https://perma.cc/2778-4B9R] (last visited Jan. 30, 2025) ("Google bought Admeld in 2011 and we've worked diligently to build Admeld's best features into the DoubleClick Ad Exchange. Now that we've completed the integration, AdX provides publishers a unique platform that combines robust, publisher-centric tools with a massive, global pool of demand."); *Google Integrates Admeld into DoubleClick AdX, Preps 'Unified' Publisher Solution*, ADEXCHANGER: AD EXCH. NEWS (Mar. 22, 2012, 4:00 PM), https://www.adexchanger.com/ad-exchange-news/google-integrates-admeld-into-doubleclick-adx-preps-unified-publisher-solution/ [https://perma.cc/SWG8-REXF].

12. Montoya, *supra* note 3.

13. Press Release, U.S. Dep't of Just., Justice Department Sues Google for Monopolizing Digital Advertising Technologies (Jan. 24, 2023), https://www.justice.gov/opa/pr/justice-department-sues-google-monopolizing-digital-advertising-technologies [https://perma.cc/X3RP-KWY5] ("Google AdExchange, which is greater than or equal to 50% of the ad exchange market share . . . .").

14. *See id.*

15. *Google Acquires Technology Assets of Outride Inc.*, GOOGLE: NEWS FROM GOOGLE (Sept. 20, 2001), https://googlepress.blogspot.com/2001/09/google-acquires-technology-assets-of.html [https://perma.cc/Y8FT-EYU8] ("Google Inc. today announced the company's acquisition of the intellectual property, including patent rights, source code, trademarks, and associated domain names, from Outride Inc., a Redwood City, Calif.-based developer of online information retrieval technologies."); *Google Acquires Kaltix Corp.*, GOOGLE: NEWS FROM GOOGLE (Sept. 30, 2003), https://googlepress.blogspot.com/2003/09/google-acquires-kaltix-corp.html [https://perma.cc/ZJ8W-SM2C] ("Google Inc. today announced it acquired Kaltix Corp., a Palo Alto, Calif.-based search technology start-up.").

---

16. Ed Sim, *Personalized Search*, BEYONDVC (Mar. 31, 2004), https://www.beyondvc.com/2004/03/ [https://perma.cc/HJW8-ELLV] ("Fast forward to now and Google is bringing this back into the market, although it is using its latest acquisition, Kaltix, as the basis for its search. This one is based on profiles rather than behavior. . . . It will be interesting to see how personalization and the search wars play out over the next couple of years. I, for one, am a big fan of the original Outride model based on user behavior.").

17. *Google Buys YouTube for $1.65 Billion*, NBC NEWS (Oct. 9, 2006, 10:54 AM), https://www.nbcnews.com/id/wbna15196982 [https://perma.cc/MF52-35KJ] ("Internet search leader Google is snapping up YouTube for $1.65 billion, brushing aside copyright concerns to seize a starring role in the online video revolution."); BROOKE AUXIER & MONICA ANDERSON, PEW RSCH. CTR., SOCIAL MEDIA USE IN 2021 3 (2021) ("A majority of Americans say they use YouTube and Facebook, while use of Instagram, Snapchat and TikTok is especially common among adults under 30."); EMILY A. VOGELS, RISA GELLES-WATNICK & NAVID MASSARAT, PEW RSCH. CTR., TEENS, SOCIAL MEDIA AND TECHNOLOGY 2022 3 (2022) ("YouTube tops the 2022 teen online landscape among the platforms covered in the Center's new survey, as it is used by 95% of teens.").

18. Chris Fox, *Google Shuts Failed Social Network Google+*, BBC (Apr. 1, 2019), https://www.bbc.com/news/technology-47771927 [https://perma.cc/DW9L-P43F].

19. *See* Chris Morris, *10 Notable Google Acquisitions*, CNBC (Aug. 9, 2012, 12:40 PM), https://www.cnbc.com/id/48569184 [https://perma.cc/NFT5-6AFG] ("In 2003, Danish brothers Lars and Jens Rasmussen founded a small mapping technology company, but had grander plans to revolutionize how people got directions. When Google heard those plans—and saw the prototype the Rasmussens and two associates had created—it quickly bought the company. The result was Google Maps, which has gone on to become one of the company's most popular features."); *Where 2 Technologies Acquired by Google*, CRUNCHBASE, https://www.crunchbase.com/acquisition/google-acquires-where2—7f71b983 [https://perma.cc/LT73-TWRQ] (last visited Jan. 31, 2025); Matt Hines, *Google Buys Satellite Image Firm Keyhole*, CNET (Oct. 27, 2004, 7:46 AM), https://www.cnet.com/tech/services-and-software/google-buys-satellite-image-firm-keyhole/ [https://perma.cc/M42M-QJZW] ("The acquisition of Keyhole underscores Google's efforts to widen its search capabilities beyond basic Web page results, as competition in the search sector heats up."); *Google Acquires ZipDash*, MERGR, https://mergr.com/alphabet-acquires-zipdash [https://perma.cc/3PLB-VQ9H] (last visited Jan. 31, 2025) ("On September 1, 2004, Google acquired software company ZipDash.").

20. Ingrid Lunden, *Google Bought Waze for $1.1B, Giving a Social Data Boost to Its Mapping Business*, TECHCRUNCH (June 11, 2013, 8:37 AM), https://techcrunch.com/2013/06/11/its-official-google-buys-waze-giving-a-social-data-boost-to-its-location-and-mapping-business [https://perma.cc/MQ84-4X79] ("After months of speculation, the fate of Waze, the social-mapping-location-data startup, is finally decided: Google is buying the company, giving the search giant a social boost to its already-strong mapping and mobile businesses.").

Like the advertising and analytics acquisitions discussed above, the strategic acquisitions of these publisher services steadily positioned Google as an ever more dominant force in various segments of the publisher market including in search engines, maps services, and video streaming. Most notably, Google Search is by far the most popular search engine, with approximately 90 percent market share,[23] which was recently classified as a monopoly by the US District Court for the District of Columbia.[24] Similarly, Gmail is the most popular website in the email category,[25] Google Drive is the most popular file sharing service,[26] and Google Workspace is the most popular office suite.[27] In fact, Google owns an outsized fraction of the top websites: the top two most visited websites in the world, google.com and youtube.com.[28]

While Google's dominance as an advertiser and publisher has been challenged in the courts,[29] there is another market that has

---

21. Morris, *supra* note 19 ("In the days leading up to its IPO, Google bought this [Picasa] online photo manager in an effort to maintain its lead over Yahoo and MSN, which were still viable competitors to its primary search business.").

22. *See* Michael Arrington, *Google Acquires Docverse to Further Office Arms Race*, TECHCRUNCH (Mar. 5, 2010, 11:03 AM) [hereinafter Arrington, *Google Acquires Docverse*], https://techcrunch.com/2010/03/05/google-acquires-docverse-to-further-office-arms-race/?guccounter=1 [https://perma.cc/V2JZ-L76L] ("With DocVerse Google will have a direct software connection to Microsoft Office, allowing users to collaborate real time on documents.").

23. *See Search Engine Market Share Worldwide, Dec 2023 - Dec 2024*, STATCOUNTER, https://gs.statcounter.com/search-engine-market-share [https://perma.cc/DZE8-2HYL] (last visited Jan. 31, 2025).

24. *See* United States v. Google LLC, No. 20-cv-3010, 2024 WL 3647498, at *4 (D.D.C. Aug. 5, 2024) ("Specifically, the court holds that (1) there are relevant product markets for general search services and general search text ads; (2) Google has monopoly power in those markets; (3) Google's distribution agreements are exclusive and have anticompetitive effects; and (4) Google has not offered valid procompetitive justifications for those agreements.").

25. *See Number of Email Users Worldwide 2024: Demographics & Predictions*, FINANCESONLINE, https://financesonline.com/number-of-email-users/ [https://perma.cc/26AK-JK3K] (last visited Jan. 31, 2025) ("Gmail is the most popular email provider, with 1.5 billion active users worldwide.").

26. *See File Sharing Software Market Share*, DATANYZE, https://www.datanyze.com/market-share/file-sharing—198 [https://perma.cc/C9DH-6FKE] (last visited Jan. 31, 2025).

27. *See Office Suites*, 6SENSE, https://6sense.com/tech/office-suites [https://perma.cc/AH6K-WP53] (last visited Jan. 31, 2025) ("Google Workspace with 79.41% market share (1,520,830 customers), Microsoft Office with 10.74% market share (205,607 customers), Google Sheets with 5.94% market share (113,828 customers).").

28. *See Top Websites Ranking*, SIMILARWEB, https://www.similarweb.com/top-websites/ [https://perma.cc/TH5Y-MFK6] (last visited Jan. 31, 2025).

29. *See* Complaint at 4, United States v. Google LLC, No.1:23-cv-00108 (E.D. Va. Jan. 24, 2023) ("The United States and Plaintiff States bring this action for violations of the Sherman Act to halt Google's anticompetitive scheme, unwind Google's monopolistic grip on the market, and garnered much less attention: Google's dominance in the market for browsers used by consumers to access the web on desktop computers and mobile devices.[30] Google launched the Chrome browser in 2008 by combining WebKit, an open-source web engine originally developed by Apple for its Safari browser,[31] with its own V8 JavaScript engine.[32]

---

restore competition to digital advertising."); Complaint at 2, United States v. Google LLC, No. 1:20-cv-03010 (D.D.C. Oct. 20, 2020) ("The United States of America, acting under the direction of the Attorney General of the United States, and the States of Arkansas, Florida, Georgia, Indiana, Kentucky, Louisiana, Mississippi, Missouri, Montana, South Carolina, and Texas, acting through their respective Attorneys General, bring this action under Section 2 of the Sherman Act, 15 U.S.C. § 2, to restrain Google LLC (Google) from unlawfully maintaining monopolies in the markets for general search services, search advertising, and general search text advertising in the United States through anticompetitive and exclusionary practices, and to remedy the effects of this conduct."); Complaint at 8, Texas v. Google LLC, No. 4:20-CV-957 (E.D. Tex. Dec. 16, 2020) ("As a result of Google's anticompetitive conduct, including its unlawful agreement with Facebook, Google has violated and continues to violate Sections 1 and 2 of the Sherman Act, 15 U.S.C. §§ 1, 2. Plaintiff States bring this action to remove the veil of Google's secret practices and end Google's abuse of its monopoly power in online advertising markets.").

30. *See* Brian Dean, *Web Browser Market Share: 85+ Browser Usage Statistics*, BACKLINKO, https://backlinko.com/browser-market-share [https://perma.cc/QH7P-VV4J] (Feb. 7, 2024). While a web browser is used on both mobile phone and desktops, there exists an interesting fundamental difference in how they are used. On mobile phones, most users prefer to use separate applications to access different services (e.g. social media and content streaming). *See* Nitin Deshdeep, *Mobile App or Website? 10 Reasons Why Apps Are Better*, VWO, https://vwo.com/blog/10-reasons-mobile-apps-are-better/ [https://perma.cc/N9KN-5RMG] (Jan. 29, 2025). On desktop, all these activities take place through the web browser. *See* Jeremy Holcombe, *Mobile vs Desktop: Where Are Your Website Visitors Coming From?*, GREENGEEKS, https://www.greengeeks.com/blog/mobile-desktop/ [https://perma.cc/Z5WR-DSSZ] (last visited Jan. 31, 2025). This gives the web browser a much bigger role on desktop as compared to mobile. *See id.* Hence this Article focuses on the desktop browser market and how Google is using its monopoly through Chrome to its advantage across parallel markets such as the publisher and advertisement markets.

31. William L. Hosch, *Chrome*, BRITANNICA, https://www.britannica.com/technology/Chrome [https://perma.cc/A3S8-7XWH] (Jan. 31, 2025). A browser engine is the core of each web browser and is responsible for communicating with websites and visualizing them. *See* Kitakabee, *What Is a Browser? How Does It Work?*, BROWSERSTACK (Oct. 25, 2024), https://www.browserstack.com/guide/what-is-browser [https://perma.cc/NX2E-CYLC]. This means that, to begin with, Chrome was mainly a container for the WebKit browser engine. Initially, Google differentiated Chrome with a different user interface compared to Safari and a deeper integration of Google's own services such as integration of Google Search, cross device tab syncing, and more. *See Understanding the Key Differences Between Safari and Google Chrome*, ASK, https://www.ask.com/news/understanding-key-differences-safari-google-chrome [https://perma.cc/CVX9-H75F] (Jan. 12, 2024).

32. *See* Hosch, *supra* note 31. Most websites on the internet are made up of three parts: HTML, JavaScript, and CSS. *See* Richard Fong, *Why Web Developers Use HTML, CSS, and Javascript*, BLISS DRIVE SEO (July 8, 2023), https://www.blissdrive.com/seo/why-do-web-developers-use-html-css-and-javascript/ [https://perma.cc/QBK4-73ZF]. HTML contains the layout

Later in 2013, Chrome started creating its own version of WebKit called Blink, citing that the rest of the WebKit partners were "slowing everybody down."[33] Chrome also benefited from several strategic acquisitions by Google, which included Reqwireless and its mobile browser, and GreenBorder, which provided state-of-the-art sandboxing functionality for Chrome on launch.[34] As such, Chrome's market share has steadily increased since its inception and it is now the most dominant browser, with more than 68 percent market share on desktop and mobile devices, mirroring Google's dominance in advertising and publishing markets.[35]

Google pours immense resources into the development of Chrome.[36] Yet, Chrome is available for free and lacks an independent revenue stream and business model.[37] Instead, Chrome's publicly

---

and the website content. *Id.* CSS contains the visual elements, such as colors and spacing between the website elements. *Id.* JavaScript contains interactive computer code that describes how a website is supposed to change in response to user interactions, like changing the layout, design, or content of a website without navigating to a different website. *Id.*

33. Frederic Lardinois, *Google Forks WebKit and Launches Blink, a New Rendering Engine That Will Soon Power Chrome and Chrome OS*, TECHCRUNCH (Apr. 3, 2013, 2:00 PM), https://techcrunch.com/2013/04/03/google-forks-webkit-and-launches-blink-its-own-rendering-engine-that-will-soon-power-chrome-and-chromeos/ [https://perma.cc/T6GX-ANE9] ("Having to integrate Google's way of doing things with WebKit and what the rest of the WebKit partners were doing was 'slowing everybody down,' Komoroske said.").

34. *See* Elinor Mills, *Google Buys Canadian Wireless-Software Company*, CNET (Jan. 9, 2006, 3:50 PM), https://www.cnet.com/culture/google-buys-canadian-wireless-software-company/ [https://perma.cc/ANB8-2SN8] ("Google has acquired Reqwireless, a small Canadian company that makes Web browser and e-mail software for use on wireless devices . . . ."); Hines, *supra* note 19 ("Google has jumped into the anti-malware market, snatching up browser-based security software maker GreenBorder Technologies for an undisclosed amount of money."). Sandboxing here refers to the technical concept of creating isolated environments to securely run different computer programs. *See What Is Sandboxing?*, CHECK POINT, https://www.checkpoint.com/cyber-hub/threat-prevention/what-is-sandboxing/ [https://perma.cc/ND73-HTSR] (last visited Jan. 31, 2025). Sandboxing is used in web browsers, such as Chrome, to isolate different browser windows or tabs. *See Security You Never Have to Think About*, GOOGLE: SAFETY CTR., https://safety.google/intl/en_us/chrome/ [https://perma.cc/8B68-QY5Q] (last visited Jan. 31, 2025).

35. *Global Market Share Held by Leading Internet Browsers from January 2012 to May 2023*, STATISTA [hereinafter *Global Market Share*], https://www.statista.com/statistics/268254/market-share-of-internet-browsers-worldwide-since-2009/ [https://perma.cc/8CZ6-XWLA] (last visited Jan. 27, 2025); *Browser Market Share Worldwide, Jan 2009 – Dec 2024*, STATCOUNTER [hereinafter *Browser Market Share Worldwide*], https://gs.statcounter.com/browser-market-share#monthly-200901-202412 [https://perma.cc/EU6Q-99UK] (last visited Jan. 27, 2025).

36. *See* Emma Roth, *Google Launhes a 'Neutral' Chromium Development Fund*, VERGE (Jan. 9, 2025, 4:55 PM), https://www.theverge.com/2025/1/9/24340196/google-neutral-chromium-development-fund [https://perma.cc/3NC5-7GCD].

37. *See* Vanessa Page, *How Mozilla Firefox and Google Chrome Make Money*, INVESTOPEDIA, https://www.investopedia.com/articles/investing/041315/how-mozilla-firefox-and-google-chrome-make-money.asp [https://perma.cc/YPQ6-Z577] (Jan. 6, 2023) ("Examining Google Chrome's revenue is much harder since Google doesn't list the revenue and expenses for all of its stated mission from the onset has centered on increasing adoption of Google's advertising and publisher services.[38] While prior research and regulatory action focused on Google's abuse of its dominance in advertising and publishing markets, far less attention has been given to how Google leverages Chrome's dominance in these markets to unfairly advance its services.[39]

This Article demonstrates that Chrome is the key to Google's dominance as an advertiser and publisher. First, Google leverages its dominance as a publisher to reinforce Chrome's dominance using (1) subtle coercion, like employing dark patterns, and (2) undermining web standards.[40] Second, Google leverages Chrome to reinforce Google's dominance as a publisher and advertiser using (3) self-preferencing and (4) "privacy controls."[41] These strategies demonstrate that Chrome is not merely a neutral gateway to the web but is instead an instrument for Google to gain and maintain an unfair advantage over its competitors. Finally, Google employs (5) "pay-to-play," a strategy where it uses its revenue primarily generated from its dominant advertising business for strategic acquisitions of more publishing and advertising

---

services. Chrome falls under Google Services, which makes most of its money through advertising.").

38. *See* Ben Goodger, *Welcome to Chromium*, GOOGLE: CHROMIUM BLOG (Sept. 2, 2008), https://blog.chromium.org/2008/09/welcome-to-chromium_02.html [https://perma.cc/66SR-LDLT] ("To be clear, improving the web in this way also has some clear benefits for us as a company. With a richer set of APIs, we can build more interesting apps allowing people to do more online. The more people do online, the more they can use our services.").

39. *See* Lauren Feiner, *Breaking Down the DOJ's Plan to End Google's Search Monopoly*, VERGE (Nov. 27, 2024, 10:50 AM) [hereinafter Feiner, *Breaking Down*], https://www.theverge.com/2024/11/27/24302415/doj-google-search-antitrust-remedies-chrome-android [https://perma.cc/84D3-H2C5]. Throughout this Article, the term *advertiser* is used to refer to an entity which facilitates the relationship between publishers and brands by identifying user interests, determining the appropriate brand for the user, and displaying the advertisement for the brand on the publisher's website. For example, Google Ads. Unless otherwise specified, advertiser is also used as an umbrella term for analytics services such as Google Analytics, which determine the quantity and qualities of the users visiting a website. Throughout this Article, the term *publisher* is used to refer to an application, website, or a service, which a user visits to get access to a specific functionality. Google Search, YouTube, Gmail, and Google Drive are some examples of Google's business products where it acts as a publisher.

40. *See* Bill Toulas, *Browser Developers Push Back on Google's "Web DRM" WEI API*, BLEEPINGCOMPUTER (July 29, 2023, 10:11 AM), https://www.bleepingcomputer.com/news/google/browser-developers-push-back-on-googles-web-drm-wei-api/ [https://perma.cc/L4G6-MP39].

41. *See* Adam Clark Estes, *You Deserve a Better Browser Than Google Chrome*, VOX (Nov. 22, 2024, 1:20 PM), https://www.vox.com/technology/387375/google-chrome-antitrust-privacy-android?utm_source [https://perma.cc/MU3T-BXSB].

services and to pay competitors to give prominence to Google Search.[42] This results in a vicious cycle of cross-market abuse in one market, be it browser, advertising, or publishing, to cement its position in the other markets.

Google's dominance of the web (through Google Chrome), publishing (through Google Search, YouTube, Gmail and many other online services), and advertising (through Google Ads and Google Analytics) markets raises concerns about fair competition in the digital economy.[43] Google's multifaceted approach—combining strategic acquisitions with tactics that disadvantage its competitors—underlines the complexity of Google's dominance and the need for regulatory scrutiny. Drawing upon historical antitrust actions, we discuss three potential types of remedies: (1) behavioral remedies, (2) structural remedies, and (3) divestiture. First, as a relatively mild measure, this Article discusses the imposition of behavioral remedies on Google. These restrictions would aim to specifically mitigate deceptive practices, such as the use of dark patterns, and prevent the coercive integration of users into Google's broader service ecosystem. Second, this Article considers an internal restructuring within Google to enforce a clear demarcation between its advertising and browser divisions. Such separation is crucial to prevent the advertising interests of Google from unduly influencing the operation and development of Chrome. However, these measures, while addressing certain aspects of Google's anticompetitive behavior, may not sufficiently tackle its extant market dominance. In this context, the extensive and successful history of antitrust law, particularly its role in disbanding monopolies like the AT&T and Bell System, becomes pertinent.[44] Thus, third, we discuss how Chrome might need to be divested from its parent company, potentially as a public utility. The divestiture of Chrome into an independent organization would ensure that it serves as a truly neutral gateway to the web. Moreover, it would not only diminish Google's ability to leverage Chrome as a tool for market capture but also foster innovation in the online advertising and publishing markets.

---

42. Alphabet Inc., Annual Report, *supra* note 2 ("Google Services generates revenues primarily by delivering both performance and brand advertising that appears on Google Search & other properties, YouTube, and Google Network partners' properties ('Google Network properties').").

43. *See* Feiner, *Breaking Down*, *supra* note 39.

44. *See* United States v. Am. Tel. & Tel. Co., 552 F. Supp. 131, 223–25 (D.D.C. 1982), *aff'd sub nom.* Maryland v. United States, 460 U.S. 1001 (1983).

## II. PAST CHALLENGES TO THE DOMINANCE OF CHROME AND GOOGLE

As a leading player across multiple markets, Google—and the Chrome browser in particular—has faced numerous challenges that threatened to restrict its control over technology infrastructure and its access to vast amounts of data, both of which are key to its business model.[45] This section examines some of the key challenges and explores how they have influenced Google's business strategies and practices. Specifically, it explores self-regulatory initiatives and the past use of privacy and competition law to rein in Google's dominance.

### *A. Self-Regulation*

Google's advertising business relies on identifying and tracking users to determine their interests, thus enabling advertisers to target specific segments of users based on these interests.[46] This means that online tracking is crucial for Google's targeted advertising business.[47]

Countermeasures against online tracking have evolved over the years. In 2007, prominent consumer advocacy groups such as the Consumer Federation of America and the Consumers Union petitioned the Federal Trade Commission (FTC) to establish a regulatory framework addressing online advertising's intrusive nature.[48] This initiative sought the creation of a "Do-Not-Track" (DNT) list, which would be conceptually like the "Do-Not-Call" registry, to opt out of receiving telemarketing phone calls.[49] After several years of relatively limited progress, in 2009 researchers Christopher Soghoian and Sid Stamm proposed the first DNT standard, designed to empower web browser users to opt out of online tracking.[50] This proposal entailed the transmission of a user's tracking preferences through browser signals (called HTTP headers), thereby passing these preferences to website operators.[51] The website operators and any embedded third parties could then choose to ignore or honor the user's preferences.[52] With no technical restrictions or legal requirements for honoring DNT, adoption was low.[53]

---

45. *See* Ken Mingis & Jon Gold, *Google US Antritrust Trials: A Timeline*, COMPUTERWORLD (Sept. 23, 2024), https://www.computerworld.com/article/1635715/googles-us-antitrust-trials-a-timeline.html [https://perma.cc/W6GW-RDNG].

46. *See Audience Targeting: Reaching the Right People With Your Google Ads*, NOBLE DESKTOP (Sept. 2, 2024), https://www.nobledesktop.com/learn/google-ads/audience-targeting-reaching-the-right-people-with-your-google-ads [https://perma.cc/8SJH-LFR2].

47. *See Case Study: Google*, NEW AM., https://www.newamerica.org/oti/reports/special-delivery/case-study-google/ [https://perma.cc/MR4D-Z8HC] (last visited Jan. 28, 2025).

48. *See* Letter from Ari Schwartz, Deputy Dir., Ctr. for Democracy & Tech., Linda Sherry, Dir., Nat'l Priorities Consumer Action, Mark Cooper, Dir. Rsch., Consumer Fed'n Am., Lee Tien, Senior Staff Att'y, Elec. Fontier Found., Deborah Pierce, Exec. Dir., Priv. Activism, Daniel Brandt, President, Pub. Info. Rsch., Robert Ellis, Publisher, Priv. J., Beth Givens, Dir., Priv. Rts. Clearinghouse & Pam Dixon, Exec. Dir., World Priv. F. to Donald S. Clark, Sec'y, Fed. Trade Comm'n 2, https://cdt.org/wp-content/uploads/privacy/20071031consumerprotectionsbehavioral.pdf [https://perma.cc/D7NS-U2F3] ("Specifically, we urge the U.S. Federal Trade Commission (FTC) to take proactive steps to adequately protect consumers as online *behavioral tracking* and *targeting* become more ubiquitous."); Diane Bartz, *Consumer Groups Urge 'Do Not Track' Registry*, NBC NEWS (Apr. 15, 2008, 5:00 PM), https://www.nbcnews.com/id/wbna24138328 [https://perma.cc/5H9D-BRF6] ("While companies like Google are trying to put pretty good practices in place, we don't want to rely on the good graces of the companies because they might change their minds." (quoting Chris Murray, Senior Counsel for Consumers Union)).

At the same time, the FTC urged the advertising industry to develop self-regulatory measures for online advertising, which resulted in the launch of the "AdChoices" program in 2010.[54] This industry-led initiative allowed individuals to opt out of targeted advertising through a dedicated website.[55] The initiative, however, never reached widespread trust and adoption in part because it does not aim to stop online tracking and (somewhat counterintuitively) involves letting advertisers set cookies in the browser to store opt-out preferences.[56]

---

49. *National Do Not Call Registry*, FED. TRADE COMM'N, https://www.donotcall.gov/index.html [https://perma.cc/43WV-23KY] (last visited Jan. 28, 2025) ("The National Do Not Call Registry gives you a choice about whether to receive telemarketing calls.").

50. Christopher Soghoian, *The History of the Do Not Track Header*, SLIGHT PARANOIA BLOG (Jan. 21, 2011), http://paranoia.dubfire.net/2011/01/history-of-do-not-track-header.html [https://perma.cc/78L4-8LWY] ("In mid July 2009, the Future of Privacy Forum organized a meeting and conference call in which I pitched the header concept to a bunch of industry players, public interest groups, and other interested parties. I was perhaps slightly over-dramatic when I told them that the 'day of reckoning was coming', for opt out cookies, and that it was time to embrace a header based mechanism.").

51. *See id.*

52. *See id.*

53. *See Tracking Do Not Track: New Ad Network Data Shows That 8 Percent of Users Have DNT On*, FUTURE PRIV. F. BLOG, https://fpf.org/blog/tracking-do-not-track-new-ad-network-data-shows-that-8-percent-of-users-have-dnt-on/ [https://perma.cc/E9XV-5GLS] (Aug. 31, 2020) ("A sample of Chitika's data shows that currently over 8 percent of users across all browsers are transmitting a DNT signal indicating a preference not to be tracked.").

54. *See Self-Regulatory Program for Online Behavioral Advertising Factsheet*, IAB, https://www.iab.com/wp-content/uploads/2015/06/OBA_OneSheet_Final.pdf [https://perma.cc/9EX2-W8QD] (last visited Jan. 28, 2025); YOURADCHOICES, https://youradchoices.com [https://perma.cc/2KK8-Q3FS] (last visited Jan. 28, 2025); *Digital Advertising Alliance (DAA) Announces 'Your AdChoices' Consumer Education Campaign*, PR NEWSWIRE (Jan. 20, 2012, 8:30AM), https://www.prnewswire.com/news-releases/digital-advertising-alliance-daa-announces-your-adchoices-consumer-education-campaign-137749828.html [https://perma.cc/UBH9-9F3L] ("The DAA Self-Regulatory Program for Online Behavioral Advertising was launched in 2010 by the Digital Advertising Alliance (DAA), a consortium of the nation's largest media and marketing associations including the American Association of Advertising Agencies (4A's), the Association of National Advertisers (ANA), the American Advertising Federation (AAF), the Direct Marketing Association (DMA), the Interactive Advertising Bureau (IAB) and the Network Advertising Initiative (NAI).").

55. *See* YOURADCHOICES, *supra* note 54.

Recognizing the potential of DNT to enhance online privacy, in 2010, the FTC endorsed its adoption with some success.[57] By the end of 2012, all major web browsers implemented DNT functionality.[58] This included Chrome, though it was the last to adopt DNT in 2012.[59] Although it had a much smaller market share at the time, it was already the most used web browser. Chrome reached a market share of about 31 percent by the end of 2012, while Microsoft's Internet Explorer had dropped to 27 percent and Mozilla's Firefox to 19 percent.[60]

Interestingly, like many other advertising companies, Google did not honor DNT on its own first-party websites nor on its third-party services like Google Analytics, reflecting a significant gap in the adoption of privacy measures by major online websites.[61] While there

---

56. *See* Aaron Sankin, *I Tried to Use the Ad Tech Industry's Tool to Opt Out of Personalized Ads. Did It Work?*, MARKUP (Mar. 25, 2021, 8:00 AM), https://themarkup.org/privacy/2021/03/25/i-tried-to-use-the-ad-tech-industrys-tool-to-opt-out-of-personalized-ads-did-it-work [https://perma.cc/5SBH-X2BQ] ("DAA's opt-out works basically the same way that online advertisements do. It installs a cookie (a small text file that connects your device with a profile of your activity) on your web browser.").

57. Press Release, Fed. Trade Comm'n, FTC Testifies on Do Not Track Legislation (Dec. 2, 2010), https://www.ftc.gov/news-events/news/press-releases/2010/12/ftc-testifies-do-not-track-legislation [https://perma.cc/G8WE-HFX2] ("[T]he agency supports giving consumers a 'Do Not Track' option because the practice is largely invisible to consumers, and they should have a simple, easy way to control it.").

58. *See Do Not Track in Google Chrome*, GOOGLE OPERATING SYS. (Nov. 8, 2012), https://googlesystem.blogspot.com/2012/11/do-not-track-in-google-chrome.html [https://perma.cc/GN6N-82U2] ("Chrome is actually the last major browser that adds support for 'Do Not Track', a feature that is already available in Firefox, Opera, Safari and Internet Explorer.").

59. *Id.*

60. *Global Market Share*, *supra* note 35.

61. *Turn "Do Not Track" On or Off*, GOOGLE: GOOGLE CHROME HELP, https://support.google.com/chrome/answer/2790761?hl=en&co=GENIE.Platform%3DDesktop [https://perma.cc/E82Q-J3D6] (last visited Jan. 28, 2025) ("Most websites and web services, *including Google's*, don't change their behavior when they receive a Do Not Track request. Chrome doesn't provide details of which websites and web services respect Do Not Track requests and how websites interpret them." (emphasis added)); Kevin Dees, *Adding Google Analytics to Your Website While Respecting "Do Not Track"*, KEVIN DEES BLOG (Apr. 13, 2020), https://kevdees.com/adding-google-analytics-to-your-website-while-respecting-do-not-track/ [https://perma.cc/8AJK-E5R6] ("You see, Google Analytics does not automatically handle 'privacy' for you."); Kashmir Hill, *'Do Not Track,' the Privacy Tool Used by Millions of People, Doesn't Do Anything*, GIZMODO (Oct. 15, 2018), https://gizmodo.com/do-not-track-the-privacy-tool-used-by-millions-of-peop-1828868324 [https://perma.cc/34B9-6MH8] ("From the department of irony, Google's Chrome browser offers was initially cautious support for the DNT standard in the advertising industry, it vanished after Microsoft announced in 2012 that its Internet Explorer—still widely used at the time—would enable the DNT signal by default.[62] It was only three years later, in 2015, that Microsoft switched back to an opt-out rather than opt-in model for DNT.[63] Nonetheless, the damage was done and the DNT initiative ultimately faltered, culminating into the dissolution of the relevant web standards working group in 2019.[64] Following the dissolution of the Tracking Protection Working Group, Apple removed DNT support from Safari in 2019, signaling a retreat from the once-promising DNT initiative.[65]

While the DNT initiative on its own did not prove to be successful,[66] industry players such as Mozilla and Apple have since pioneered anti-tracking countermeasures. Apple's Safari browser introduced Intelligent Tracking Prevention in 2017, a feature meant to curb cross-site tracking.[67] Mozilla followed suit with Firefox's Enhanced

---

users the ability to turn off tracking, but Google itself doesn't honor the request, a fact Google added to its support page some time in the last year.").

62. *See Internet Explorer 10 Released for Windows 7*, PCMAG (Nov. 13, 2012), https://www.pcmag.com/archive/internet-explorer-10-released-for-windows-7-304943 [https://perma.cc/5SSE-L7X3] ("On the security front, IE10 includes the 'do not track' technology, so advertisers cannot secretly monitor your activity in order to serve up targeted ads. Those who want it, however, can disable 'do not track.'").

63. Gregg Keizer, *Microsoft Rolls Back Commitment to Do Not Track*, CSO (Apr. 3, 2015), https://www.csoonline.com/article/551084/microsoft-rolls-back-commitment-to-do-not-track.html [https://perma.cc/Q3LL-GGGF] ("Microsoft today rolled back its commitment to the nearly-dead 'Do Not Track' (DNT) standard, saying that it would no longer automatically switch on the signal in its browsers.").

64. *See Tracking Protection Working Group*, W3C, https://www.w3.org/2011/tracking-protection/ [https://perma.cc/E7W9-CF6K] (last visited Jan. 28, 2025) ("This working group is currently closed. It closed on 17 January 2019.").

65. *See Safari 12.1 Release Notes*, APPLE: DEV., https://developer.apple.com/documentation/safari-release-notes/safari-12_1-release-notes [https://perma.cc/2HU7-6WHY] (last visited Jan. 28, 2025) ("Removed support for the expired Do Not Track standard to prevent potential use as a fingerprinting variable.").

66. *See* W3C Working Group, *Tracking Preference Expression (DNT)*, W3C (Jan. 17, 2019), https://www.w3.org/TR/tracking-dnt/ [https://perma.cc/D9VP-67VT] ("Since its last publication as a Candidate Recommendation, there has not been sufficient deployment of these extensions (as defined) to justify further advancement, nor have there been indications of planned support among user agents, third parties, and the ecosystem at large. The working group has therefore decided to conclude its work and republish the final product as this Note, with any future addendums to be published separately.").

67. John Wilander, *Intelligent Tracking Prevention*, WEBKIT BLOG (June 5, 2017), https://webkit.org/blog/7675/intelligent-tracking-prevention/ [https://perma.cc/6HXU-KGBX] ("Intelligent Tracking Prevention is a new WebKit feature that reduces cross-site tracking by further limiting cookies and other website data.").

Tracking Protection in 2019.[68] As part of these initiatives, Safari and Firefox started blocking third-party cookies in 2020 and 2022 respectively.[69] These actions signify a major shift in the browser industry toward actively prioritizing user privacy rather than reliance on industry standards or regulation. Other efforts, such as browser extensions like AdBlock Plus, uBlock Origin, and Disconnect.me, have also shaped this landscape by blocking tracking scripts and giving users greater control over their online privacy.[70]

Google's approach to online tracking and user privacy stands in stark contrast with the actions of other major browser vendors, such as Mozilla and Apple. While the other browser vendors were attempting to curtail tracking through third-party cookies, Google faced several lawsuits concerning its online tracking.[71] A lawsuit in the United Kingdom alleged that Google misled users of the Safari browser by circumventing privacy settings and placing tracking cookies without their consent, in breach its duties as a data controller under the United

---

68. Dave Camp, *Firefox Now Available with Enhanced Tracking Protection by Default Plus Updates to Facebook Container, Firefox Monitor and Lockwise*, MOZILLA: DISTILLED (June 4, 2019), https://blog.mozilla.org/en/products/firefox/firefox-now-available-with-enhanced-tracking-protection-by-default/ [https://perma.cc/T8UY-45SB] ("Firefox will be rolling out this feature, Enhanced Tracking Protection, to all new users on by default, to make it harder for over a thousand companies to track their every move.").

69. John Wilander, *Full Third-Party Cookie Blocking and More*, WEBKIT BLOG (Mar. 24, 2020), https://webkit.org/blog/10218/full-third-party-cookie-blocking-and-more/ [https://perma.cc/WC5K-2AAF] ("Cookies for cross-site resources are now blocked by default across the board. This is a significant improvement for privacy since it removes any sense of exceptions or 'a little bit of cross-site tracking is allowed.'"); *Firefox Rolls Out Total Cookie Protection by Default to More Users Worldwide*, MOZILLA: DISTILLED (June 14, 2022), https://blog.mozilla.org/en/products/firefox/firefox-rolls-out-total-cookie-protection-by-default-to-all-users-worldwide/ [https://perma.cc/T47N-8UDQ] ("Any time a website, or third-party content embedded in a website, deposits a cookie in your browser, that cookie is confined to the cookie jar assigned to only that website.").

70. *See* ADBLOCKPLUS, https://adblockplus.org [https://perma.cc/T7GP-8U8N] (last visited Jan. 28, 2025); UBLOCK ORIGIN, https://ublockorigin.com [https://perma.cc/G3LB-57BF] (last visited Jan. 28, 2025); DISCONNECT, https://disconnect.me [https://perma.cc/HPR7-S8KV] (last visited Jan. 28, 2025).

71. *See generally* Lloyd vs Google LLC [2021] UKSC 50, [1], [2019] EWCA (Civ) 1599 (appeal taken from Eng.) ("Mr Richard Lloyd—with financial backing from Therium Litigation Funding IC, a commercial litigation funder—has issued a claim against Google LLC, alleging breach of its duties as a data controller under section 4(4) of the Data Protection Act 1998 ('the DPA 1998'). The claim alleges that, for several months in late 2011 and early 2012, Google secretly tracked the internet activity of millions of Apple iPhone users and used the data collected in this way for commercial purposes without the users' knowledge or consent.").

Kingdom Data Protection Act of 1998.[72] In 2012, Google settled a similar charge with the FTC for $22.5 million, the largest civil penalty ever at that time.[73]

The divergence between Google and other major browser vendors became particularly evident in 2019 when, in response to the industry's shift towards enhanced privacy measures, Google announced—contrary to all other major browser vendors—its intention not to phase out third-party cookies in Chrome.[74] Instead, Google proposed to refine the classification of cookies to better balance privacy concerns with the needs of web publishers and advertisers.[75]

Google's reluctance to limit online tracking, potentially to preserve the status quo that benefits its advertising business, has drawn significant scrutiny. In August 2019, Google announced the Privacy Sandbox initiative, which included Federated Learning of Cohorts (FLoC) as a new method for privacy-preserving online advertising.[76] Despite this progressive step, Google simultaneously expressed reservations about the complete elimination of cookies,

---

72. *See Vidal-Hall v. Google Inc. [2015]* EWCA (Civ) 311 [3], [2014] QB 13 (Eng.) The United Kingdom Data Protection Act of 1998 implements the 1995 European Union Data Protection Directive (DPD), the predecessor of the 2016 European Union General Data Protection Regulation (GDPR). *The History of the General Data Protection Regulation*, EUR. DATA PROT. SUPERVISOR, https://www.edps.europa.eu/data-protection/data-protection/legislation/history-general-data-protection-regulation_en [https://perma.cc/63VF-FUG7] (last visited Feb. 7, 2025).

73. *See* Press Release, Fed. Trade Comm'n, Google Will Pay $22.5 Million to Settle FTC Charges it Misrepresented Privacy Assurances to Users of Apple's Safari Internet Browser (Aug. 9, 2012), https://www.ftc.gov/news-events/news/press-releases/2012/08/google-will-pay-225-million-settle-ftc-charges-it-misrepresented-privacy-assurances-users-apples https://www.ftc.gov/news-events/news/press-releases/2012/08/google-will-pay-225-million-settle-ftc-charges-it-misrepresented-privacy-assurances-users-apples [https://perma.cc/8WHV-2XJM] ("Google Inc. has agreed to pay a record $22.5 million civil penalty to settle Federal Trade Commission charges that it misrepresented to users of Apple Inc.'s Safari Internet browser that it would not place tracking 'cookies' or serve targeted ads to those users, violating an earlier privacy settlement between the company and the FTC.").

74. *See* Ben Galbraith & Justin Schuh, *Improving Privacy and Security on the Web*, GOOGLE: CHROMIUM BLOG (May 7, 2019), https://blog.chromium.org/2019/05/improving-privacy-and-security-on-web.html [https://perma.cc/K96U-C458] ("Because of this, blunt solutions that block all cookies can significantly degrade the simple web experience that you know today, while heuristic-based approaches—where the browser guesses at a cookie's purpose—make the web unpredictable for developers.").

75. *Id.* ("In the coming months, Chrome will require developers to use this mechanism to access their cookies across sites. This change will enable users to clear all such cookies while leaving single domain cookies unaffected, preserving user logins and settings. It will also enable browsers to provide clear information about which sites are setting these cookies, so users can make informed choices about how their data is used.").

76. Natalia Figas, *The Evolution of Google's Privacy Sandbox*, CLEARCODE BLOG (Oct. 21, 2024), https://clearcode.cc/blog/google-privacy-sandbox-evolution/ [https://perma.cc/L2NN-FSJH].

advocating instead for a more sophisticated classification.[77] Its argument hinged on the financial repercussions for publishers, predicting a substantial average decline in revenue of 52 percent if third-party cookies were phased out.[78] Indeed, Google's FLoC was widely criticized as potentially hampering online privacy protections even further, rather than delivering on its promises to increase privacy protections.[79] It took another six months for Google's policy to further shift. By 2020, Google announced that it would phase out third-party cookies in Chrome by 2022.[80] This plan was delayed to 2023 in 2021,[81]

77. *Id.*

78. Justin Schuh, *Building a More Private Web*, GOOGLE: KEYWORD (Aug. 22, 2019), https://blog.google/products/chrome/building-a-more-private-web/ [https://perma.cc/3NDU-VYBS] ("So today, we are announcing a new initiative to develop a set of open standards to fundamentally enhance privacy on the web. . . . [L]arge scale blocking of cookies undermine people's privacy by encouraging opaque techniques such as fingerprinting. . . . Recent studies have shown that when advertising is made less relevant by removing cookies, funding for publishers falls by 52% on average.").

79. Bennett Cyphers, *Google's FLoC Is a Terrible Idea*, ELEC. FRONTIER FOUND. (Mar. 3, 2021), https://www.eff.org/deeplinks/2021/03/googles-floc-terrible-idea [https://perma.cc/828X-ACL4] ("The technology [FLoC] will avoid the privacy risks of third-party cookies, but it will create new ones in the process. It may also exacerbate many of the worst non-privacy problems with behavioral ads, including discrimination and predatory targeting."); Alex Berke & Dan Calacci, *Privacy Limitations of Interest-Based Advertising on The Web: A Post-Mortem Empirical Analysis of Google's FLoC*, *in* CCS '22: PROCEEDINGS OF THE 2022 ACM SICSAC CONFERENCE ON COMPUTER AND COMMUNICATIONS SECURITY 337, 337, 342 (2022) ("We show how FLoC cohort ID sequences observed over time can provide this unique identifier to trackers, even with third-party cookies disabled. We estimate the number of users in our dataset that could be uniquely identified by FLoC IDs is more than 50% after 3 weeks and more than 95% after 4 weeks.").

80. Justin Schuh, *Building a More Private Web: A Path Towards Making Third Party Cookies Obsolete*, GOOGLE: CHROMIUM BLOG (Jan. 14, 2020), https://blog.chromium.org/2020/01/building-more-private-web-path-towards.html [https://perma.cc/4MLN-DY6T] ("[W]e plan to phase out support for third-party cookies in Chrome. Our intention is to do this within two years.").

81. *See* Vinay Goel, *An Updated Timeline for Privacy Sandbox Milestones*, GOOGLE: KEYWORD (June 24, 2021) [hereinafter Goel, *Updated Timeline*], https://blog.google/products/chrome/updated-timeline-privacy-sandbox-milestones/ [https://perma.cc/8GAK-LJS3] ("Chrome will phase out support for third-party cookies over a three month period finishing in late 2023.").

and to 2024 in 2023.[82] In 2024, Google announced that it would not phase out third-party cookies in Chrome after all.[83]

The history of DNT and its failure underscores the complexities of regulating online privacy, the challenges of effective industry self-regulation, and the need for alternative measures, such as legal action combined with effective industry action. It also reveals Google's challenges in protecting the privacy of Chrome users and users of its other services. Since most of Google's revenue comes from online advertising, the company has adopted a more cautious approach towards rolling out privacy-enhancing measures in Chrome that would limit flow of user data crucial for targeted advertising.[84] This creates a fundamental conflict of interest, as Chrome—as a browser—is inherently tied to the business priorities of its parent company, Google.[85] As a result, this conflict undermines user privacy, leaving Chrome users with a less private browsing experience.

### *B. Regulatory Action*

#### 1. Privacy and Data Protection Law

Google's dominance in online advertising has been at odds with self-regulatory attempts to curb online tracking. The FTC, in a set of guidelines released in 2009, urged self-regulation for online behavioral advertising with the hope that as online advertising and tracking become more commonplace, companies with significant stake in the industry would behave in a responsible manner and adopt a strong self-regulatory framework.[86] However, the FTC later realized it was a mistake to leave a topic as sensitive as user's privacy and security to entities whose financial interests were tied to the selling and monetization of data.[87] Other policymakers and legislators around the world have had a similar realization and have increasingly shifted their attention to the enactment of novel laws and regulations aimed at curbing excesses of data collection practices online.[88] These actions have direct consequences for large tech companies, such as Google, whose primary business model is tightly coupled to access to user data.[89]

---

82. *See* Anthony Chavez, *The Next Step Toward Phasing Out Third-Party Cookies in Chrome*, GOOGLE: KEYWORD (Dec. 14, 2023), https://blog.google/products/chrome/privacy-sandbox-tracking-protection/ [https://perma.cc/6CVC-BQ4T] ("We'll roll this out to 1% of Chrome users globally, a key milestone in our Privacy Sandbox initiative to phase out third-party cookies for everyone in the second half of 2024 . . . .").

83. *See* Anthony Chavez, *A New Path for Privacy Sandbox on the Web*, GOOGLE: PRIV. SANDBOX (July 22, 2024) [hereinafter Chavez, *New Path*], https://privacysandbox.com/intl/en_us/news/privacy-sandbox-update/ [https://perma.cc/FT2U-6HUV] ("In light of this, we are proposing an updated approach that elevates user choice. Instead of deprecating third-party cookies, we would introduce a new experience in Chrome that lets people make an informed choice that applies across their web browsing . . . .").

84. *See* Stephen Morris, *Google Abandons Plan to Remove Cookies from Chrome Browser*, FIN. TIMES (July 22, 2024), https://www.ft.com/content/cf56a351-9729-4b4c-8936-c512828186e6 [https://perma.cc/CK8C-JP24].

85. *See* Konrad Kollnig, *Chrome Is the Forgotten Fulcrum of Google's Dominance*, PROMARKET (Aug. 8, 2024) [hereinafter Kollnig, *Forgotten Fulcrum*], https://www.promarket.org/2024/08/08/chrome-is-the-forgotten-fulcrum-of-googles-dominance/ [https://perma.cc/D6CC-Q8R6].

In this domain, the arguably most impactful legal change was the adoption of the European Union General Data Protection Regulation (GDPR) in 2016 and its subsequent coming into force in 2018.[90] The GDPR, with its stringent data protection standards, has led to notable consequences due to the lack of legal ground for excessive data collection under Article 6 of the GDPR.[91] These standards have also affected Google. For example, the French data protection authority,

---

86. *See* FED. TRADE COMM'N, STAFF REPORT: SELF-REGULATORY PRINCIPLES FOR ONLINE BEHAVIORAL ADVERTISING iv, 47–48 (2009), https://www.ftc.gov/sites/default/files/documents/reports/federal-trade-commission-staff-report-self-regulatory-principles-online-behavioral-advertising/p085400behavadreport.pdf [https://perma.cc/9KCJ-WDGL] ("Some companies and industry groups have begun to develop new privacy policies and self-regulatory approaches, but more needs to be done to educate consumers about online behavioral advertising and provide effective protections for consumers' privacy.").

87. *See* Samuel Levine, Dir., Bureau Consumer Prot., Fed. Trade Comm'n, Surveillance in the Shadows – Third-Party Data Aggregation and the Threat to Our Liberties 2 (Sept. 21, 2023), https://www.ftc.gov/system/files/ftc_gov/pdf/cdia-sam-levine-9-21-2023.pdf [https://perma.cc/3832-U9BV] ("But I believe it was a serious mistake to favor self-regulation over establishing baseline but binding protections for the American public. And although the Commission later came to endorse privacy legislation,5 that took a decade – by which time powerful interests were already lined up against laws that could limit their ability to monetize data.").

88. *See* Craig Riddell, *International Data Privacy Laws: A Guide*, NETWRIX (Sept. 18, 2023), https://blog.netwrix.com/2023/09/18/international-data-privacy-laws/ [https://perma.cc/H8XZ-U7JN].

89. *See* Mohsin Ali Farhad, *Consumer Data Protection Laws and Their Impact on Business Models in the Tech Industry*, TELECOMM. POL'Y, Oct. 2024, at 1, 2 (2024).

90. *The History of the General Data Protection Regulation*, *supra* note 72 ("In 2016, the EU adopted the General Data Protection Regulation (GDPR), one of its greatest achievements in recent years. It replaces the 1995 Data Protection Directive which was adopted at a time when the internet was in its infancy.").

91. *See* Ben Wolford, *What Is the GDPR, the EU's New Data Protection Law?*, GDPR.EU, https://gdpr.eu/what-is-gdpr/ [https://perma.cc/224X-BECJ] (last visited Jan. 28, 2025). Article 6 of the GDPR requires a legal ground for every processing of personal data. *Id.* There are six legal grounds that may apply: consent, contractual obligation, legal obligation, vital interests of the data subject, public interest, or legitimate interest of the data controller. *Id.*

Commission Nationale de l'Informatique et des Libertés (CNIL), has imposed fines totaling 200 million euros on Google over GDPR violations related to a lack of transparency in its advertising practices.[92] The CNIL fined Google another 150 million euros around Google's consent practices regarding tracking cookies under the 2009 EU ePrivacy Directive, another EU data protection law separate from the GDPR.[93] Google has also been fined ten million euros by Spanish data protection authority Agencia Española de Protección de Datos (AEPD) and five million euros by Swedish data protection authority Datainspektionen for similar data malpractices.[94]

Rather than privacy, the GDPR—like its 1995 predecessor—puts its primary focus on the rights and control that residents of the European Economic Area (EEA) have over their own data.[95] Data protection in the European Union is a fundamental right protected under the EU Charter and exists alongside the right to privacy.[96] Both rights have somewhat independent legal histories.[97] The right to privacy arose from the right to protection of the home, which ultimately dates to Samuel Warren and Louis Brandeis's seminal 1890 paper *The Right to Privacy*.[98] Meanwhile, the right to data protection emerged to protect individuals against the overreach of state actors following Europe's experiences with the Nazi regime in the Second World War.[99] In 1930s Nazi Germany, census data was used to isolate citizens based on their nationality, native language, profession, and ethnicity.[100] The data, collected on punch cards, was counted through machines manufactured by the German subsidiary of IBM.[101] In response to a complaint regarding the 1983 German Census, the right to data protection was first formulated by the highest German court, the Bundesverfassungsgericht (Federal Constitutional Court of Germany).[102] In the *Census* judgment (Volkszählungsurteil), the court ruled that German citizens have a right to informational self-determination arising directly from the right to human dignity in the age of the data-centric society.[103]

---

92. *The CNIL's Restricted Committee Imposes a Financial Penalty of 50 Million Euros Against GOOGLE LLC*, EDPB (Jan. 21, 2019), https://edpb.europa.eu/news/national-news/2019/cnils-restricted-committee-imposes-financial-penalty-50-million-euros_en [https://perma.cc/Q2QW-W642] ("The company GOOGLE states that it obtains the user's consent to process data for ads personalization purposes. However, the restricted committee considers that the consent is not validly obtained . . . . The CNIL restricted committee publicly imposes a financial penalty of 50 Million euros against GOOGLE.").

93. Mathieu Rosemain, *Google Hit With 150 mln Euro French Fine for Cookie Breaches,* REUTERS (Jan. 6, 2022, 3:46 PM), https://www.reuters.com/world/europe/france-imposes-fines-facebook-ireland-google-2022-01-06/ [https://perma.cc/Q8MP-76VN] ("[W]hile Google and Facebook provided a virtual button to allow the immediate acceptance of cookies, there was no equivalent to refuse them as easily."); Konrad Kollnig, Reuben Binns, Pierre Dewitte, Max Van Kleek, Ge Wang, Daniel Omeiza, Helena Webb & Nigel Shadbolt, A Fait Accompli? *An Empirical Study into the Absence of Consent to Third-Party Tracking in Android Apps*, *in* PROCEEDINGS OF THE SEVENTEENTH SYMPOSIUM ON USABLE PRIVACY AND SECURITY 181, 181, 186 (2021) ("EU and UK data protection law, however, requires consent, both 1) to access and store information on users' devices and 2) to legitimate the processing of personal data as part of third-party tracking . . . .").

94. *See* Tim Rollins, *Data Privacy Alert: Spanish DPA Fines Google €10 Million*, SC MEDIA (Sept. 21, 2022), https://www.scworld.com/native/data-privacy-alert-spanish-dpa-fines-google-e10-million [https://perma.cc/NV2V-HSZQ].

95. *See* Council Directive 95/46, 1995 O.J. (L 281) ¶ 30 (EC) ("Whereas, in order to be lawful, the processing of personal data must in addition be carried out with the consent of the data subject or be necessary for the conclusion or performance of a contract binding on the data subject, or as a legal requirement, or for the performance of a task carried out in the public interest or in the exercise of official authority, or in the legitimate interests of a natural or legal person, provided that the interests or the rights and freedoms of the data subject are not overriding . . . .").

96. *See* Charter of Fundamental Rights of the European Union, art. 7, 8, 2012 O.J. (C 326).

97. A good account is provided by Orla Lynskey in her 2016 book on the subject. *See* ORLA LYNSKEY, THE FOUNDATIONS OF EU DATA PROTECTION LAW (2016).

---

98. *See* Samuel D. Warren & Louis D. Brandeis, *The Right to Privacy*, 4 HARV. L. REV. 193, 220 (1890). Samuel D. Warren, a prominent Boston attorney, and Louis D. Brandeis, a Harvard Law graduate who later became an Associate Justice of the U.S. Supreme Court, are widely credited with defining the modern concept of the right to privacy in their 1890 Harvard Law Review article. Benjamin E. Bratman, *Brandeis and Warren's* The Right to Privacy *and the Birth of the Right to Privacy*, 69 TENN. L. REV. 623, 624 (2002). They argued for a "right to be let alone" in response to technological and societal changes, particularly the emergence of intrusive journalism and photography. *Id.* This work laid the foundation for privacy law in the United States. *Id.* at 624–25.

99. *See* Olivia B. Waxman, *The GDPR Is Just the Latest Example of Europe's Caution on Privacy Rights. That Outlook Has a Disturbing History*, TIME (May 24, 2018, 7:12 PM), https://time.com/5290043/nazi-history-eu-data-privacy-gdpr/ [https://perma.cc/VT8W-YE9Z] ("We trace them back to World War II and the atrocities of the Nazis, who systematically abused private data to identify Jews and other minority groups . . . ." (quoting Anu Bradford, Professor of Law at Columbia Law School)).

100. *See id.*

101. *Id.* ("In 1930s Germany, census workers went door to door filling out punch cards that indicated residents' nationalities, native language, religion and profession. The cards were counted by the early data processors known as Hollerith machines, manufactured by IBM's German subsidiary at the time, Deutsche Hollerith Maschinen GmbH (Dehomag).").

102. *See id.*

103. *See* Grundgesetz [GG] [Basic Law], art. I(1), translation at https://www.gesetze-im-internet.de/englisch_gg/englisch_gg.html [https://perma.cc/268Q-7KAG] ("Human dignity shall be inviolable. To respect and protect it shall be the duty of all state authority."). With decreasing storage costs private and state actors increasingly adopted database systems in the 1980s. *See A Timeline of Database History & Database Management*, QUICKBASE, https://www.quickbase.com/articles/timeline-of-database-history [https://perma.cc/VP9G-UA2Q] (last visited Apr. 1, 2025). This spurred the first iteration of data protection and privacy laws at While the full impact of EU data protection legislation—and the GDPR in particular—is challenging to quantify,[104] its implementation has substantially elevated the importance of data privacy globally, prompting similar legal initiatives in Brazil,[105] China,[106] and several US states, including California.[107] The introduction of the 2018 California Consumer Privacy Act (CCPA) marked a significant step in data privacy regulation in the United States, especially given that the US Constitution does not explicitly grant a right to privacy.[108]

An important development arising from the CCPA is the creation of the Global Privacy Control (GPC), which is an advancement over the earlier DNT initiative.[109] While DNT represented an industry-led effort of self-regulation in response to growing data privacy concerns, it fell short because of its lack of enforceability, regulatory backing, and clarity over implementation as an "opt-in" or "opt-out" the time for Europe and beyond. BVerfG, 1 BvR 209, 269, 362, 420, 440, 484/83, Dec. 15, 1983, https://www.bundesverfassungsgericht.de/SharedDocs/Entscheidungen/DE/1983/12/rs19831215_1bvr020983.html [https://perma.cc/4FH6-GFAF].

104. There is a wealth of economic studies on the matter. However, these studies only focus on short-term economic effects and do not usually factor in fundamental rights considerations as externalities in their models. This omission is problematic because ignoring fundamental rights can lead to policy recommendations that prioritize immediate economic gains at the expense of long-term social welfare, equity, and justice. By treating fundamental rights as externalities, economic analyses risk undervaluing or disregarding impacts on important factors such as right to privacy and data governance. *See* Jian Jia, Ginger Zhe Jin & Liad Wagman, *The Short-Run Effects of GDPR on Technology Venture Investment* 4 (Nat'l Bureau of Econ. Rsch., Working Paper No. w25248, 2018), https://www.nber.org/papers/w25248 [https://perma.cc/NE9C-C6G2].

105. *See* Lei No. 13.709, de 14 de Agosto de 2018, Diário Oficial da União [D.O.U.] de 8.15.2018 (Braz.).

106. *See* Ken (Jianmin) Dai & Jet (Zhisong) Deng, *China's Personal Information Protection Law (PIPL)*, BLOOMBERG L. (Apr. 12, 2022), https://pro.bloomberglaw.com/brief/china-personal-information-protection-law-pipl-faqs/ [https://perma.cc/587D-WZEA].

107. *See* CAL. CIV. CODE §§ 1798.100–199 (West 2022).

108. *See* Assemb. B. 375, 2017–2018 Leg., Reg. Sess. (Cal. 2018) ("This bill would enact the California Consumer Privacy Act of 2018. Beginning January 1, 2020, the bill would grant a consumer a right to request a business to disclose the categories and specific pieces of personal information that it collects about the consumer, the categories of sources from which that information is collected, the business purposes for collecting or selling the information, and the categories of 3rd parties with which the information is shared."); Griswold v. Connecticut, 381 U.S. 479, 485 (1965).

109. *See* W3C Editor, *Global Privacy Control (GPC)*, W3C (Jan. 16, 2025), https://privacycg.github.io/gpc-spec/ [https://perma.cc/2BWR-QK69] ("This document defines a signal, transmitted over HTTP and through the DOM, that conveys a person's request to websites and services to not sell or share their personal information with third parties. This standard is intended to work with existing and upcoming legal frameworks that render such requests enforceable.").

option in web browsers.[110] In contrast, GPC builds upon the legal foundations set out by the CCPA, offering a more robust framework.[111] Under the CCPA, each business collecting users' personal data must allow consumers to submit requests to opt out of the sale and sharing through an opt-out preference signal, which allows users to communicate their preference not to be tracked across the internet through a universal and legally recognized signal.[112] This mechanism is designed to be more effective than DNT because it leverages the legally binding nature of the CCPA's privacy protections to ensure user choices are respected and implemented.[113] It remains, however, an opt-out signal; thus, it may not provide the vast majority of users with real protections given the known stickiness of defaults in behavioral economics.[114]

---

110. Konrad Kollnig, *Do-Not-Track Is Dead. Long Live Do-Not-Track!*, HUM. CENTRED COMPUTING (May 31, 2020), https://hcc.cs.ox.ac.uk/news/2020/05/31/do-not-track.html [https://perma.cc/NE2V-U7S4] ("The weakest point of DNT is its reliance on the tracking industry. Implementation in all major browsers is not enough. Websites using tracking must also respect the user's DNT setting."); *see* Sebastian Zimmeck, *Background*, GITHUB: PRIV. CMTY. GRP. (Apr. 6, 2020), https://github.com/privacycg/proposals/issues/10/ [https://perma.cc/4ZVS-4AAU] ("Previously, the Tracking Protection Working Group developed the Tracking Preference Expression (DNT). There are certainly lots of learnings that can be taken from that effort for the question here. Though, a big difference is that recipients of a DNT signal are not required to comply with it.").

111. *Take Control of Your Privacy*, GLOB. PRIV. CONTROL, https://globalprivacycontrol.org/ [https://perma.cc/JD9C-VSSM] (last visited Feb. 11, 2025) ("Under the CCPA, the GPC signal will be intended to communicate a Do Not Sell request from a global privacy control, as per CCPA-REGULATIONS §999.315 for that browser or device, or, if known, the consumer.").

112. *See California Consumer Privacy Act (CCPA)*, OFF. ATT'Y GEN., https://oag.ca.gov/privacy/ccpa [https://perma.cc/N2FE-3EUB] (Mar. 13, 2024) ("Right to opt-out of sale or sharing: You may request that businesses stop selling or sharing your personal information ('opt-out'), including via a user-enabled global privacy control. Businesses cannot sell or share your personal information after they receive your opt-out request unless you later authorize them to do so again.").

113. *See Take Control of Your Privacy*, *supra* note 111.

114. *See* United States v. Google LLC, No. 20-cv-3010, 2024 WL 3647498, at *131 (D.D.C. Aug. 5, 2024). The same discussion can also lie at the heart of this lawsuit, in which Google paid billions of dollars to be set as the standard search engine in most major web browsers. *Id.*; Ryan Bubb & Richard H. Pildes, *How Behavioral Economics Trims Its Sails and Why*, 127 HARV. L. REV. 1593, 1617 (2014) ("[B]ehavioralists have convincingly demonstrated the powerful consequences of where the default is set. In a famous study of a company that adopted automatic enrollment, Professor Brigitte Madrian and Dennis Shea found that 86% of a cohort of newly hired employees was enrolled in the company's 401(k) under automatic enrollment. In contrast, for those under the prior and more conventional opt-in approach, the comparable figure was only 37%.").

As of this writing, the GPC initiative has garnered support from several prominent browsers and extensions, including Firefox,[115] Brave,[116] DuckDuckGo,[117] and Disconnect.[118] However, the most widely used web browser, Chrome, has not implemented the functionality for Chrome users to emit the GPC signal.[119] Although CCPA regulations mandate businesses to respect the GPC, browsers are not required to implement it yet.[120] This has allowed Chrome to resist implementing the GPC despite its fast adoption as the standard privacy tool on other browsers. Indeed, Chrome's representatives have voiced relatively univocal opposition to the GPC standard becoming an official standard supported by the World Wide Web Consortium.[121] This hesitation is reminiscent of Google's previous stance on the phase-out of third-party cookies and on the DNT initiative, where its lack of support was a crucial factor in DNT's ultimate failure.[122]

---

115. *See Implementing Global Privacy Control*, MOZILLA: OPEN POL'Y & ADVOC., https://blog.mozilla.org/netpolicy/2021/10/28/implementing-global-privacy-control/ [https://perma.cc/JR4D-MLZK] (Dec. 2021) ("UPDATE, December 2021: Global Privacy Control is now available in the general release version of Firefox (Firefox 95).").

116. Peter Snyder & Anton Lazarev, *Global Privacy Control, a New Privacy Standard Proposal*, BRAVE: WEBSTANDARDS@BRAVE, https://brave.com/web-standards-at-brave/4-global-privacy-control/ [https://perma.cc/3X9U-ZLFV] (Sept. 8, 2023) ("We are also excited to announce our implementation of the GPC proposal, available today in the Nightly channel of our Desktop browser and in our Android browser beta release.").

117. *See Global Privacy Controls (GPC) in DuckDuckGo*, DUCKDUCKGO, https://duckduckgo.com/duckduckgo-help-pages/privacy/gpc/ [https://perma.cc/LC34-ZKPH] (last visited Jan. 29, 2025) ("In order to provide additional protection for situations where the websites otherwise sell or share your data with other companies that may profit or benefit from it (such as selling data to advertisers or data brokers after your visit), we decided to help pioneer the Global Privacy Control (GPC) standard.").

118. *See Introducing Global Privacy Control*, DISCONNECT: DISCONNECT ANNOUNCEMENTS (Oct. 7, 2020), https://blog.disconnect.me/introducing-global-privacy-control/ [https://perma.cc/ZYW3-QLFN].

119. *Is Your Site Ready for New Regulations Going Into Effect this Month?*, RUBYLAW (Jan. 20, 2023), https://www.rubylaw.com/innovate/client-alert-ca-202301-CPRA-and-Global-Privacy-Controls-GPC.html [https://perma.cc/CCM5-G4CH] ("[I]t's important to note that Google Chrome (which accounts for ~65% of the total browser market share), does not yet support the GPC signal by default.").

120. *See Understanding Global Privacy Control (GPC): What It Is and Why It Matters*, USERCENTRICS (July 10, 2024), https://usercentrics.com/knowledge-hub/what-is-global-privacy-control/ [https://perma.cc/6D6C-W8VX] ("Browsers do not currently have to have the functionality built in, and some websites or apps may not be capable of enabling the GPC to function."); Jon Brodkin, *Calif. Governor Vetoes Bill Requiring Opt-Out Signals for Sale of User Data*, ARS TECHNICA (Sept. 24, 2024), https://arstechnica.com/tech-policy/2024/09/calif-gov-vetoes-attempt-to-require-new-privacy-option-in-browsers-and-oses/ [https://perma.cc/8B2B-6JKW].

121. *See* Priv. Cmty. Grp., 2023-08-24 Privacy CG Meeting Minutes (Aug. 25, 2023), https://github.com/privacycg/meetings/blob/9bfd2fcdfe76ac9393ba8a046513bde6118c329f/2023/telcons/08-24-minutes.md [https://perma.cc/6HEF-ASG7]. The World Wide Web Consortium is one of the main standardization organizations for technical standards followed by the web browsers. Alexander S. Gillis, *W3C (World Wide Web Consortium)*, TECHTARGET, https://www.techtarget.com/whatis/definition/W3C-World-Wide-Web-Consortium#:~:text=The%20W3C%20(World%20Wide%20Web%20Consortium)%20is%20an%20international%20organizat

Google's reluctance to embrace GPC without regulatory compulsion raises concerns about the initiative's future success. Given Google's substantial influence in the browser market, its support for GPC is pivotal.[123] Without Google's participation, GPC may struggle to achieve widespread impact, potentially facing a similar fate as DNT. This scenario underscores the broader challenge of aligning Google's business interests with the increasing demand for stronger privacy protections. Unless mandated by regulation, Google has limited incentives to alter its current data practices and prioritize user privacy.

### 2. Antitrust and Competition Law

#### *a. Academic Literature*

Current data protection and privacy laws have had a limited impact, and in the absence of a comprehensive federal privacy law in the US, many scholars suggest turning to other long-established legal frameworks such as antitrust law for renewed focus.[124]

In their 2020 report "Roadmap for a Digital Advertising Monopolization Case Against Google," economist Fiona M. Scott Morton and attorney David C. Dinielli focus on the traditional price effects paradigm of antitrust law.[125] They assert that Google has established

---

ion,guidelines%20for%20web%20technologies%20worldwide [https://perma.cc/EK4T-U95Z] (Aug. 2022).

122. *Google Phase Out of Third-Party Cookies: Impacts and Solutions*, USERCENTRICS (May 7, 2024), https://usercentrics.com/knowledge-hub/google-third-party-cookies/ [https://perma.cc/JUS6-6FLV].

123. *See id.*

124. *See* STEPHEN P. MULLIGAN & CHRIS D. LINEBAUGH, CONG. RSCH. SERV., R45631, DATA PROTECTION LAW: AN OVERVIEW 30 (2019); *Antitrust and Cartels*, EUR. COMM'N, https://competition-policy.ec.europa.eu/antitrust-and-cartels_en [https://perma.cc/GS7R-NE27] (last visited Jan. 29, 2025) ("Antitrust rules prohibit agreements between market operators that would restrict competition, and the abuse of dominance. European Antitrust policy is developed from two central rules set out in the Treaty on the Functioning of the European Union . . . .").

125. FIONA M. SCOTT MORTON & DAVID C. DINIELLI, ROADMAP FOR A DIGITAL ADVERTISING MONOPOLIZATION CASE AGAINST GOOGLE 3 (2020), https://omidyar.com/wp-content/uploads/2020/09/Roadmap-for-a-Case-Against-Google.pdf [https://perma.cc/5C4Z-YTH8] ("The end result is that, in the digital advertising market, virtually all roads lead through Google. Google now performs every function that connects advertisers to publishers. Using the insurmountable data advantage it derives from its search engine and other properties as well as extensive control over the AdTech stack, enabling the company to capture a substantial portion of advertising budgets, with estimates between 40–60%.[126] According to their analysis, Google's dominance is a result of various anticompetitive practices designed to stifle rival participation in AdTech.[127] These tactics include impeding interoperability, leveraging its search engine dominance to coerce advertisers into using Google's display products, and granting exclusive access to YouTube's ad inventory solely through Google's tools.[128] Scott Morton and Dinielli's analysis concludes that Google's practices have resulted in wide-ranging detrimental effects on the digital market.[129] Advertisers grapple with inflated costs, publishers face declining revenues, competitors are marginalized, and consumers suffer from increased prices for goods and services, stifled innovation, lowered-quality content, and eroded privacy.[130] Their viewpoint offers a practical lens for examining Google's influence in online advertising.[131]

Professors Reuben Binns and Elettra Bietti provide a critical analysis of the mergers and acquisitions (M&A) landscape within the

---

contract and design choices, Google has made it nearly impossible for publishers and advertisers to do business with each other except through Google.").

126. *Id.* at 10 ("The SSP market is somewhat less concentrated, but Google's AdX product has a significant share, which the CMA estimates at 40–60%.").

127. *Id.* at 17 ("When viewed collectively, the conduct suggests a long-term strategy to occupy, through acquisitions, the entirety of the ad tech stack that connects buyers to sellers, and then to use its presence across the stack, its data, and its control of the flow of payments to exclude and prevent entry of competitors, raise rivals' costs, and force buyers and sellers to rely on Google services to effectuate sales. Google has used exclusivity and the denial of interoperability, and leveraged power across the stack to disadvantage competitors and advantage itself. Google's opacity keeps many of the details of its conduct secret, even from customers, which suppresses competition and helps Google to maintain dominance.").

128. *Id.* at 22 ("Google makes its valuable YouTube inventory available to buyers using the ad tech stack exclusively through its own demand side services. This is a contractual way to deny interoperability.").

129. *Id.* at 31.

130. *Id.* at 35 ("Google's conduct denies competitive payments to publishers and others who provide traffic. These underpayments undermine the incentive of publishers—news organizations, for example—to produce valuable and high-quality content. Because consumers themselves do not directly pay for most content on the internet (it is free), those content providers support their businesses through advertising. The returns from advertising directly incentivize more and higher quality content. Consumers directly benefit from more and higher quality content. Therefore, when Google's conduct suppresses publisher returns, consumers are harmed.").

131. *Id.* at 36 ("The competitive harms the CMA describes, and that we outline above, are not mere academic matters of interest only to economists and technology insiders. . . . Consumers deserve the best these types of services can offer, as do future consumers, who should inherit functioning digital markets, the conditions of which we can change today.").

AdTech and online tracking industry.[132] Their research contributed novel quantitative evidence and brought to light the increasing market concentration.[133] Among other aspects, the study showed that this increase in market concentration was fueled by Google's acquisitions of platforms like DoubleClick, YouTube, Firebase, and AdMob.[134] By 2018, this trend of consolidation, as revealed in related work, had already reached a level potentially warranting scrutiny by EU competition authorities.[135] Yet, Binns and Bietti note that many of these significant transactions have escaped scrutiny by both EU and US competition regulators.[136] Traditionally, regulators did not extensively consider privacy concerns and data concentration in their assessments, often focusing instead on traditional market dynamics.[137] Binns and Bietti advocate for a more comprehensive approach to antitrust law in digital markets, one that transcends the conventional consumer welfare standard primarily based on price effects.[138] The insights from Binns and Bietti's work are particularly pertinent in the context of ongoing investigations and legal actions against dominant companies like Google. Their research calls for a recalibration of antitrust frameworks to better address the multifaceted challenges and implications of digital market dominance.[139]

---

132. Reuben Binns & Elettra Bietti, *Dissolving Privacy, One Merger at a Time: Competition, Data and Third Party Tracking*, 36 COMPUT. L. & SEC. REV. 1, 1, 17 (Apr. 24, 2020) [hereinafter Binns & Bietti, *Dissolving Privacy*] ("In the years since the acquisition, the Alphabet companies (including Google and DoubleClick) have expanded the reach of their third party tracking capability to encompass the majority of all websites and apps on the Android platform."). "Reuben Binns is an Associate Professor of Human Centred Computing [at the University of Oxford] . . . focusing on data protection, machine learning, and [technology regulation]." *Reuben Binns*, U. OXFORD, https://www.cs.ox.ac.uk/people/reuben.binns/ [https://perma.cc/8SZK-P8WR] (last visited Feb. 11, 2025). He has served as a Postdoctoral Research Fellow in AI at the UK's Information Commissioner's Office. *Id.* Elettra Bietti is an Assistant Professor of Law and Computer Science at Northeastern University, specializing in the regulation of digital technologies, data, and digital platforms. *Elettra Bietti*, NE. SCH. L., https://law.northeastern.edu/faculty/bietti/ [https://perma.cc/M7RT-RSNA] (last visited Feb. 11, 2025). She is also a Faculty Associate at Harvard's Berkman Klein Center for Internet and Society. *Id.*

133. Rueben Binns & Elettra Bietti, *Dissolving Privacy, One Merger at a Time: Competition, Data and Third Party Tracking*, DLI@CORNELLTECH (Oct. 7, 2019) [hereinafter Binns & Bietti, DLI@CORNELLTECH], https://www.dli.tech.cornell.edu/post/dissolving-privacy-one-merger-at-a-time-competition-data-and-third-party-tracking [https://perma.cc/7A42-MKF6] ("We came to these conclusions in the paper through an empirical exercise. First, we analysed 10,000 websites and apps to uncover which third party trackers were present on them. Then we analysed the firms behind those trackers and their corporate ties to one another.").

134. *Id.* ("We found that the most commonly found trackers on web and mobile included those owned by Alphabet / Google (e.g. Youtube, Firebase, Admob, and Google Ads (previously DoubleClick)), Facebook, Twitter, Verizon and Microsoft. Second, after excluding any third party trackers which were present on less than 5 websites or apps, we identified 42 mergers and acquisitions between the remaining firms.").

135. Reuben Binns, Jun Zaho, Max Van Kleek & Nigel Shadbolt, *Measuring Third-Party Tracker Power Across Web and Mobile*, 18 ACM TRANSACTIONS ON INTERNET TECH., no. 52, 2018, at 1, 1 ("Third-party networks collect vast amounts of data about users via websites and mobile applications. Consolidations among tracker companies can significantly increase their individual tracking capabilities, prompting scrutiny by competition regulators.").

136. Binns & Bietti, DLI@CORNELLTECH, *supra* note 133 ("We found that only 21 transactions had been scrutinized by one or more of these five competition authorities. Only five of the 42 transactions were the subject of in-depth competition law investigations by one or more of these authorities: seven full-merits decisions in total. In only four of these seven decisions we found noteworthy analyses on questions surrounding data and market power.").

Binns and Bietti's work falls within the Neo-Brandeisian school of thought, which emerged in the 2010s and was significantly influenced by the works of legal scholars like FTC Commissioner Lina Khan and legal scholar Tim Wu.[140] Khan's seminal paper, *Amazon's Antitrust Paradox*, challenges the traditional antitrust framework focused mainly on consumer welfare through pricing.[141] Khan argues for a broader, more holistic approach to antitrust laws—one that accounts for the unique challenges posed by digital market behemoths like Amazon and Google.[142] Khan emphasizes the need to consider factors such as market dominance, the role of data in reinforcing market power, and the implications for competition and innovation.[143] This perspective is particularly relevant in the context of Google, a company whose extensive market reach and data practices raise similar concerns to those Khan identifies in Amazon.[144] Khan's advocacy for a broader interpretation of antitrust laws, to include aspects like data privacy and market concentration, aligns with the growing scrutiny of Google's dominance in various digital markets and its broader implications for consumer welfare in the digital era.[145]

---

137. *Id.* ("In our analysis of these decisions, we found a progressive evolution of antitrust authorities' approach toward greater concern for the effects of data concentrations, and yet an insufficient consideration – or in most cases a complete disregard – of third party tracking and their effects on consumers.").

138. Binns & Bietti, *Dissolving Privacy*, *supra* note 132, at 16 ("By overtly omitting privacy considerations, the Commission in fact has failed to fulfill its mandate to protect consumer welfare (which, as argued above, must be understood as encompassing more than just price and quality of search), while also protecting and celebrating potentially harmful data reliant business models in the advertising ecosystem."); *see* U.S. DEP'T OF JUST. & FED. TRADE COMM'N, MERGER GUIDELINES 1–2 (2023), https://www.ftc.gov/system/files/ftc_gov/pdf/2023_merger_guidelines_final_12.18.2023.pdf [https://perma.cc/EF38-RHUA]. The new merger guidelines released by the DOJ and FTC attempt to broaden the scope of agency-led merger investigations. *See id.* at 1–4. Now the agencies are expected to go far and beyond price effects—they must look to competition, innovation, and the general effect of the merger on other platforms. *See id.*

139. *See* Binns & Bietti, DLI@CORNELLTECH, *supra* note 133.

140. TIM WU, THE CURSE OF BIGNESS: ANTITRUST IN THE NEW GILDED AGE 127–40 (2018).

141. Lina M. Khan, Note, *Amazon's Antitrust Paradox,* 126 YALE. L.J. 710, 710, 737 (2017) ("This Note argues that the current framework in antitrust—specifically its pegging competition to 'consumer welfare,' defined as short-term price effects—is unequipped to capture the architecture of market power in the modern economy. . . . The current framework in antitrust fails to register certain forms of anticompetitive harm and therefore is unequipped to promote real competition—a shortcoming that is illuminated and amplified in the context of online platforms and data-driven markets.").

142. *Id.* at 717 ("Rather than pegging competition to a narrow set of outcomes, this approach would examine the competitive process itself. Animating this framework is the idea that a company's power and the potential anticompetitive nature of that power cannot be fully understood without looking to the structure of a business and the structural role it plays in markets.").

### *b. Regulators and Courts*

Khan's academic work contributed to antitrust regulators around the world acting against market power.[146] In a notable development against Google in 2019, the UK's Competition and Markets Authority (CMA) launched a market study into online platforms and digital advertising.[147] This investigation primarily focused on Google and Facebook, with a goal of investigating concerns regarding potential market power abuses.[148] At its conclusion in 2020, the study scrutinized Google's initiative to phase out third-party

143. *Id.* at 746 ("Attention to structural concerns and the competitive process are especially important in the context of online platforms, where price-based measures of competition are inadequate to capture market dynamics, particularly given the role and use of data.").

144. *Id.* at 764 ("[T]he types of consumer behavior that internet firms can access—how long you hover your mouse on a particular item, how many days an item sits in your shopping basket before you purchase it, or the fashion blogs you visit before looking for those same items through a search engine—is uncharted ground.").

145. *Id.* at 746 ("In practice, adopting this approach would involve assessing a range of factors that give insight into the neutrality of the competitive process and the openness of the market. These factors include: (1) entry barriers, (2) conflicts of interest, (3) the emergence of gatekeepers or bottlenecks, (4) the use of and control over data, and (5) the dynamics of bargaining power.").

146. *See id.* at 721–22.

147. *See* Competition and Markets Authority, *Online Platforms and Digital Advertising Market Study*, GOV.UK, https://www.gov.uk/cma-cases/online-platforms-and-digital-advertising-market-study [https://perma.cc/U49Y-35ED] (July 1, 2020).

148. COMPETITION & MKTS. AUTH., ONLINE PLATFORMS AND DIGITAL ADVERTISING MARKET STUDY: STATEMENT OF SCOPE 2 (2019), https://assets.publishing.service.gov.uk/media/5d1b297e40f0b609dba90d7a/Statement_of_Scope.pdf [https://perma.cc/ETR8-6GME] ("Two suppliers in particular, Google and Facebook (and their respective subsidiaries, such as YouTube and Instagram) hold leading positions in the market for online advertising in the UK, with the majority of digital advertising revenue in the UK split between these two businesses. In turn, digital advertising comprises the substantial majority of the revenues of both of these companies.").

cookies.[149] Among its findings, the CMA report emphasized the third-party cookie phase-out's substantial immediate impact on publisher revenues, with a potential reduction of up to 70%.[150]

Prompted by its earlier findings, in 2021, the CMA initiated a formal review of Google's planned phase-out of third-party cookies.[151] This move by the CMA underscored the regulatory concerns surrounding Google's influence on the advertising market.[152] In a quick turn of events, Google announced postponement of the third-party cookie phase-out to 2023, a decision reflective of the mounting regulatory and industry pressures.[153] Further evolving its approach, Google replaced the FLoC with the Topics API in 2022.[154]

Subsequently, in February 2022, the CMA accepted a set of commitments from Google concerning the third-party cookie phase-out.[155] These commitments were designed to address key concerns, including supporting publishers' revenue generation, and enhancing user transparency and control over data.[156] To ensure compliance, the CMA appointed a monitoring trustee in March 2022.[157] However, in a further adjustment to its timeline, Google announced in 2022 an additional delay in the third-party cookie phase-out, extending it to 2024.[158] In 2024, Google announced that it would not phase-out third-party cookies in Chrome after all.[159]

---

149. COMPETITION & MKTS. AUTH., ONLINE PLATFORMS AND DIGITAL ADVERTISING MARKET STUDY: MARKET STUDY FINAL REPORT 109, 434 (2020), https://assets.publishing.service.gov.uk/media/5fa557668fa8f5788db46efc/Final_report_Digital_ALT_TEXT.pdf [https://perma.cc/2QPK-KWZR].

150. *See* COMPETITION & MKTS. AUTH., ONLINE PLATFORMS AND DIGITAL ADVERTISING MARKET STUDY: APPENDIX F: THE ROLE OF DATA IN DIGITAL ADVERTISING F54 (2020), https://assets.publishing.service.gov.uk/media/5fe495438fa8f56af97b1e6c/Appendix_F_-_role_of_data_in_digital_advertising_v.4_WEB.pdf [https://perma.cc/4KGL-AVJ2].

151. *See* Competition and Markets Authority, *Investigation into Google's 'Privacy Sandbox' Browser Changes*, GOV.UK [hereinafter Competition and Markets Authority, *Investigation*], https://www.gov.uk/cma-cases/investigation-into-googles-privacy-sandbox-browser-changes#case-timetable [https://perma.cc/8PN4-LJRL] (Dec. 20, 2024).

152. Press Release, Competition & Mkts. Auth., CMA to Investigate Google's 'Privacy Sandbox' Browser Changes (Jan. 8, 2021), https://www.gov.uk/government/news/cma-to-investigate-google-s-privacy-sandbox-browser-changes [https://perma.cc/5MBF-XK6W] ("The CMA has received complaints including from Marketers for an Open Web Limited, a group of newspaper publishers and technology companies, which allege that, through the proposals, Google is abusing its dominant position.").

153. Goel, *Updated Timeline*, *supra* note 81 ("Subject to our engagement with the United Kingdom's Competition and Markets Authority (CMA) and in line with the commitments we have offered, Chrome could then phase out third-party cookies over a three month period, starting in mid-2023 and ending in late 2023.").

154. Vinay Goel, *Get To Know the New Topics API for Privacy Sandbox*, GOOGLE: KEYWORD (Jan. 25, 2022) [hereinafter Goel, *Topics API*], https://blog.google/products/chrome/get-know-new-topics-api-privacy-sandbox/ [https://perma.cc/G9WN-FATL] ("[W]e're announcing Topics, a new Privacy Sandbox proposal for interest-based advertising. Topics was informed by our learning and widespread community feedback from our earlier FLoC trials, and replaces our FLoC proposal."). API is short for Application Programming Interface and refers to a set of pre-defined and shared instructions to interact between different pieces of software. Michael Goodwin, *What Is an API (Application Programming Interface)?*, IBM (Apr. 9, 2024), https://www.ibm.com/think/topics/api [https://perma.cc/2ZSV-SDGN]. As such, APIs are akin to glue that pieces together software products from different developers and allows them to interact. *See id.* In the case of the Topics API, this API enables website developers to access a set of user-specific topics stored within the Chrome browser and learn about characteristics of the Chrome or website user in a more privacy-preserving manner. Goel, *Topics API*, *supra*.

In the United States, Google is currently subject to several investigations and legal battles that put to test both traditional and Neo-Brandeisian legal theories. In October 2020, the US Department of Justice (DOJ), along with eleven state Attorney Generals, initiated a lawsuit against Google, alleging the company unlawfully maintains monopolies in search and search advertising through anticompetitive and exclusionary practices.[160] These practices, as stated by the DOJ,

---

155. *See* Competition and Markets Authority, Privacy Sandbox Google Commitments Offer 4 (Feb. 4, 2022), https://assets.publishing.service.gov.uk/media/62052c6a8fa8f510a204374a/100222_Appendix_1A_Google_s_final_commitments.pdf [https://perma.cc/4QTD-VSKX].

156. *Id.* at 54 ("In this regard, the CMA considers that the concerns that third parties have expressed to it regarding the impact that the Privacy Sandbox Proposals are likely to have in the future, reflect in part: (a) the asymmetry of information between Google and third parties regarding the development of the Privacy Sandbox Proposals, including the criteria that Google will use to assess different design options and evidence relating to their effectiveness against these criteria; and (b) a lack of confidence on the part of third parties regarding Google's intentions in developing and implementing the Privacy Sandbox Proposals, given the commercial incentives that Google faces in developing Google's Proposals and the lack of independent scrutiny of Google's Proposals.").

157. *See* Competition and Markets Authority, CMA Appoints Monitoring Trustee to Supervise Commitments in Relation to Google's 'Privacy Sandbox' Browser Changes 1 (Mar. 23, 2022), https://assets.publishing.service.gov.uk/media/6239a8468fa8f540f5c3c068/220323_-_CMA_Appointment_of_Monitoring_Trustee.pdf [https://perma.cc/AH6K-WT4W] ("The CMA has today (23 March 2022) approved under paragraph 32(b) of the commitments the appointment by Google of ING Bank N.V. as Monitoring Trustee to monitor compliance with the abovementioned provisions of the commitments.").

158. Anthony Chavez, *Expanding Testing for the Privacy Sandbox for the Web*, GOOGLE: KEYWORD (July 27, 2022), https://blog.google/products/chrome/update-testing-privacy-sandbox-web/ [https://perma.cc/WWG4-T22P] ("[W]e now intend to begin phasing out third-party cookies in Chrome in the second half of 2024.").

159. Chavez, *New Path*, *supra* note 83 ("Instead of deprecating third-party cookies, we would introduce a new experience in Chrome that lets people make an informed choice that applies across their web browsing, and they'd be able to adjust that choice at any time.").

160. *See* Press Release, U.S. Dep't of Just., Justice Department Sues Monopolist Google for Violating Antitrust Laws (Oct. 20, 2022), https://www.justice.gov/opa/pr/justice-department-sues-monopolist-google-violating-antitrust-laws [https://perma.cc/5FLX-AX4H].

include exclusivity agreements that stifle competition.[161] A similar lawsuit was filed in December 2020 by a coalition of states led by Colorado and Nebraska, accusing the tech giant of monopolizing internet search and search advertising, as well as engaging in anticompetitive behavior.[162] After the merger of the two lawsuits for pretrial and discovery purposes,[163] the trial focused on Google's use of contractual agreements to perpetuate its market dominance and its strategies to make Google the default search engine on mobile devices.[164] Following a nine-week bench trial, the court ruled that Google illegally maintained a monopoly over search services and search advertisement in violation of Section 2 of the Sherman Act.[165] In light

---

> Today, the Department of Justice — along with eleven state Attorneys General — filed a civil antitrust lawsuit in the U.S. District Court for the District of Columbia to stop Google from unlawfully maintaining monopolies through anticompetitive and exclusionary practices in the search and search advertising markets and to remedy the competitive harms. . . .
>
> . . . .
>
> As alleged in the Complaint, Google has entered into a series of exclusionary agreements that collectively lock up the primary avenues through which users access search engines, and thus the internet, by requiring that Google be set as the preset default general search engine on billions of mobile devices and computers worldwide and, in many cases, prohibiting preinstallation of a competitor. In particular, the Complaint alleges that Google has unlawfully maintained monopolies in search and search advertising . . . .

*Id.*

161. *Id.* ("These and other anticompetitive practices harm competition and consumers, reducing the ability of innovative new companies to develop, compete, and discipline Google's behavior.").

162. William Padmore, *Nebraska, 30 Other States, File Antitrust Lawsuit Against Google*, NEB. PUB. MEDIA (Dec. 17, 2020), https://nebraskapublicmedia.org/es/news/news-articles/nebraska-30-other-states-file-antitrust-lawsuit-against-google/ [https://perma.cc/X428-9TXG].

163. United States v. Google LLC, No. 20-cv-3010, 2024 WL 3647498, at *33 (D.D.C. Aug. 5, 2024); *see Timeline on Monopoly Lawsuit Regarding Search and Search Advertising Market*, AM. ECON. LIBERTIES PROJECT [hereinafter *Timeline*], https://www.economicliberties.us/colorado-v-google/ [https://perma.cc/L49E-KRKF] (last visited Jan. 29, 2025) ("January 8 . . . A D.C. federal judge consolidates the States' case with that of the Dept. of Justice. The cases are combined for pretrial purposes, including discovery.").

164. *Timeline*, *supra* note 163 ("Like the Justice Department's case, the States allege that Google prevents consumers from using search competitors through 'exclusionary agreements' that also deny competitors access to search distribution. In these agreements, the States say, Google makes billion-dollar payoffs to device manufacturers that force its search engine to be the default on most mobile devices.").

165. United States v. Google LLC, No. 20-cv-3010, 2024 WL 3647498, at *3 (D.D.C. Aug. 5, 2024) ("After having carefully considered and weighed the witness testimony and evidence, the court reaches the following conclusion: Google is a monopolist, and it has acted as one to maintain its monopoly. It has violated Section 2 of the Sherman Act."). *See generally The Antitrust Laws*, of this ruling, the US DOJ submitted its suggested behavioral and structural remedies against Google's monopolistic behavior.[166] The Court's ruling mirrored the approach taken in the 2022 Google Android ruling by the Court of Justice of the European Union, in which Google was found guilty of abusing its dominant market position with Android to reinforce its dominance in general internet search services.[167] This ruling was a confirmation of prior investigations by the European Commission (EC) and imposed a €4.125 billion fine on Google.[168]

Beyond these cases, Google faces a spectrum of additional legal challenges. In 2023, the US DOJ escalated its actions against Google with a groundbreaking lawsuit seeking the breakup of Google's advertising business.[169] This move follows a similar lawsuit filed by a coalition of ten states led by Texas in 2020 and marks a significant step toward a more aggressive regulatory stance in antitrust enforcement.[170]

---

FED. TRADE COMM'N, https://www.ftc.gov/advice-guidance/competition-guidance/guide-antitrust-laws/antitrust-laws [https://perma.cc/KE3U-G835] (last visited Jan. 29, 2025) ("Congress passed the first antitrust law, the Sherman Act, in 1890 as a 'comprehensive charter of economic liberty aimed at preserving free and unfettered competition as the rule of trade.' In 1914, Congress passed two additional antitrust laws: the Federal Trade Commission Act, which created the FTC, and the Clayton Act. With some revisions, these are the three core federal antitrust laws still in effect today. . . . The Sherman Act outlaws 'every contract, combination, or conspiracy in restraint of trade,' and any 'monopolization, attempted monopolization, or conspiracy or combination to monopolize.'").

166. The proposed judgment seeks remedies including divestment of Chrome and Android, restrictions on Google to sign exclusivity agreements, and data sharing with competitors. Siladitya Ray, *DOJ's Proposal to Stop Google's Search Monopoly Includes Forced Sale of Chrome, Changes to Android Search*, FORBES, https://www.forbes.com/sites/siladityaray/2024/11/21/force-sale-of-chrome-and-big-changes-to-android-search-the-dojs-proposal-to-crackdown-on-googles-search-monopoly/ [https://perma.cc/LS6F-W3SR] (Nov. 21, 2024, 7:20 AM); Barry Schwartz, *Google Files Its Proposed Remedies in DOJ's Monopoly Case*, SEARCH ENGINE LAND (Dec. 21, 2024, 7:45 PM), https://searchengineland.com/google-files-its-proposed-remedies-in-dojs-monopoly-case-449743 [https://perma.cc/4KPU-LQJP]; *see also infra* Part 4.

167. *See* Nicholas Levy, Henry Mostyn, Paul Stuart & Patrick Todd, *General Court Partially Annuls European Commission Decision in Google Android*, CLEARY GOTTLIEB (Oct. 20, 2022), https://www.clearygottlieb.com/news-and-insights/publication-listing/general-court-partially-annuls-european-commission-decision-in-google-android [https://perma.cc/887L-WXRP].

168. *See* Press Release, Eur. Union Ct. of Just., The General Court Largely Confirms the Commission's Decision that Google Imposed Unlawful Restrictions on Manufacturers of Android Mobile Devices and Mobile Network Operators in Order to Consolidate the Dominant Position of its Search Engine (Sept. 14, 2022), https://curia.europa.eu/jcms/upload/docs/application/pdf/2022-09/cp220147en.pdf [https://perma.cc/35V5-D8SK] ("In order better to reflect the gravity and duration of the infringement, the General Court considers it appropriate however to impose a fine of €4.125 billion on Google . . . .").

169. Complaint at 4, United States v. Google LLC, No.1:23-cv-00108 (E.D. Va. Jan. 24, 2023).

170. Complaint at 8, Texas v. Google LLC, No. 4:20-CV-957 (E.D. Tex. Dec. 16, 2020).

In July 2023, Gannett, the media conglomerate owning USA Today and various other news outlets, filed a lawsuit against Google for its monopolization of advertising technology markets and deceptive commercial practices.[171] Their lawsuit revolves around the discrepancy between the lucrative growth of the digital advertising business and unprofitable publishers.[172] Despite the digital advertising business growing to $200 billion and rapidly increasing numbers of online news readers, publishers are unable to reap the benefits due to Google's monopolistic practices.[173] Gannett alleges that Google controls how publishers sell their ad slots, to whom they sell it to, and dictates the price at which the ad slots are sold.[174] This results in a revenue model that negatively affects publisher revenues to such a degree that Gannett was forced to shutter more than 170 publications across the United States.[175]

The lawsuits do not end there—Google has been and continues to be subject to a variety of lawsuits targeting its privacy practices.[176] These cases reflect increasing concerns over data protection and privacy in the digital age and underscore the broader societal and legal scrutiny regarding how tech companies, particularly those with significant market power like Google, handle user data.

---

171. *See Gannett Files Federal Lawsuit Against Google*, GANNETT (June 20, 2023), https://gannett.com/pr/gannett-files-federal-lawsuit-against-google/ [https://perma.cc/VG2Q-G9ZL] ("Google has monopolized market trading to their advantage and at the expense of publishers, readers and everyone else." (quoting Michael Reed, Gannett Chairman and Chief Executive Officer)).

172. *Id.* ("In 2022, Google made upwards of $30 billion in revenue from the sale of ad space on publishers' websites which was six times the digital advertising revenue of all U.S. news publications, combined.").

173. Complaint for Damages & Injunctive Relief at 2, Gannett Co. v. Google LLC, No. 1:23-cv-05177 (S.D.N.Y. June 20, 2023) ("Today, online digital advertising is a $200 billion business — a nine-fold increase since 2009. Yet, despite the opportunity for publishers to produce more news content and earn more revenue, news publications' advertising revenue has declined by nearly 70% over the same timeframe. As a result, newspaper newsroom employment has dropped by more than half, and more than 20% of all newspapers have closed. The circulation of daily and weekly newspapers has decreased by more than 40%.").

174. *Id.* ("Google controls how publishers sell their ad slots, and it forces publishers to sell growing shares of that ad space to Google at depressed prices. The result is dramatically less revenue for publishers and Google's ad-tech rivals, while Google enjoys exorbitant monopoly profits.").

175. *Id.* at 2, 6 ("Gannett has not been spared. Since 2019 — just the past four years — over 170 Gannett publications have been shuttered. For Gannett's largest remaining publications, average daily circulation fell by nearly 20% between 2020 and 2021 *alone*. The result is less news where it is needed most. Communities throughout the United States now do not have a suitable local paper to advise on local events, hold local officials to account, or encourage the civic bonds that are paramount in an increasingly polarized country. . . . Gannett brings this antitrust action for compensation and for injunctive relief to restore competition in the monopolized markets and safeguard news content for readers.").

176. *See, e.g.*, Gaos v. Google Inc., No. 5:10-CV-4809, 2012 WL 1094646 (N.D. Cal. Mar. 29, 2012); Patacsil v. Google, Inc., No. 3:18-cv-05062, 2018 WL 3957362 (N.D. Cal. Aug. 17, 2018); Rodriguez v. Google LLC, No. 20-cv-04688, 2024 WL 38302 (N.D. Cal. Jan. 3, 2024); Brown v. Google, LLC, No. 4:20-cv-3664, 2023 WL 4336718 (N.D. Cal. May 1, 2023).

In the European Union, Google's advertising business is under intense scrutiny, mirroring similar concerns as in the United States. In addition to the €4.125 penalty imposed in the *Google Android* case, the EC has also put forth a proposal expressing the necessity of dismantling Google's online advertising operations.[177] This radical proposal stems from the Commission's assessment of Google's conflict of interest arising from its control of the smartphone market, dominance in online search services, and its overwhelming market power, which, in their view, stifles free competition within the advertising sector.[178] The EC's critique centers on Google's dual role: it operates not only as a seller of digital advertising space, exemplified by platforms like YouTube, but also as an intermediary between advertisers and advertising space.[179] Google's dominance in the EU market is pronounced in two areas. First, it is pronounced in publisher ad servers, where its "Google Ad Manager" service holds a significant position.[180] Second, it is dominant in the realm of programmatic ad buying tools for the open web, facilitated by "Google Ads" and "Google Display & Video 360."[181] The Commission contends that Google's practices, which include favoring its own ad exchange in Google Ads and Google Display & Video 360, constitute anticompetitive conduct, reinforcing the need for a potential breakup to restore market balance.[182]

---

177. *See* European Commission Press Release IP/23/3207, Antitrust: Commission Sends Statement of Objections to Google Over Abusive Practices in Online Advertising Technology (June 13, 2023), https://ec.europa.eu/commission/presscorner/detail/en/ip_23_3207 [https://perma.cc/DJ2M-BV97] ("The Commission's preliminary view is therefore that only the mandatory divestment by Google of part of its services would address its competition concerns.").

178. *Id.*

179. *Id.* ("The Commission preliminarily finds that, in this particular case, a behavioural remedy is likely to be ineffective to prevent the risk that Google continues such self-preferencing conducts or engages in new ones. Google is active on both sides of the market with its publisher ad server and with its ad buying tools and holds a dominant position on both ends. Furthermore, it operates the largest ad exchange. This leads to a situation of inherent conflicts of interest for Google.").

180. *Id.*

181. *Id.* ("The Commission preliminarily finds that Google is dominant in the European Economic Area-wide markets: (i) for publisher ad servers with its service 'DFP'; and (ii) for programmatic ad buying tools for the open web with its services 'Google Ads' and 'DV360'.").

182. *Id.* ("Favouring its ad exchange AdX in the way its ad buying tools Google Ads and DV360 place bids on ad exchanges. For example, Google Ads was avoiding competing ad exchanges and mainly placing bids on AdX, thus making it the most attractive ad exchange.").

### 3. Platform Law

As a third pillar to challenging the dominance of digital platforms, there is an increasing body of laws targeting the dominance of online platforms directly. These laws are rooted in the observations that other areas of law (such as data protection and antitrust) cannot keep up with the fast-moving technological change.[183] Online platforms have become primary spheres for day-to-day life and should thus have similar protections for civil liberties as the offline sphere.[184] Currently, however, they do not protect civil liberties.[185] This has been highlighted, for example, by the alleged spurring of genocide of the Rohingya Muslims by Facebook's algorithms or Instagram's addictive design that hampers teens' mental health.[186]

The European Union has arguably always been at the forefront of developing platform laws.[187] One of its earliest legislative platform laws was the 2019 Business-to-Platform (B2P) regulation, which aims to protect businesses using online platforms like Amazon to sell their products.[188] These rules were complemented by revised copyright rules in the European Union that, for the first time, explicitly set out

183. *See* Ursula von der Leyen, A Union That Strives for More: My Agenda for Europe 13, https://commission.europa.eu/system/files/2020-04/political-guidelines-next-commission_en_0.pdf [https://perma.cc/268A-QYG2] (last visited Jan. 31, 2025) ("A new Digital Services Act will upgrade our liability and safety rules for digital platforms, services and products, and complete our Digital Single Market.").

184. *See* Mary Robinson, Opinion, *Protecting Fundamental Freedoms, Online and Offline*, ELDERS (Dec. 10, 2012), https://theelders.org/news/protecting-fundamental-freedoms-online-and-offline [https://perma.cc/CG5X-B4RZ].

185. *See id.*

186. Chad de Guzman, *Meta's Facebook Algorithms 'Proactively' Promoted Violence Against the Rohingya, New Amnesty International Report Asserts*, TIME (Sept. 28, 2022, 9:13 PM), https://time.com/6217730/myanmar-meta-rohingya-facebook/ [https://perma.cc/2ACQ-RF4E] ("Amnesty claims that Facebook's algorithms 'proactively amplified' anti-Rohingya content. It also alleges that Meta ignored civilians' and activists' pleas to curb hate-mongering on the social media platform while profiting from increased engagement."); Nicole Westman, *Facebook's Whistleblower Report Confirms What Researchers Have Known for Years*, VERGE (Oct. 6, 2021, 12:28 PM), https://www.theverge.com/2021/10/6/22712927/facebook-instagram-teen-mental-health-research [https://perma.cc/YJA8-TQ89] ("But for researchers who study social media, the internal study that sparked the controversy was mostly confirmation of what they already knew — that Instagram makes teen girls feel worse about their bodies, and that they blame the platform for anxiety, depression, and suicidal thoughts.").

187. *See* Chris Riley, *EU Advances Groundbreaking Law for Online Platforms – U.S. Lawmakers Should Pay Attention*, R ST. INST. (Dec. 14, 2020), https://www.rstreet.org/commentary/eu-advances-groundbreaking-law-for-online-platforms-u-s-lawmakers-should-pay-attention/ [https://perma.cc/N5AF-VSBX].

188. *See* Commission Regulation 2019/1150, 2019 O.J. (L 186) 57 (EU).

obligations for online platforms such as YouTube to implement measures reducing the dissemination of illegal content.[189]

In 2022, the European Union adopted two more platform laws: the Digital Services Act (DSA) and the Digital Markets Act (DMA).[190] Both statutes include a set of specific obligations for gatekeepers and online platforms.[191] While the DMA tries to complement existing EU competition law, the DSA mainly aims at strengthening existing EU consumer protection laws.[192] To this end, the DMA has several obligations to mitigate an abuse of market dominance through interoperability and transparency requirements.[193] Meanwhile, the DSA aims to make the moderation of content on online platforms and search engines more transparent and accountable to tackle the spread of hate speech and disinformation.[194] The DSA classifies online platforms and search engines with 45 million monthly active users (about 10 percent of the EU population) as Very Large Online Platforms (VLOPs) or Very Large Online Search Engines (VLOSEs), which face stringent obligations.[195] The DMA has a similar classification for gatekeepers, albeit with additional conditions.[196] At the time of this writing, Google Search, Google Play, Google Maps, and

---

189. *See* Council Directive 2019/789, 2019 O.J. (L 130) 82 (EU).

190. *See generally* Commission Regulation 2022/2065, 2022 O.J. (L 277) (EU); Commission Regulation 2022/1925, 2022 O.J. (L 265) 1 (EU).

191. *See generally* Commission Regulation 2022/2065, 2022 O.J. (L 277) (EU); Commission Regulation 2022/1925, 2022 O.J. (L 265) 1 (EU).

192. *The Digital Services Act*, EUR. COMM'N, https://commission.europa.eu/strategy-and-policy/priorities-2019-2024/europe-fit-digital-age/digital-services-act_en#:~:text=Digital%20Services%20Act%20(DSA)%20overview&text=Its%20main%20goal%20is%20to,and%20open%20online%20platform%20environment [https://perma.cc/QEX9-6CZG] (last visited Jan. 31, 2025) ("Its main goal is to prevent illegal and harmful activities online and the spread of disinformation. It ensures user safety, protects fundamental rights, and creates a fair and open online platform environment.").

193. *About the Digital Markets Act*, EUR. COMM'N, https://digital-markets-act.ec.europa.eu/about-dma_en [https://perma.cc/49B2-H5VP] (last visited Jan. 31, 2025) ("The Digital Markets Act is the EU's law to make the markets in the digital sector fairer and more contestable. In order to do so, the Digital Markets Act ('DMA') establishes a set of clearly defined objective criteria to identify 'gatekeepers'.").

194. *See The Digital Services Act*, *supra* note 192.

195. *DSA: Very Large Online Platforms and Search Engines*, EUR. COMM'N, https://digital-strategy.ec.europa.eu/en/policies/dsa-vlops [https://perma.cc/K7BK-4SUN] (last visited Jan. 31, 2025).

196. *See About the Digital Markets Act*, *supra* note 193.

Google Shopping are classified as VLOPs or VLOSEs under the DMA.[197] Meanwhile, Google Search, Google Maps, Google Play, Google Shopping, Google Ads, Google Chrome, and Android are classified as gatekeepers in their respective markets.[198]

Under both laws, substantial fines may be imposed for infringements. The DMA can levy fines up to 10 percent of a firm's global turnover.[199] In cases of repeated infringements, fines can increase to as much as 20 percent.[200] Meanwhile, the DSA can lead to fines of up to 6 percent of global turnover.[201]

Platform regulation has also been considered in other countries.[202] For example, Gukhoe, the South Korean National Assembly, passed a law in 2021 limiting Apple's and Google's ability to charge commission on mobile app store transactions.[203] Yet, so far, this law seems to have had a limited effect on commission charged by app stores, leading the Korea Communications Commission (KCC) to threaten fines.[204]

---

197. *See* European Commission Press Release IP/23/2413, Digital Services Act: Commission Designates First Set of Very Large Online Platforms and Search Engines (Apr. 24, 2023), https://ec.europa.eu/commission/presscorner/detail/en/IP_23_2413 [https://perma.cc/YJU5-6TDA].

198. *See* European Commission Press Release IP/23/4328, Digital Markets Act: Commission Designates Six Gatekeepers (Sept. 6, 2023), https://ec.europa.eu/commission/presscorner/detail/en/ip_23_4328 [https://perma.cc/L4K4-4S27]. According to Article 3(1) of the Digital Markets Act (DMA), a company is presumed to be a gatekeeper if it has an annual turnover in the European Economic Area (EEA) of at least €7.5 billion in the last three financial years or a market capitalization of at least €75 billion in the last financial year, provides a core platform service with at least 45 million monthly active end users and 10,000 yearly active business users in the EU, and holds or is expected to hold an entrenched and durable market position. Commission Regulation 2022/1925, 2022 O.J. (L 265) 30 (EU).

199. European Commission Press Release IP/23/2413, *supra* note 198.

200. *Id.*

201. *The Enforcement Framework Under the Digital Services Act*, EUR. COMM'N, https://digital-strategy.ec.europa.eu/en/policies/dsa-enforcement [https://perma.cc/CAY9-EJPN] (Jan. 22, 2025).

202. *See* YASMIN AFINA, MARJORIE BUCHSER, ALEX KRASODOMSKI, JACQUELINE ROWE, NIKKI SUN & ROWAN WILKINSON, TOWARDS A GLOBAL APPROACH TO DIGITAL PLATFORM REGULATION (2024), https://www.chathamhouse.org/2024/01/towards-global-approach-digital-platform-regulation/03-regulatory-pathways-and-potential [https://perma.cc/ZM8R-FC22].

203. *See* Sayuri Umeda, *South Korea: Amended Telecommunications Business Act Will Ban App Payment Monopolies*, LIBR. CONG. (2021), https://www.loc.gov/item/global-legal-monitor/2021-09-16/south-korea-amended-telecommunications-business-act-will-ban-app-payment-monopolies/ [https://perma.cc/REJ9-F8TC] (Sept. 17, 2021).

204. Kate Park, *Google, Apple Face Fines in South Korea for Breaching In-App Billing Rules*, TECHCRUNCH (Oct. 6, 2023, 10:10 AM), https://techcrunch.com/2023/10/06/google-apple-face-fines-in-south-korea-for-breaching-in-app-billing-rules/ [https://perma.cc/UCC6-U2PK] ("South Korea's telecommunication regulator, the Korea Communications Commission (KCC), said Friday that it plans to levy fines on Google and Apple, which could total up to $50.5 million, for violating the country's in-app payment law.").

In the United States, the debate around platform regulation is at the top of the agenda.[205] At the core of the debate is the Digital Services Oversight and Safety Act (DSOSA), a proposed law, and Section 230(c) of the Communications Decency Act (CDA), which was enacted into law in 1996.[206] DSOSA aims to establish a Bureau of Digital Services Oversight and Safety at the FTC to hold online companies accountable for their policies, internal processes, and safety features.[207] Meanwhile, debate around the CDA is centered around whether Section 230(c) of the CDA—which exempts online platforms from liability regarding shared content—should be revised.[208] The CDA was motivated by two conflicting court rulings, which disagreed over whether content intermediaries would face liability for the content that they distributed.[209] Although the CDA has been challenged many times, courts have generally upheld Section 230(c) of the CDA.[210] Recently, support for Section 230(c) has started to crumble. For example, in 2023, the United States District Court for the Northern District of California ruled against application of this immunity based on the First

205. *See Regulators and Economic Experts Debate on Possible Challenges in Platform Regulation*, PYMNTS (Jan. 28, 2022), https://www.pymnts.com/news/regulation/2022/regulators-and-economic-experts-debate-on-possible-challenges-in-platform-regulation/ [https://perma.cc/PKR5-JWN7].

206. *See Section 230*, ELEC. FRONTIER FOUND., https://www.eff.org/issues/cda230 [https://perma.cc/C48N-E4LV] (last visited Feb. 13, 2025); Frank Konkel, *House Bill Would Create FTC Bureau to Oversee Online Platforms*, NEXTGOV FCW (Feb. 23, 2022), https://www.nextgov.com/policy/2022/02/house-bill-would-create-ftc-bureau-oversee-online-platforms/362327/ [https://perma.cc/4RN9-EQQM].

207. *See* Press Release, Lori Trahan, Congresswoman, House of Representatives, Trahan Unveils Comprehensive Online Transparency Legislation (Feb. 22, 2022), https://trahan.house.gov/news/documentsingle.aspx?DocumentID=2389 [https://perma.cc/PGK4-S2JC] ("Congresswoman . . . unveiled the Digital Services Oversight and Safety Act (DSOSA), comprehensive transparency legislation to establish a Bureau of Digital Services Oversight and Safety at the Federal Trade Commission that would have the authority and resources necessary to hold powerful online companies accountable for the promises they make to users, parents, advertisers, and enforcers.").

208. *See* Danielle Draper, *Summarizing the Section 230 Debate: Pro-Content Moderation vs Anti-Censorship*, BIPARTISAN POL'Y CTR. BLOG (July 5, 2022), https://bipartisanpolicy.org/blog/summarizing-the-section-230-debate-pro-content-moderation-vs-anti-censorship/ [https://perma.cc/W2AE-JY5N].

209. *See* Stratton Oakmont, Inc. v. Prodigy Servs. Co., No. 31063/94, 1995 WL 323710, at *3 (N.Y. Sup. Ct. May 24, 1995); Cubby, Inc. v. CompuServe, Inc., 776 F. Supp. 135, 142 (S.D.N.Y. 1991).

210. *See Section 230: Key Legal Cases*, ELEC. FRONTIER FOUND., https://www.eff.org/issues/cda230/legal [https://perma.cc/JFJ6-YGCB] (last visited Jan. 31, 2025).

Amendment and Section 230(c).[211] Both these defenses, protection of free speech under the First Amendment and Section 230(c) of the CDA, were used by social media companies to distance themselves from content published on their platforms.[212] The district court's rejection of these defenses opens up the possibility of future lawsuits against social media companies on the basis of the content shared on these platforms.

## III. GOOGLE'S CYCLICAL USE OF DOMINANCE

This section explores how Google uses its leading position across various markets to gain an unfair advantage over its competitors, resulting in a vicious cycle where dominance in one market helps Google dominate the others as well. Specifically, this section will explore how Chrome's leading position is used as a bridge between different markets to perpetuate Google's dominance. This advantage flows in three different directions: from Google as a publisher to Chrome, from Chrome to Google as an advertiser and publisher, and from Google as an advertiser to Google as a publisher.

211. *See* Jonathan Stempel & Nate Raymond, *Social Media Companies Must Face Youth Addiction Lawsuits, US Judge Rules*, REUTERS (Nov. 14, 2023, 5:35 PM), https://www.reuters.com/legal/judge-says-social-media-companies-must-face-lawsuits-over-harm-children-2023-11-14/ [https://perma.cc/VQK3-V58Z] ("In her 52-page ruling, Rogers rejected arguments that the companies were immune from being sued under the U.S. Constitution's First Amendment and a provision of the federal Communications Decency Act.").

212. *See id.*

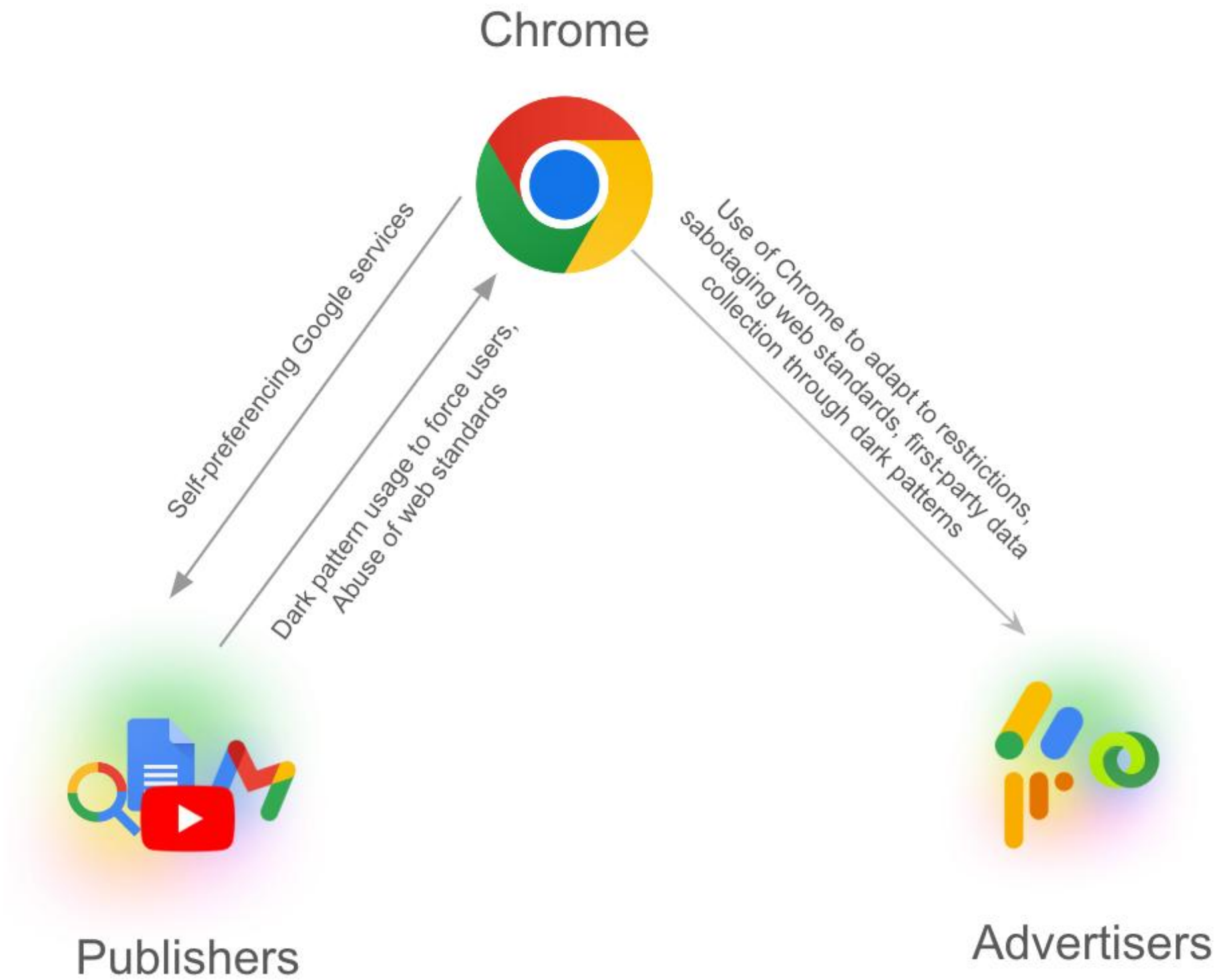

Figure 1. Google's dominance of browser, publisher, and advertising markets.

Google abuses its position as a web browser market leader to augment its publishing and advertising business, while its publisher business works to increase Chrome's market share.

### *A. What Are Publishers and Advertisers?*

To understand Google's overreach across different markets, it is important to understand the difference between Google as an advertiser and Google as a publisher. This Article defines a publisher as a first-party website or application visited by a user to gain access to a particular service or functionality. For example, when a user visits Google Search or Google Maps, Google is acting as a publisher by providing the user with a search and navigation functionality, respectively.

On the other hand, an advertiser, in partnership with a publisher, shows the visiting users advertisements to attract the user towards other products or services that are related to the user's past or current activities. For example, while visiting BBC's website, a user might see advertisements to purchase sports shoes. While the advertisement shows a sport shoe company, it is shown to a user through a third-party advertising company such as Google Ads, which performs the technical labor involved in identifying the user, their interests and activities, and delivering the actual advertisement on the BBC website.

Google is a dominant force as a publisher and advertiser, and it uses its dominance in browser market share to further solidify its position and gain unfair advantage over its competitors.[213] This results in a vicious cycle of Google's dominance across multiple markets.[214]

### *B. Flow of Dominance from Publisher to Browser Market*

First, Google uses its publisher market share to coerce users into switching to Chrome. Google is a dominant publisher in several different key areas, which include, but are not limited to, search, navigation, video streaming, and email.[215] Google makes use of this dominant position to push users towards using its Chrome browser, solidifying its position as a browser market leader.[216] It pushes users to Chrome through use of dark patterns and abuse of web standards.[217]

#### 1. Dark Pattern Usage by Google to Force Users to Switch to Chrome

To push users towards Chrome, Google employs user interface strategies—often termed "dark patterns" or "deceptive patterns"—to encourage a switch to Chrome.[218]

Google's employment of dark patterns to nudge users towards Chrome is evidenced through persistent prompts across its services. Users of non-Chrome browsers are shown pop-ups recommending Chrome installation when they visit Google services like Google Search and Google Docs.[219] These prompts, designed to be unavoidable and recurring, even permeate into the user experience of Google services accessed through competitors' browsers, such as Microsoft Edge, where users might find suggestions to switch to Chrome within their email notifications.[220]

---

213. *See* Katharine Gemmell & Samuel Stolton, *Google Is Abusing Dominance in Advertising Technology, UK Regulator Warns*, BLOOMBERG (Sept. 6, 2024, 4:38 AM), https://www.bloomberg.com/news/articles/2024-09-06/google-is-abusing-dominance-in-tech-ads-uk-regulator-warns [https://perma.cc/JST4-TGRW].

214. *See id.*

215. *See* David Goldman, *Google's Search Dominance Is Unwinding*, CNN BUS. (Oct. 8, 2024, 5:00 AM), https://www.cnn.com/2024/10/08/business/google-search-amazon-nightcap/index.html [https://perma.cc/6ME6-W2ZK] ("Google is the fourth-most-valuable public company on the planet. At $2.1 trillion, its market value trails only Apple, Microsoft and AI chip darling Nvidia. And it remains dominant in other markets, including display ads, which it leads alongside Facebook's parent company Meta, and video ads through YouTube.").

216. *See* Kollnig, *Forgotten Fulcrum*, *supra* note 85.

217. *See id.*

218. *Id.*

While Microsoft engages in similar practices by promoting Edge within its Bing search engine, Edge holds a modest market share of 5 percent, which does not confer the same level of market dominance.[221]

The persistent push from both Google and Microsoft to direct users to their respective services underscores the substantial economic value derived from user data and the promotion of proprietary digital services.[222] With the release of version 69 of Chrome in 2018, users found themselves automatically signed into the browser when accessing Google services like YouTube and Gmail, a design choice that was somewhat mitigated in Chrome 70 by offering an opt out.[223] Yet, the default setting persists and most users tend to stick with defaults.[224] These default settings combined with jaded users result in Google obtaining a larger amount of data from users who use Chrome to visit their services versus other browsers.[225]

Furthermore, Chrome's synchronization feature, which allows seamless access to browsing history, open tabs, passwords, and more across Chrome browsers running on different devices, would have a

219. *Id.*

220. *See infra* Figures 2–5.

221. *See Global Desktop Browser Market Share*, KINSTA, https://kinsta.com/browser-market-share/ [https://perma.cc/Y9AD-MLDX] (last visited Feb. 1, 2025).

222. *See infra* Figures 2–5.

223. *See* Matthew Green, *Why I'm Done with Chrome*, CRYPTOGRAPHIC ENG'G BLOG (Sept. 23, 2018), https://blog.cryptographyengineering.com/2018/09/23/why-im-leaving-chrome/ [https://perma.cc/C2WH-N65W] ("From now on, every time you log into a Google property (for example, Gmail), Chrome will automatically *sign the browser into your Google account* for you. It'll do this without asking, or even explicitly notifying you."); Zach Koch, *Product Updates Based on Your Feedback*, GOOGLE: KEYWORD (Sept. 26, 2018), https://blog.google/products/chrome/product-updates-based-your-feedback/ [https://perma.cc/9TA2-GUSS] ("We've heard—and appreciate—your feedback. We're going to make a few updates in the next release of Chrome (Version 70, released mid-October) to better communicate our changes and offer more control over the experience.").

224. *See* Geoffrey A. Fowler, *Google Spent $26 Billion to Hide This Phone Setting from You*, WASH. POST, https://www.washingtonpost.com/technology/2023/11/08/google-search-default-iphone-samsung/ [https://perma.cc/X69V-E8ZX] (Nov. 8, 2023) ("We're getting an inside view of how Google exploits this behavioral science, sometimes called the 'power of defaults.' The idea is that defaults can nudge people's choices one way or another, because most people are too distracted or confused to change them.").

225. *See id.*

lock-in effect.[226] This synchronized sign-in mechanism across multiple devices enables Google to track user activities across devices and more effectively tailor advertisements.[227] Therefore, a synchronized and signed-in Chrome user is more valuable to Google than a user on another browser.[228] Even if the user is not using Google Chrome, a user visiting Google Search is nudged toward signing in.[229]

---

226. *See* HAROLD ØVERBY & JAN ARILD AUDESTAD, INTRODUCTION TO DIGITAL ECONOMICS 177 (2d ed. 2021) ("Lock-in to a technology or supplier implies that it is expensive for the customer to switch to a competing technology or supplier. The expenses may be monetary, psychological, or associated with loss of intangible assets.").

227. *See* Julia Angwin, *Google Has Quietly Dropped Ban on Personally Identifiable Web Tracking*, PROPUBLICA (Oct. 21, 2016, 8:00 AM), https://www.propublica.org/article/google-has-quietly-dropped-ban-on-personally-identifiable-web-tracking [https://perma.cc/9W44-N32C] ("Google substituted new language that says browsing habits 'may be' combined with what the company learns from the use [of] Gmail and other tools. . . .The practical result of the change is that the DoubleClick ads that follow people around on the web may now be customized to them based on your name and other information Google knows about you. It also means that Google could now, if it wished to, build a complete portrait of a user by name, based on everything they write in email, every website they visit and the searches they conduct.").

228. *See [GA4] Activate Google Signals for Google Analytics Properties*, GOOGLE: ANALYTICS HELP, https://support.google.com/analytics/answer/9445345 [https://perma.cc/32ZU-ZX6K] (last visited Feb. 1, 2025) ("Google signals are session data from sites and apps that Google associates with users who have signed in to their Google accounts, and who have turned on Ads Personalization. This association of data with these signed-in users is used to enable cross-device remarketing, and cross-device key events export to Google Ads.").

229. *See infra* Figure 6.

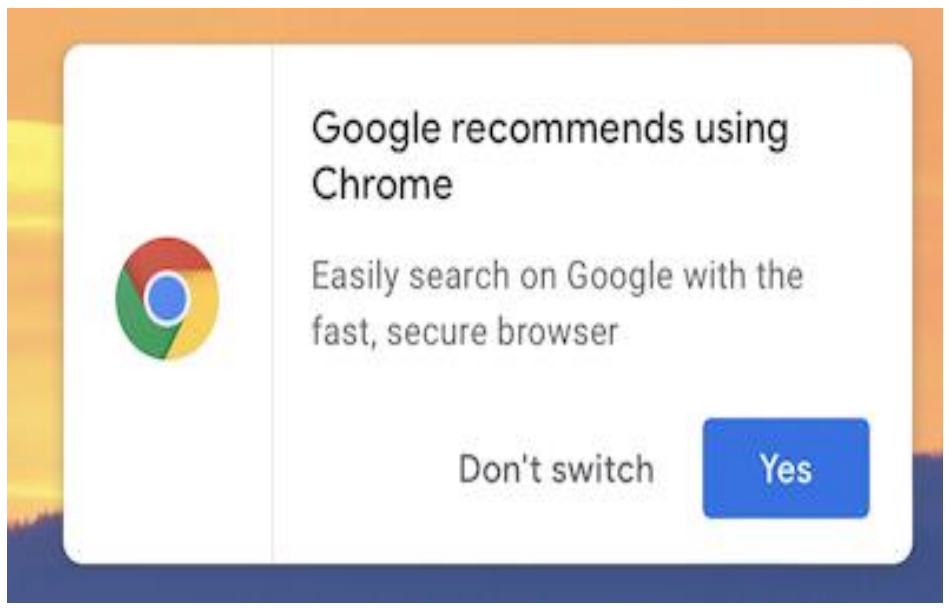

Figure 2. Pop-up on Google service, suggesting user to switch to Chrome.

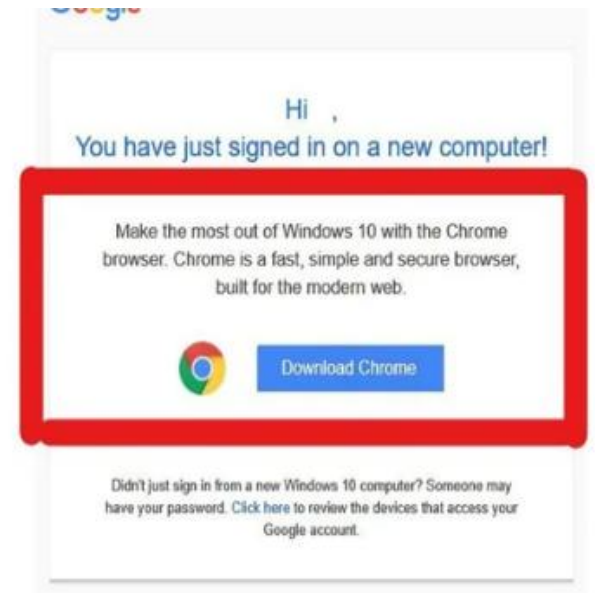

Figure 3. Email after logging in to Google in Microsoft Edge browser.

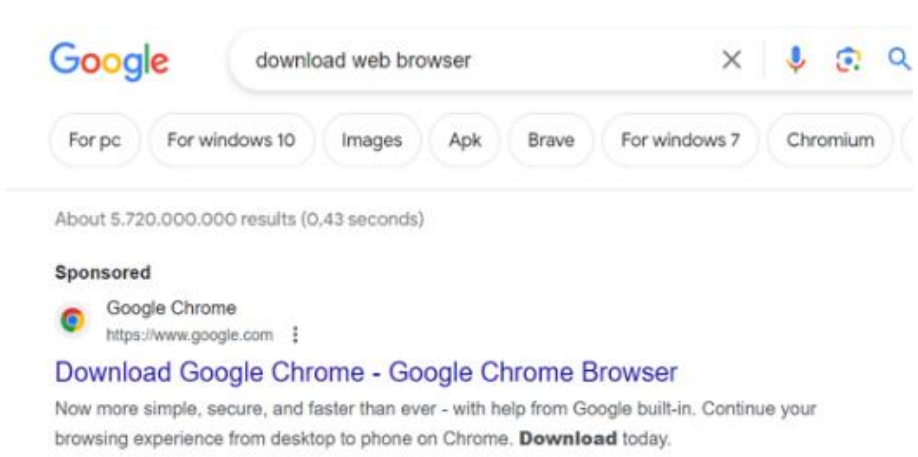

Figure 4. Google Ad on Google Search promoting Google Chrome, with no visual distinction from other search results.

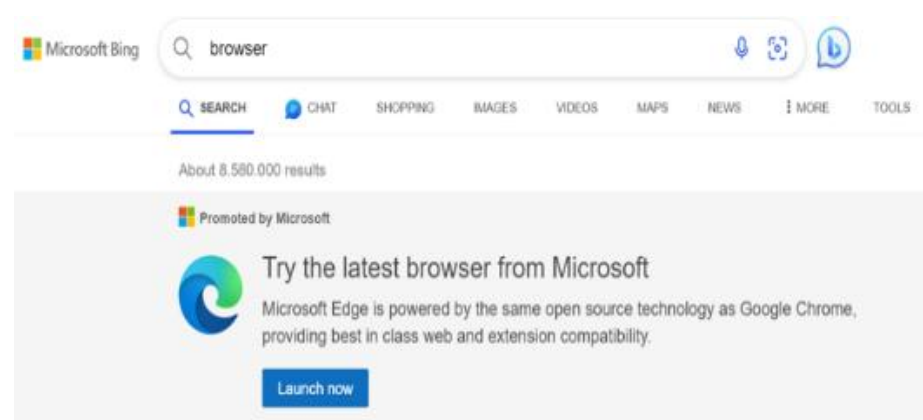

Figure 5. Microsoft promoting Edge in Bing. In contrast to Google Search, the ad does not have the format of usual ads on Bing.

Google leverages its advertising dominance to promote its Chrome browser directly through its search engine.[230] When users search for "web browser" on Google Search, they are often met with a top-placed Google Ad advocating for Chrome.[231] This scenario illustrates a circular flow of funds within Google's corporate

230. *See* Kollnig, *Forgotten Fulcrum*, *supra* note 85.

231. *See id.*

structure: the company effectively pays itself to ensure Chrome's ad outbids competitors on its own search platform.[232] As these internal transactions occur within Google's subsidiaries, the company possesses the capacity to consistently outspend other advertisers, forfeiting only the potential ad revenue from the next highest bid.[233]

This strategy raises significant antitrust concerns, particularly since Google is the default search engine on major browsers like Safari and Firefox.[234] Users on these platforms encounter the same Google-centric ads when searching for web browsers, subtly nudging them towards Google's browser.[235] Such practices underscore the intricate ways in which Google can use its integrated business model to maintain and extend its market influence, potentially at the expense of fair competition and consumer choice.

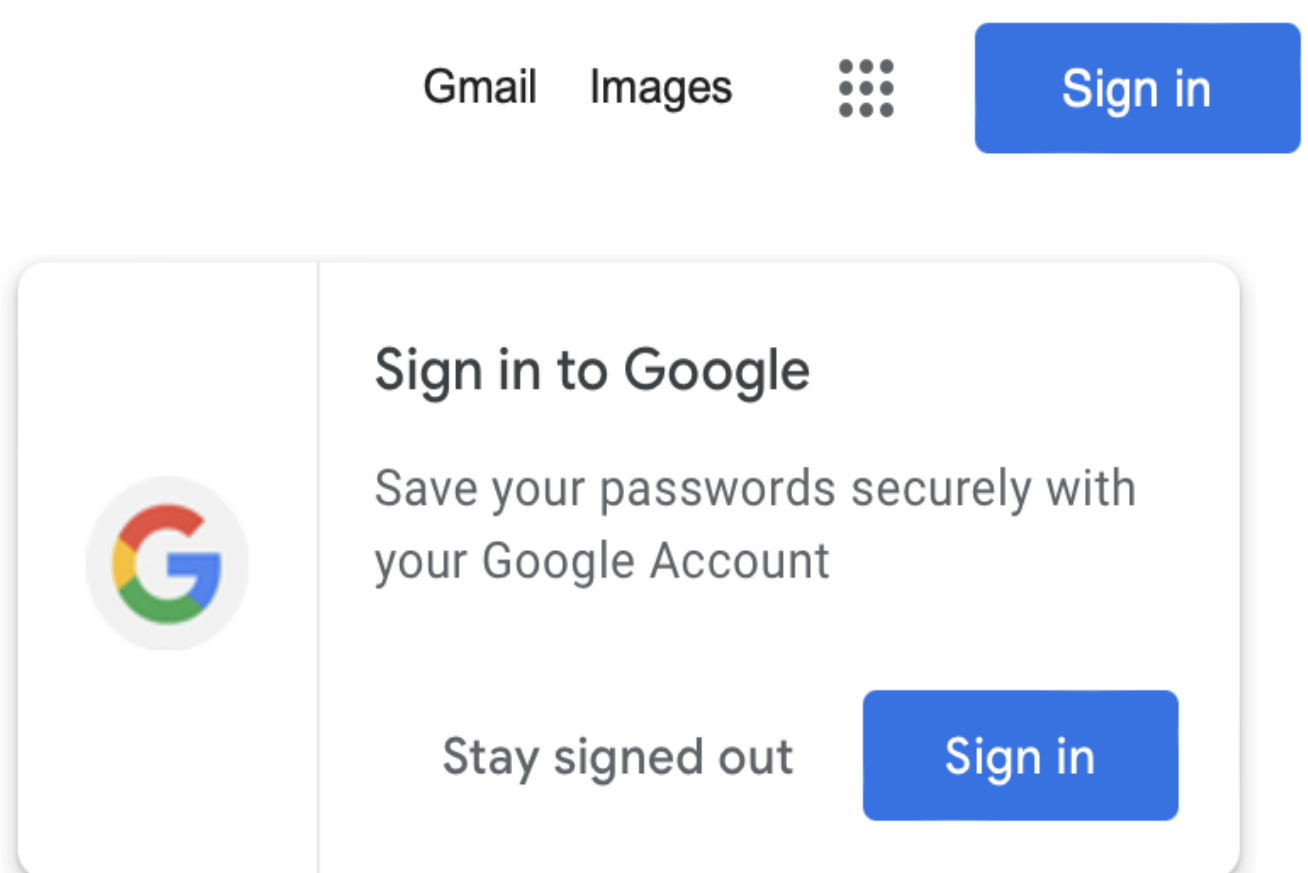

Figure 6. Visiting Google Search on any browser results in a suggestion to sign in to Google.

2. Abuse of Web Standards

Google's strategy to force users to switch to Chrome extends to promoting proprietary standards that are not widely adopted by other browsers. A pertinent case is Google's acquisition and development of Widevine DRM, a digital rights management solution for premium

232. *See id.*
233. *See id.*
234. *See id.*
235. *See id.*

media content.[236] Widevine's compatibility issues have recurrently resulted in subpar user experiences across various media platforms.[237] A notable instance occurred in 2017 when Spotify updated its web player to incorporate Widevine DRM.[238] This update led to Safari users encountering messages suggesting they switch browsers or download the Spotify mobile app, illustrating how Google's control over key technologies can indirectly compel users to migrate to its products, such as Chrome, to avoid disruptions in service.[239] This practice not only highlights Google's influence over web standards but also raises questions about its impact on user choice and market competition.

Over time, Google's use of nonstandard APIs has led to several instances where certain services or websites were optimized exclusively for Chrome. Services such as Google Meet, Google Earth, and YouTube TV were initially accessible only through Chrome or with certain

236. *See Widevine Technologies Acquired by Google*, CRUNCHBASE, https://www.crunchbase.com/acquisition/google-acquires-widevine—1d323873 [https://perma.cc/HF36-GLRL] (last visited Feb.1, 2025); *Digital Rights Management*, GOOGLE: GOOGLE DEVS., https://developers.google.com/widevine/drm/overview [https://perma.cc/QEQ4-TEET] (last visited Feb. 1, 2025) ("Widevine DRM is Google's content protection system for premium media. It is used by major partners around the world such as Google Play, YouTube, Netflix, Disney+, Amazon Prime Video, HBO Max, Hulu, Peacock, Discovery+, Paramount+ and many more.")

237. *See* @akaforty_1, SPOTIFY: CMTY. (April 26, 2017, 10:31 AM), https://community.spotify.com/t5/Other-Podcasts-Partners-etc/New-Web-Player-not-working-due-to-Widevine/td-p/1658251 [https://perma.cc/5K7Z-TPCC].

238. *See id.*

239. *See* @riegelstamm, SPOTIFY: CMTY. (Sept. 7, 2017, 3:52 PM), https://community.spotify.com/t5/Other-Podcasts-Partners-etc/Safari-No-Longer-Supported/td-p/1975103 [https://perma.cc/6BXQ-EZGR] ("Apparently it has something to do with the Google Widevine content decryption module, which Apple doesn't support because it's not very secure. I tried enabling the Widevine plugin and got the attached message. Looks like Apple is having a pissing contest with Google, Spotify, and anyone else who uses Widevine. In the meantime, we users are caught in the crossfire."); @AITech, LINUS TECH TIPS (Sept. 11, 2017), https://linustechtips.com/topic/833410-spotify-web-player-no-longer-supports-safari-as-they-move-to-use-widevine-cdm-in-web-player-music-drm-in-your-browser/ [https://perma.cc/H3PA-6VZ4] ("Spotify recently moved their recently updated Web Player to use Widevine CDM, a controversial mechanism aimed at combating piracy. Widevince CDM has been supported in Edge, Firefox, Chrome, and Opera for a while. Apple hasn't implemented Widevine into Safari as of this [sic] yet and thus Safari users will be unable to play back content in Spotify Web Player."); Andrew Liptak, *Spotify's Web Player No Longer Works on Safari*, VERGE (Sept. 10, 2017, 10:15 AM), https://www.theverge.com/2017/9/10/16283494/spotify-web-player-safari-browser-support-apple [https://perma.cc/5KGB-SZR2] ("The company's system requirements page now states that it only supports Chrome, Firefox, Edge, and Opera.").

features exclusively available on Chrome.[240] But this preferential treatment is not confined to Google's services; other companies like Airbnb have also been observed recommending users to switch to Chrome for an optimized experience.[241]

Chrome's dominant share in the browser market explains why such companies prefer to develop only for one browser.[242] This approach effectively discriminates against other browsers, underscoring a broader concern about Google's influence in shaping user choices and reinforcing its own market position at the potential expense of fair competition and browser diversity.

The issue of websites favoring Chrome to the exclusion of other browsers has become pervasive enough to prompt Mozilla to create Webcompat.com, a platform dedicated to users reporting websites that lack cross-browser compatibility.[243] A significant number of issues reported on Webcompat.com pertain to websites functioning exclusively with Chrome. Prominent examples include Microsoft Teams and Snapchat, where users are prompted to switch to Chrome to access certain features like video calling.[244] Over time, Webcompat.com has

---

240. *See* Rajesh Pandey, *Google Meet's New Picture-in-Picture Mode for Chrome Is Almost an App of Its Own*, ANDROID POLICE (June 9, 2023), https://www.androidpolice.com/google-meets-new-pip-mode-chrome-almost-an-app/ [https://perma.cc/3DCK-YMWC] ("Google is taking advantage of the new Document Picture-in-Picture API, which first debuted with Chrome 111, to deliver enhanced picture-in-picture capabilities in Google Meet. Sadly, this also means you must use Chrome to enjoy the improvements, as the API is not available on other browsers."); Rich McCormick, *Redesigned Google Earth Brings Guided Tours and 3D View to Chrome Browsers and Android Devices*, VERGE (Apr. 18, 2017, 5:00 AM), https://www.theverge.com/2017/4/18/15337646/google-earth-redesign-update-guided-tours [https://perma.cc/TMJ5-2UDR] ("The revamped Google Earth — which the company says was two years in the making — is now available in Chrome or on Android, and will be coming to iOS and other browsers in the future."); Chaim Gartenberg, *YouTube TV Now Works in Firefox*, VERGE (Apr. 5, 2018, 1:43 PM), https://www.theverge.com/2018/4/5/17203192/youtube-tv-google-firefox-browser-support-chrome-update-streaming [https://perma.cc/2H5Y-2XS3] ("Google is expanding YouTube TV to support Firefox, as spotted by YourTechExplained, marking the first browser to work with Google's over-the-top streaming service that isn't Chrome (which you'll recall is also owned by Google.) [sic]").

241. *See* @AirbnbHelp, TWITTER (July 12, 2016, 6:37 AM), https://twitter.com/AirbnbHelp/status/752829250198245376 [https://perma.cc/92YW-HXP7].

242. *See Understanding Browser Market Share: Which Browsers to Test on in 2024*, BROWSERSTACK (June 30, 2024), https://www.browserstack.com/guide/understanding-browser-market-share [https://perma.cc/E2SK-D7NC].

243. *See About*, WEBCOMPAT, https://webcompat.co\]m/about [https://perma.cc/KRL5-NUMG] (last visited Feb. 1, 2025) ("Webcompat.com is developed by volunteers and supported by Mozilla. This site is an open invitation for all web users, developers, and browser vendors to get involved in the web compatibility effort. Our goal is to make it easy to report and view compatibility problems for any part of the web.").

244. *See* @digitarald, WEBCOMPAT (Jan. 29, 2019), https://webcompat.com/issues/25070 [https://perma.cc/AS9K-ZN7B] ("Video calls on MS Teams only work in Chrome, Edge or the amassed numerous reports of websites that are solely operational with Chrome, highlighting a growing trend that raises significant concerns about browser diversity and the principles of an open and competitive web.[245]

The recurrent issues of browser incompatibility stem largely from Chrome's reliance on non-standard APIs, the use of those APIs by Google-owned services, and the influence on web developers to prioritize Chrome due to its substantial market share.[246] This prioritization often results in subpar or non-functional experiences for users of other browsers, presenting them with a stark choice: contend with these limitations or switch to Chrome for a smoother experience.[247] Most users, opting to avoid the inconvenience, choose the latter.[248] This trend further entrenches Chrome's dominance in the web browser market, highlighting how Google uses its dominant position in the publisher market to force users to switch to Chrome.

### *C. Flow of Dominance from Browser to the Advertising and Publishing Markets*

#### 1. From Browser to Advertising Market

Before discussing how Google uses Chrome to dominate its advertising business, it is important to understand the distinction

---

Desktop App (aka an Electron wrapper)."); @webcompat-bot, WEBCOMPAT (July 19, 2022), https://webcompat.com/issues/107613 [https://perma.cc/HU6X-AEXF] ("Snapchat for web says only chrome and edge are supported. . .this is so rude!").

245. *See* @adamopenweb, WEBCOMPAT (Sept. 18, 2018), https://webcompat.com/issues/18922 [https://perma.cc/QFC9-539Y] ("Site says issues with 3D content in Firefox, recommends Chrome."); @raffaem, WEBCOMPAT (Jan. 13, 2021), https://webcompat.com/issues/65496 [https://perma.cc/9JVZ-8NVV] (sharing a screenshot that states "this web browser isn't supported yet. Please join from Google Chrome"); @denschub, WEBCOMPAT (May 19, 2022), https://webcompat.com/issues/104596 [https://perma.cc/JL6H-KNPD] ("The site shows a 'please use Chrome' banner. This can be dismissed, but it's very intrusive.").

246. *See* Ken Bellows, *Chromium and the Browser Monoculture Problem*, DEV CMTY. (June 12, 2019), https://dev.to/kenbellows/chromium-and-the-browser-monoculture-problem-420n [https://perma.cc/KY8K-NGFX].

247. *See* Tom Warren, *Chrome Is Turning Into the New Internet Explorer 6*, VERGE (Jan. 4, 2018, 8:30 AM), https://www.theverge.com/2018/1/4/16805216/google-chrome-only-sites-internet-explorer-6-web-standards [https://perma.cc/X4LD-FJ7S] ("Either way, Chrome now has the type of dominance that Internet Explorer once did, and we're starting to see Google's own apps diverge from supporting web standards much in the same way Microsoft did a decade and a half ago.").

248. *See* Katie Ritter, *4 Reasons to Use Google Chrome over Other Internet Browsers*, TALK TECH WITH ME BLOG (Feb. 21, 2015), https://talktechwithme.com/2015/02/21/4-reasons-to-use-google-chrome-over-other-internet-browsers/ [https://perma.cc/5CV3-5ZRV].

between third- and first-party data collection, and how it affects online advertisers.

*a. Difference Between a First-Party and a Third-Party Data Collector*

An online website typically utilizes two distinct types of resources: first-party and third-party.[249] These resources can encompass elements like JavaScript code, images, and videos, and are differentiated by their source domain.[250] For instance, on a website such as nytimes.com, resources that originate from the same domain (nytimes.com) are considered first-party. In contrast, resources which are included from a different domain, like doubleclick.net (an advertising company acquired by Google), are deemed third-party.[251] These resources are used to serve content which is not owned by the main website.[252] The classification as a first-party or third-party resource has significant implications for the privileges granted to the resources within a webpage. One key privilege is the capability to store information, often in the form of cookies, in a user's web browser.[253] Cookies set directly by the domain the user is visiting are termed first-party cookies, whereas those set by external, third-party domains are referred to as third-party cookies.[254]

---

249. *See* Griffin LaFleur, *First-Party vs. Third-Party Cookies: What's the Difference?*, TECHTARGET (July 29, 2024), https://www.techtarget.com/searchcustomerexperience/tip/First-party-vs-third-party-cookies-Whats-the-difference [https://perma.cc/HB2J-KNQK].

250. *See What Are 3rd Party Scripts?*, SOURCE DEF., https://sourcedefense.com/glossary/3rd-party-scripts/ [https://perma.cc/4DUQ-CHCC] (last visited Feb. 1, 2025) ("Third-party scripts are often JavaScript code used to add additional functionality or features to a website or application, such as tracking analytics, displaying ads, or providing social media integration. It is also used to leverage code that has been developed and tested by others rather than having to write it from scratch.").

251. *See* Quan Chen, Michalis Polychronakis, Panagiotis Ilia & Alexandros Kapravelos, *Cookie Swap Party: Abusing First-Party Cookies for Web Tracking*, *in* WWW '21: PROCEEDINGS OF THE WEB CONFERENCE 2021 2117, 2118 (2021) [hereinafter Chen et al., *Cookie Swap Party*] ("HTTP cookies can be categorized as first-party or third-party, depending on their domain of origin. The cookies set when visiting a website are considered as first-party, while those set by other domains as a result of loading external resources are considered as third-party. Consequently, if the same third-party resource (e.g., a popular JavaScript library) is present on multiple websites, it enables cross-site tracking: any third-party domain that host resources referenced by multiple websites can track users across these sites.").

252. *See id.*

253. *See What Are Cookies?*, KASPERSKY, https://usa.kaspersky.com/resource-center/definitions/cookies?srsltid=AfmBOoonM-JQYXgvwxQVd4aDGrzzeqDhtHdR-WxZi83te7C1jtz-JWTI [https://perma.cc/B9H9-KJRE] (last visited Feb. 1, 2025).

254. *See Client-Side Storage*, W3C, https://www.w3.org/2001/tag/2010/09/ClientSideStorage.html [https://perma.cc/AHF7-8TDL] (last visited Feb. 1, 2025) ("The most important use of cookies however, and the most controversial, is to use cookies for tracking where you go and what

Third-party cookies play a crucial role for advertisers because they allow a user to be recognized across all websites where the third party is active.[255] Taking the previous example of doubleclick.net, imagine it places a cookie with a unique user identifier on nytimes.com. The same identifier can then be picked up by doubleclick.net when the same user visits another website, like washingtonpost.com, if it also uses resources from doubleclick.net. This identifier would then enable third-party websites to follow and record a person's online behavior across various websites, including which sites they visit, the actions they take on those sites, and the amount of time they spend on each site. Consequently, this information is used by advertisers to determine the interests of a user and display relevant advertisements.[256]

Concerns about privacy and the tracking capabilities of third-party cookies have prompted major browsers to act. Safari started imposing restrictions in 2017 and fully blocked third-party cookies by 2020, while Firefox initiated limitations in 2018 and followed with a complete block in 2022.[257] In response, some advertisers and tracking

---

you do there. These are typically used by advertising sites but you do not visit any of the advertising websites, so how can they get their cookies into your local storage? If you look at the cookies stored on your machine you will probably find cookies from DoubleClick, a site that tracks what ads you look at. This happens because a search engine you used has a relationship with DoubleClick and allows it to set cookies in your local storage. These are called third-party cookies.").

255. *See [GA4] Activate Google Signals for Google Analytics Properties*, *supra* note 228.

256. *See* Steven Englehardt & Arvind Narayanan, *Online Tracking: A 1-Million-Site Measurement and Analysis*, *in* CCS '16: PROCEEDINGS OF THE 2016 ACM SIGSAC CONFERENCE ON COMPUTER AND COMMUNICATIONS SECURITY 1388, 1389 (2016) ("As users browse and interact with websites, they are observed by both 'first parties,' which are the sites the user visits directly, and 'third parties' which are typically hidden trackers such as ad networks embedded on most web pages. Third parties can obtain users' browsing histories through a combination of cookies and other tracking technologies that allow them to uniquely identify users, and the 'referer' header that tells the third party which first-party site the user is currently visiting. Other sensitive information such as email addresses may also be leaked to third parties via the referer header.").

257. *See* Wilander, *supra* note 69 ("Cookies for cross-site resources are now blocked by default across the board. This is a significant improvement for privacy since it removes any sense of exceptions or 'a little bit of cross-site tracking is allowed.'"); *Firefox Rolls Out Total Cookie Protection by Default to More Users Worldwide*, *supra* note 69 ("Any time a website, or third-party content embedded in a website, deposits a cookie in your browser, that cookie is confined to the cookie jar assigned to *only* that website."); Nick Nguyen, *Latest Firefox Rolls Out Enhanced Tracking Protection*, MOZILLA: DISTILLED (Oct. 23, 2018), https://blog.mozilla.org/en/products/firefox/latest-firefox-rolls-out-enhanced-tracking-protection/ [https://perma.cc/HK2A-Q8JE] ("With today's Firefox release, users will have the option to block cookies and storage access from third-party trackers. This is designed to effectively block the most common form of cross-site tracking.").

companies have started to pivot towards first-party cookies.[258] Unlike third-party cookies, creating a first-party cookie as a third party requires cooperation with the site owner and the execution of privileged JavaScript in a first-party context.[259] This approach necessitates a high degree of trust from the website owner towards the third party, incurs higher costs, and demands a time investment from both parties.[260] These barriers make it challenging for many advertisers to transition to first-party cookies. Google capitalizes on this complex situation to their advantage.[261]

Google is one of the largest advertisers and analytics service providers.[262] This section shows that Google's command over the browser market, particularly with Chrome, has enabled it to implement policies, establish controls, and set standards that disproportionately advantage its own advertising and analytics services, while not affording the same benefits to its competitors and limiting their market access over time.

---

258. *See* Shaoor Munir, Sandra Siby, Umar Iqbal, Steven Englehardt, Zubair Shafiq & Carmela Troncoso, *CookieGraph: Understanding and Detecting First-Party Tracking Cookies*, *in* CCS '23: PROCEEDINGS OF THE 2023 ACM SIGSAC CONFERENCE ON COMPUTER AND COMMUNICATIONS SECURITY 3490, 3490 (2023) ("We show that third-party cookie blocking does not significantly impact the sharing of identifiers to known tracking endpoints because major trackers are already using first-party cookies.").

259. *See id.* at 3493 ("When cookies are set by a script, their classification depends on whether the script is embedded in a first- or third-party execution context. The cookies set by third-party scripts running in the first-party context are first-party cookies."). Additionally, setting a third-party cookie can be as easy as embedding an image on a website. *See* Dan Muller, *What Is Tracking Pixel in Affiliate Marketing?*, BIXGROW (Apr. 30, 2024), https://bixgrow.com/tracking-pixel-in-affiliate-marketing/ [https://perma.cc/TK3H-2G3H]. This why tracking scripts are sometimes referred to as "pixels," even though they are usually more complex than just images these days and include sophisticated fingerprinting techniques. *See id.*; *see also Meta Pixel*, META, https://www.facebook.com/business/tools/meta-pixel [https://perma.cc/C6LQ-DVWJ] (last visited Feb. 1, 2025).

260. *See* WEBCOMPAT, *supra* note 243 ("Webcompat.com is developed by volunteers and supported by Mozilla. This site is an open invitation for all web users, developers, and browser vendors to get involved in the web compatibility effort. Our goal is to make it easy to report and view compatibility problems for any part of the web.").

261. *See* Brandon Heagle, *Google's Pivot on Third-Party Cookies: What It Means for Advertisers and the Open Web*, STELLA RISING (July 29, 2024), https://www.stellarising.com/bl7og/googles-pivot-on-third-party-cookies-and-its-impact [https://perma.cc/8NQP-RSCX].

262. *See* Tim Wambach & Katharina Bräunlich, *The Evolution of Third-Party Web Tracking*, *in* COMMUNICATIONS IN COMPUTER AND INFORMATION SCIENCE 137, 141 (2017) ("73% of all analyzed websites are covered by the top three of the most included third party trackers from 2015: according to Table 4 these are google-analytics.com, doubleclick.net, and facebook.net.").

### *b. Google's Third-Party Services*

Google administers two main third-party services: Google Ads and Google Analytics.[263] Google Ads enables website publishers to display advertisements on their sites.[264] By integrating a provided code snippet into their website, publishers grant Google the ability to gather user data and display targeted ads based on user behavior and interests.[265] This integration allows Google to act as a first-party resource, which comes with privileges such as reading and writing first-party cookies.[266] Each user is assigned a distinct identifier, allowing Google Ads to aggregate information like pages visited and links clicked at the individual level.[267] Similarly, Google Analytics is embedded into websites using a code snippet from Google that leverages user data to offer comprehensive insights into the volume and nature of traffic a website receives.[268] Recent data suggests that Google Ads and Google Analytics are both utilized by about 50 percent of all.[269]

---

263. *See Advertising and Attribution*, GOOGLE: ANALYTICS HELP, https://support.google.com/analytics/answer/9379420?hl=en#zippy=%2Cin-this-article [https://perma.cc/AX98-UCF2] (last visited Feb. 1, 2025).

264. *See About Google Ads*, GOOGLE: GOOGLE ADS HELP, https://support.google.com/google-ads/answer/6349091?hl=en [https://perma.cc/9QDG-FNYF] (last visited Feb. 1, 2025).

265. *See [GA4] Set Up Analytics for a Website and/or App*, GOOGLE: ANALYTICS HELP [hereinafter *Set Up Analytics*], https://support.google.com/analytics/answer/9304153?hl=en [https://perma.cc/JTS7-4DAZ] (last visited Feb. 13, 2025) ("On the screen, you'll see the JavaScript snippet for your account's Google tag. Your Google tag is the entire section of code that appears, beginning with: <!— Google tag (gtag.js) —> and ending with </script> Paste your Google tag immediately after the <head> on each page of your website.").

266. *See First-Party Cookies Controls Are Changing*, GOOGLE: GOOGLE ADSENSE HELP, https://support.google.com/adsense/answer/15569935?hl=en [https://perma.cc/L2KD-2SP7] (last visited Feb. 1, 2025).

267. *See About Data Segments That Use User ID to Advertise*, GOOGLE: GOOGLE ADS HELP, https://support.google.com/google-ads/answer/9199250?hl=en [https://perma.cc/2B58-EGSG] (last visited Feb. 1, 2025) ("These segments that use User ID are eligible for targeting across Google properties, including Search, Shopping, and YouTube.").

268. *See Set Up Analytics*, *supra* note 265 ("On the screen, you'll see the JavaScript snippet for your account's Google tag . . . ."); *How Google Analytics Works*, GOOGLE: ANALYTICS HELP, https://support.google.com/analytics/answer/12159447?hl=en [https://perma.cc/HF8B-6C2B] (last visited Feb. 1, 2025) ("Every time a user visits a webpage, the tracking code will collect pseudonymous information about how that user interacted with the page.").

269. *Usage Statistics and Market Share of Google Ads for Websites*, W3TECHS [hereinafter *Google Ads*], https://w3techs.com/technologies/details/ad-google [https://perma.cc/CS3M-47CA] (last visited Feb. 1, 2025) ("Google Ads is used by 99.2% of all the websites whose advertising network we know. This is 47.5% of all websites."); *Usage Statistics and Market Share of Google Analytics for Websites*, W3TECHS [hereinafter *Google Analytics*], https://w3techs.com/technologies/details/ta-googleanalytics [https://perma.cc/39PP-5QXJ] (last visited Feb. 16, 2025) ("Google Analytics is used by 81.7% of all the websites whose traffic analysis tool we know. This is 47.4% of all websites.").

Table 1. Companies' presence with first- and third-party cookies. Sorted by first-party cookie prevalence.[270]

| Organization | Percentage of sites with first-party cookies | Percentage of sites with third-party cookies |
|---|---|---|
| Google LLC | 80.74 | 73.78 |
| Facebook, Inc. | 26.23 | 5.09 |
| Microsoft Corporation | 15.78 | 22.83 |
| Amazon Technologies, Inc. | 8.64 | 10.26 |
| Hotjar Ltd | 8.26 | 0.04 |
| Yandex LLC | 6.63 | 8.32 |
| OneTrust LLC | 6.11 | 0.11 |
| Criteo SA | 5.88 | 11.06 |
| Twitter, Inc. | 5.05 | 10.63 |
| Quantcast Corporation | 3.95 | 9.82 |
| ByteDance Ltd. | 3.93 | 4.30 |
| HubSpot, Inc. | 3.70 | 3.58 |
| Adobe Inc. | 3.28 | 16.89 |
| Oracle Corporation | 3.10 | 13.39 |
| Pinterest, Inc. | 3.00 | 3.13 |
| Baidu, Inc. | 2.41 | 2.60 |
| Salesforce.com, Inc. | 2.08 | 7.13 |
| ID5 Technology Ltd | 2.07 | 6.72 |
| Lotame Solutions, Inc. | 2.01 | 8.47 |
| Taboola, Inc. | 1.82 | 7.13 |

270. *See* Munir et al., *supra* note 258, at 3502

#### *c. Use of Chrome to Adapt to Privacy Restrictions*

Historically, publishers have incorporated Google's services into their websites as third-party scripts, enabling the setting of third-party cookies to store user identifiers across various domains.[271] Concerns over privacy implications of allowing anyone to set these third-party cookies, and subsequently track user activity across different sites, have resulted in widespread criticism of third-party cookies.[272] Due to these concerns, Firefox and Safari recently blocked third-party cookies.[273]

Google has not enacted the same privacy protections as its peer browsers.[274] After initially dragging its feet on third-party cookies, Chrome revealed its intention to phase out third-party cookies by introducing the Privacy Sandbox initiative in early 2020.[275] There is some evidence that Google has advocated for the adoption of first-party cookies within its analytics and advertising tools since 2018.[276] While first-party cookies cannot directly identify users in a similar fashion to third-party cookies, Google's documentation provides evidence that their first-party cookies, in combination with other techniques, can be used to track a user's activity across different domains.[277] Google's reliance on third-party cookies and the subsequent shift in rhetoric in 2019 to phase out third-party cookies could be due to its successful shift to first-party cookies.

---

271. *See* Oliver Fich, *Third-Party Cookie Deprecation in Chrome: In-Depth Overview*, COOKIE INFO. (Apr. 3, 2024), https://cookieinformation.com/resources/blog/end-of-third-party-cookie/ [https://perma.cc/7DB9-V5KL].

272. *See id.*

273. *See id.*

274. *See supra* Section II.B.1.

275. *See* Frederic Lardinois, *Google Wants to Phase Out Support for Third-Party Cookies in Chrome Within Two Years*, TECHCRUNCH (Jan. 14, 2020, 8:00 AM), https://techcrunch.com/2020/01/14/google-wants-to-phase-out-support-for-third-party-cookies-in-chrome-within-two-years/ [https://perma.cc/ZSC8-MYPF] ("Google today announced its plans to phase out support for third-party cookies in Chrome within the next two years."); Joe Duball, *Google Ends Third-Party Cookie Phaseout Plans*, IAPP (July 23, 2024), https://iapp.org/news/a/google-ends-third-party-cookie-phaseout-plans [https://perma.cc/4ZKX-96UA] ("The company announced 22 July intentions to leave cookies available on its Chrome browser while reworking its Privacy Sandbox initiative to address a balance between consumer privacy and sustainable advertising.").

276. *See* Google Partners Livestream, *Academy on Air: Sitewide Tagging*, YOUTUBE, at 02:52 (Aug. 30, 2018), https://www.youtube.com/watch?v=8qvyP_7-PnA [https://perma.cc/Y6GU-ATAZ] ("[I]n the near future, first-party cookies will be the way to go in the industry."); *Ad Serving Settings*, GOOGLE: ADSENSE HELP, https://web.archive.org/web/20201027223352/https:/support.google.com/adsense/answer/3234887?hl=en [https://perma.cc/S6B2-R4V2] (last visited Feb. 1, 2025) ("Choose whether or not you want to allow first-party cookies from Google on your site. . . . Allowing first-party cookies from Google may increase your revenue because it enables features like frequency capping on ads and allows ads with a frequency cap to serve on your site.").

This shift towards first-party cookies is also visible in some other aspects of Chrome. In Chrome Version 112, Google updated the cookie settings page by removing the option to block all cookies, including both first-party and third-party cookies.[278] Now, in Version 112, users can only block third-party cookies.[279] The removal of the explicit option to block all cookies can be interpreted as a strategic move to discourage users from opting for a more restrictive cookie policy that would impede the functionality of first-party cookies. First-party cookies are integral to Google's business model, as they enable the company to collect vast amounts of user data directly from its services, including Chrome.[280] This data is invaluable for refining Google's targeted advertising algorithms, which represent the core of its revenue stream.[281]

---

277. *See [GA4] Set Up Cross-Domain Measurement*, GOOGLE: ANALYTICS HELP, https://support.google.com/analytics/answer/10071811?hl=en [https://perma.cc/3EJS-C3PD] (last visited Feb. 1, 2025) ("With cross-domain measurement, the cookies retain the same IDs as they are passed from one domain to another via a URL parameter (_gl) when the user navigates between domains through a link or a form. As a result, Analytics identifies just one user and one session.").

278. *See supra* Figure 7; *supra* Figure 8; User 11102988865155881675, GOOGLE: GOOGLE CHROME HELP (July 23, 2023), https://support.google.com/chrome/thread/227023954/can-no-longer-block-cookies-on-google-chrome-on-windows?hl=en [https://perma.cc/V69Y-6ZZZ] ("'Cookies and other site data' is no longer in the 'Privacy and security' menu.").

279. *See Delete, Allow and Manage Cookies in Chrome*, GOOGLE: GOOGLE CHROME HELP, https://support.google.com/chrome/answer/95647?hl=en&co=GENIE.Platform%3DDesktop [https://perma.cc/R779-XMNJ] (last visited Feb. 1, 2025).

280. *See* Munir et al., *supra* note 258, at 3501 ("These first-party tracking cookies are set by third-party embedded scripts served from 2,099 domains that include major advertising entities such as Google, Facebook, and TikTok.").

281. *See How Our Business Works*, GOOGLE: ABOUT, https://about.google/how-our-business-works/#:~:text=Ultimately%2C%20we%20earn%20most%20of,we%20make%20money%20with%20advertising [https://perma.cc/5FUF-ZL5W] (last visited Feb. 1, 2025) ("Ultimately, we earn most of our money by showing ads alongside relevant Search results on Google.com. If you're interested, you can learn more about how we make money with advertising.").

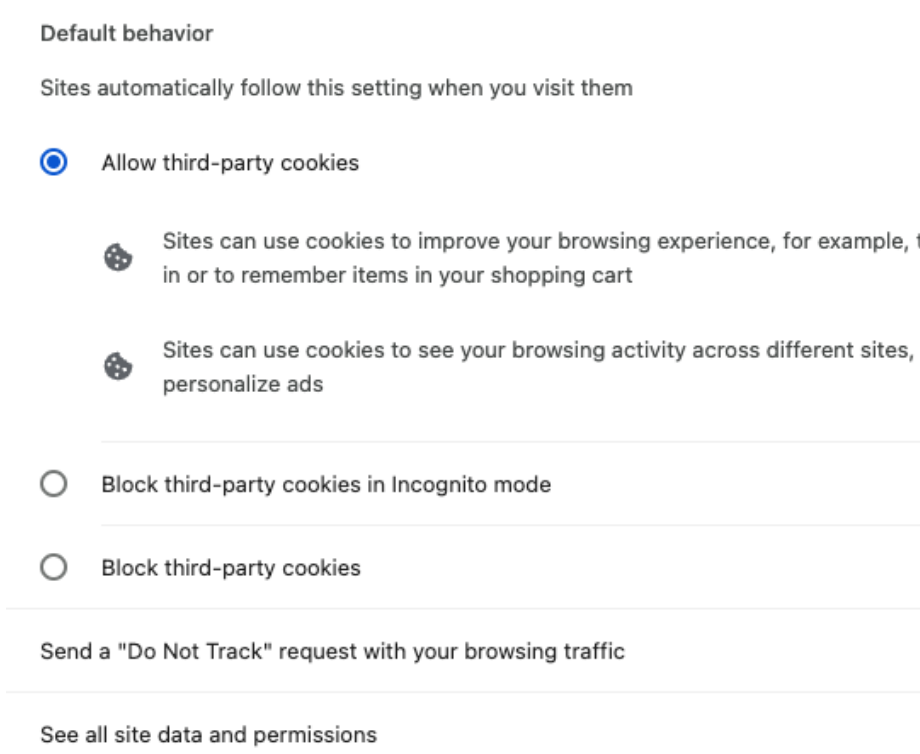

Figure 7 New cookie settings in Google Chrome, with the option to block all cookies missing.

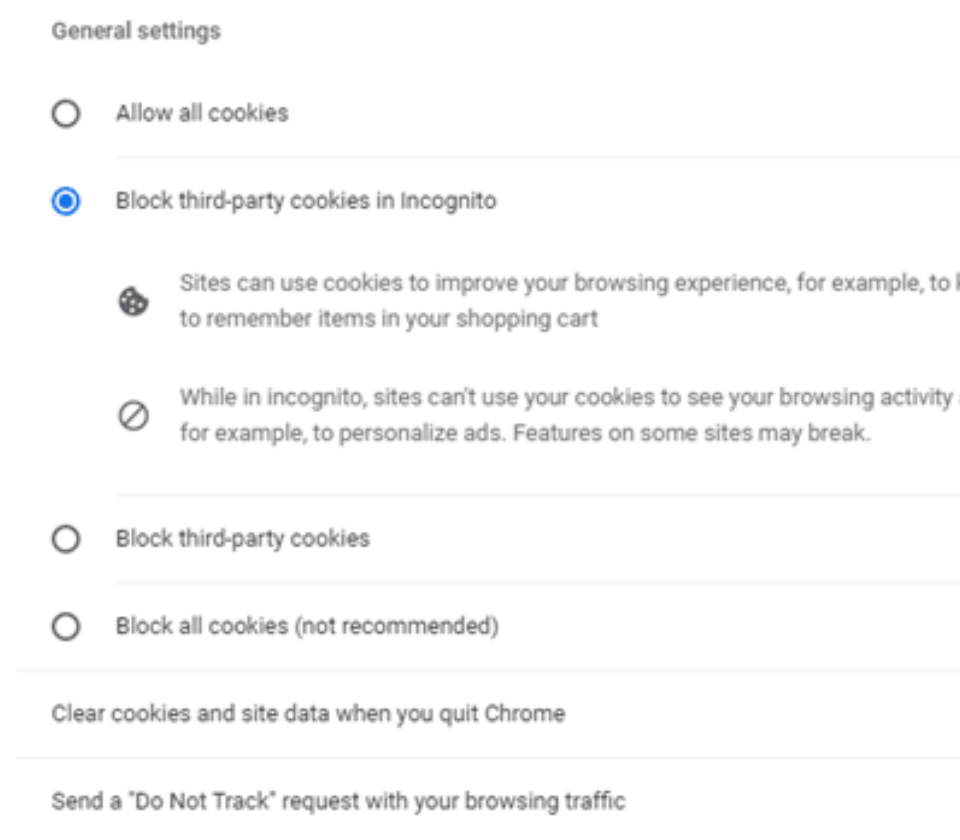

Figure 8 Older cookie settings in Google, which allowed users to block all cookies, including first-party cookies.

Another interesting aspect of Google's transition to first-party cookies is the speed of transition. Transitioning to first-party cookies requires a third party to establish a level of trust with the website publisher, due to the potential risks such as credential theft or other malicious activities if the third-party has access to the first-party context.[282] Leveraging its substantial market share and established relationships with publishers, there is evidence to suggest that Google has moved relatively swiftly to adopt first-party cookies across a vast array of websites. Recent research analyzing the use of first-party cookies across a sample of 20,000 websites from the top 100,000 most-visited sites revealed that Google has implemented first-party cookies on approximately 80% of these websites.[283] Notably, Google

282. Chen et al., *Cookie Swap Party*, *supra* note 251, at 2117 ("This cross-domain inclusion of third-party JS code poses security and privacy risks."); *see Cross-Domain JavaScript Source File Inclusion*, STACKHAWK, https://docs.stackhawk.com/vulnerabilities/10017/#:~:text=Cross%2DDomain%20JavaScript%20Source%20File%20Inclusion%20occurs%20when%20a%20web,leading%20to%20various%20security%20risks [https://perma.cc/JR7D-VESY] (last visited Feb. 1, 2025).

283. *See* Munir et al., *supra* note 258, at 3500 ("Two major advertising entities (Google and Facebook) set first-party ATS cookies on approximately a third of all sites in our dataset. CookieGraph detects _gid and _ga cookies by Google Analytics as ATS on 62.63% and 53.27% of the sites.").

owns six out of the top ten domains that were identified as setting first-party cookies in this sample.[284]

Table 1 illustrates the deployment of first-party cookies by various organizations. Google is prominently reliant on first-party cookies, significantly outpacing its competitors.[285] For instance, Facebook, which has the second-largest footprint in terms of first-party cookie use, still only achieves less than a third of Google's extensive reach and presence.[286] Other competitors, including Criteo, Adobe, Yandex, and Oracle, predominantly utilize third-party cookies and face the biggest threat in the case of complete third-party cookie blockage.[287] On the other hand, any restrictions on third-party cookies by Chrome or other web browsers is now unlikely to substantially affect Google's capacity for user tracking through first-party cookies. Google's deployment of first-party cookies demonstrates that Google used its dominance in the browser market to effectively sidestep any considerable impact on its advertising business by delaying third-party cookie blockage.[288] Now, Google is in a prime position to leverage its browser market share against its advertising competitors.

Indeed, the CMA came to the same conclusion after scrutiny of Google's proposed elimination of third-party cookies and its planned introduction of the Privacy Sandbox.[289] The CMA's findings suggest that these changes, as they stand, would significantly harm Google's competitors that rely on third-party cookies, at least in the short-run as they adapt their strategies.[290] Furthermore, the CMA deemed Google's suggested alternatives to third-party cookies insufficient for rival advertisers and analytics services.[291] According to the CMA, the

284. *See id.* at 3495 ("Six of the top-10 ATS domains, all owned by Google, show only a negligible reduction in the number of ATS requests with identifiers when third-party cookies are blocked.").

285. *See id.* at 3491.

286. *See id.* at 3500.

287. *See id.* at 3501.

288. *See* Heidi Bullock, *The Balancing Act: Google's Cookie Conundrum and the Future of Advertising*, FORBES (Aug. 23, 2024, 7:15 AM), https://www.forbes.com/councils/forbescommunicationscouncil/2024/08/23/the-balancing-act-googles-cookie-conundrum-and-the-future-of-advertising/ [https://perma.cc/L7BU-83P7].

289. *See* Duball, *supra* note 275.

290. Competition and Markets Authority, *Investigation*, *supra* note 151 ("On 7 January 2021, the CMA launched an investigation under Chapter II of the Competition Act 1998 into suspected breaches of competition law by Google. The investigation concerns Google's proposals to remove third-party cookies (TPCs) on Chrome and replace TPCs functionality with a range of 'Privacy Sandbox' tools, while transferring key functionality to Chrome.").

291. *See* Nicola Agius, *Google 'Cannot Proceed with Third-Party Cookie Deprecation'*, SEARCH ENGINE LAND (Feb. 5, 2024, 10:34 AM), https://searchengineland.com/google-cannot-proceed-third-party-cookie-deprecation-437212 [https://perma.cc/BY9A-CGF6].

transition period was not adequate for competitors to adapt without incurring substantial revenue losses, indicating up to a 70% decline in revenue following the restriction of third-party cookies.[292] The concerns highlighted by the CMA prompted Google to postpone the phasing out of third-party cookies several times.[293] Meanwhile, its peers like Firefox and Safari moved forward with the phaseout due to their noninvolvement in the advertising business.[294]

The most notable tool in Google's arsenal of mitigation measures is the Topics API, which categorizes a user's interests into general themes or "Topics."[295] Due to a mandate from the CMA, Chrome is collaborating with other advertising and analytics entities to assess the Topics API's efficacy.[296] Evaluations by Audience X and Criteo, companies dependent on third-party cookies, hold that the API is an inadequate substitute for third-party cookies.[297] Audience X's analysis suggests that the Topics API lacks the specificity of third-party cookies, underscoring the increased value of first-party data for website operators and advertisers, so much so that "website owners and advertisers will need to put a premium on acquiring and managing first-party data."[298] Criteo's trial revealed that Topics API was five times less effective than third-party cookies.[299] However, Criteo also noted that the Topics API's effectiveness could be improved when supplemented with additional inputs like first-party data and contextual cues.[300] This point is pivotal as Google's extensive collection of first-party data potentially amplifies the effectiveness of third-party cookie alternatives such as the Topics API for Google rather than for its competitors.

---

292. *See* COMPETITION & MKTS. AUTH., *supra* note 149.

293. Goel, *Updated Timeline*, supra note 81; Chavez, *New Path*, *supra* note 83.

294. *See The End of Advertising Cookie Brings Opportunities*, IO DIGITAL (Oct. 10, 2022), https://www.iodigital.com/en/insights/blogs/end-ad-cookie-opportunities [https://perma.cc/4QET-E4TS].

295. *See* Prasant Naidu, *Google Privacy Sandbox*, PN'S BLOG (Feb. 25, 2022), https://prasantnaidu.substack.com/p/google-privacy-sandbox [https://perma.cc/7XNA-7XB5] ("[The] Topics API infers coarse-grained interest signals on-device based on a user's app usage. These signals, called topics, will be shared with advertisers, supporting IBA use cases without requiring tracking of individual users across apps.").

296. *See* Competition and Markets Authority, *Investigation*, *supra* note 151; *Chrome-Facilitated Testing*, GOOGLE: PRIV. SANDBOX, https://developers.google.com/privacy-sandbox/setup/web/chrome-facilitated-testing [https://perma.cc/4GFF-TEUK] (Nov. 20, 2024) ("We are providing Chrome-facilitated testing modes that allow sites to preview how site behavior and features work without third-party cookies.").

297. *See The Impact of the Google Topics API*, AX INSIGHTS (July 7, 2022), https://audiencex.com/insights/google-topics-api/ [https://perma.cc/N4T7-UWTK].

298. *See id.* ("Since Google API reflects only broad user behavior from site visits, it will be less granular than what advertisers are used to or like.").

299. Elias Selman, *Topics API: Criteo's First Look at Google's Interest-Based Advertising Solution*, MEDIUM (Nov. 10, 2022), https://medium.com/criteo-engineering/is-googles-topics-api-a-viable-replacement-for-interest-based-advertising-297076192bd [https://perma.cc/D3AH-VSHC] ("Overall, we observe that Criteo's interest-based audiences are five times more relevant than those generated by this first iteration of the Topics API.").

While there is merit to the argument that some of the negative effects reported may only be short-lived and will be mitigated over time as firms adapt; in the meantime, Google might be able to snap up even more market share than it already has. After all, Google would have, at least for a while, some of the most attractive advertising offerings.[301]

---

300. *Id.* ("We are also aware that the signals the Topics API provides could have further utility if combined with other signals such as first-party data and contextual information, and this will be a matter for our future experimentation.").

301. *See* Todd Parsons, *Privacy Sandbox Testing Results Show Shortfalls to Meet CMA Requirements*, CRITEO, https://www.criteo.com/blog/privacy-sandbox-testing-results-show-shortfalls-to-meet-cma-requirements/ [https://perma.cc/SC4H-PDGW] (Aug. 7, 2024). This is reminiscent of the public clash between Apple and Facebook over Apple's planned roll-out of an opt-in (rather than opt-out) user tracking feature with iOS 14 from late 2020. *See* Samantha Murphy Kelly, *Facebook Fueds with Apple over Privacy Changes That Threaten Its Advertising Business*, CNN BUS., https://www.cnn.com/2020/12/16/tech/facebook-apple-ios-privacy-rules/index.html [https://perma.cc/Y9GW-A5JJ] (Dec. 16, 2020, 6:40 PM). While an opt-in to user tracking is already required by many privacy laws such as those in the EU, these laws have long been ignored by the industry. *See* Konrad Kollnig, Reuben Binns, Max Van Kleek, Ulrik Lyngs, Jun Zhao, Claudine Tinsman & Nigel Shadbolt, *Before and After GDPR: Tracking in Mobile Apps*, 10 INTERNET POL'Y REV., no. 4, 2021, at 2, 8–9. For example, it was found that more than 70% of Android apps used to send data to tracking companies once apps were opened the first time, but less than 3.5% implemented the opt-in mechanisms that are legally required under EU law. *Id.* at 8–9, 23. Tracking, if not for a good reason and communicated in a transparent manner, remains highly unpopular with end-users, most of which feel like neither private companies nor the government sufficiently give them control over their data. *See* MARY MADDEN & LEE RAINIE, PEW RSCH. CTR., AMERICANS' ATTITUDES ABOUT PRIVACY, SECURITY, AND SURVEILLANCE 3, 22 (2015). Putting users in control over tracking would also limit the industry's ability to create profiles about users and monetize those through advertising. Hence, Facebook ended up running a public campaign against Apple in leading newspapers. *See* Brandon Baum-Zepeda, *Apple vs. the Free Internet? Privacy and Antitrust in Mobile App Advertising*, 25 U.C. DAVIS BUS. L.J. 30, 33 (2025). In those ads (some of which were titled "Apple vs. the free internet"), Facebook claimed that it stood up to small businesses and helped them reach potential customers through targeted advertising, which Apple sought to undermine through the announced changes. *See id.* Regardless, Apple went ahead with those changes and many advertising companies, including Facebook, suffered a significant loss in revenues. *Id.* at 53. Indeed, it was reported that many advertisers shifted their budgets from iOS apps to Android apps, which did not face similar restrictions. *See id.* at 56. Some advertisers also shifted their marketing budget from in app solutions to Apple's own advertising systems. *See* John Koetsier, *Apple's Privacy Changes Slashed Ad ROI 38%. This Company Says They Can Fix It*, FORBES (May 18, 2022, 9:36 PM), https://www.forbes.com/sites/johnkoetsier/2022/05/18/apples-privacy-changes-slashed-ad-roi-38-this-company-says-they-can-fix-it/ [https://perma.cc/UJ92-CPUU]. Despite those concerns, Facebook revenues seem to have recovered. *See* Kali Hays, *Meta Is Recovering from Apple's Privacy*

Criteo's testing revealed that even with the new Privacy Sandbox APIs, the market share of Google Ad Manager would increase from 23% to a staggering 83% in the absence of third-party cookies.[302] These concerns by both regulators and competitors eventually forced Google to drop plans for deprecating third-party cookies in Chrome.[303]

### *d. Role of First-Party Data Collection in Google's Dominance of the Advertising Market*

The assessments conducted by Criteo and Audience X suggest that the Privacy Sandbox APIs cannot independently replace third-party cookies, forcing Google to keep third-party cookies around despite significant privacy concerns.[304] The assessments suggest that instead of relying on only Privacy Sandbox APIs, a combination of different techniques is necessary to replace third-party cookies while maintaining advertising revenue.[305] One of the techniques mentioned is the use of first-party data.[306]

Although trackers are beginning to use first-party cookies in anticipation of third-party cookie restrictions, first-party cookies have limitations for tracking purposes. Unlike third-party cookies, which can be accessed across various websites visited by a user, first-party cookies are confined to the domain where they were set.[307] This restriction limits the ability of trackers and advertisers to monitor and gather

---

*Changes, Wall Street Expects a Return to Major Business Growth*, BUS. INSIDER (Apr. 14, 2023, 1:36 PM), https://www.businessinsider.com/meta-recovers-from-apple-ad-privacy-hit-major-growth-expected-2023-4 [https://perma.cc/5PPT-TF4C]. What remains are investigations by various competition and data protection authorities against Apple's opt-in measures, a major fine against the company by the French data protection authority over Apple's own user tracking practices, and a notably increased share of iOS advertising for Apple, whose market share has tripled since. Baum-Zepeda, *supra*, at 34; *see* Laura Kayali, *Apple Fined €8M in French Privacy Case*, POLITICO (Jan. 4, 2023, 5:57 PM), https://www.politico.eu/article/apple-fined-e8-million-in-privacy-case/ [https://perma.cc/8KQP-3RL5]. This underlines the risk of market capture by large tech companies that persists even if the underlying anti-competitive mechanisms are addressed over time. Google's phase out of third-party cookies could bring similar risks. *See generally* Parsons, *supra*.

302. *See* Parsons, *supra* note 301.

303. *See* Chavez, *New Path*, *supra* note 83.

304. *See* Parsons, *supra* note 301.

305. *See id.*

306. *Id.*

307. *See* Munir et al., *supra* note 258, at 3492 ("While third-party cookies have been used extensively in cross-site tracking, i.e., where a tracker links a user's activity across sites, the mechanisms by which first-party cookies are used in cross-site tracking have not been studied so far." (emphasis omitted)).

comprehensive user data across multiple websites.[308] To circumvent the constraints of third-party cookie alternatives, trackers are increasingly focusing on aggregating more extensive collection of first-party data.[309] This data often comprises personally identifiable information, such as email addresses and phone numbers, which facilitates the recognition of users across various websites.[310] For instance, if a tracker operates on two separate websites and a user enters the same email address on both, the tracker can then link the user's activities across these sites.

#### i. First-Party Data Collection through New Tools

Google has been intensifying its efforts to gather first-party data, complementing its reliance on first-party cookies and other substitutes like the Topics API.[311] A prime example is Google Tag Manager, which enables publishers to manage first-party data for purposes such as analytics and advertising conversion tracking.[312] This tool works synergistically with Google Analytics to capture various user activities on websites and mobile apps, which allows Google to obtain a large amount of first-party data from publishers.[313] A recent study revealed that googletagmanager.com ranks as the third most popular domain, trailing only behind google-analytics.com and doubleclick.net,

---

308. *See id.* ("Previous research has also shown that it is non-trivial to generate first-party identifiers that are accessible across websites.").

309. *Id.* ("[E]xperiments show that trackers make use of identifiers like email addresses to link user activity across different sites. They make use of this knowledge to perform identity entanglement, where an attacker can make use of an email address or other identifiers to influence the advertisements shown to a victim. This sharing of additional information when third-party cookies are blocked allows trackers to track users across different sites.").

310. *See How Google Uses Customer Match Data*, GOOGLE: GOOGLE ADS HELP, https://support.google.com/google-ads/answer/6334160?hl=en#:~:text=Google%20doesn%27t%20receive%20actual,is%20not%20unencrypted%20by%20Google [https://perma.cc/6GMT-62SC] (Apr. 2021) ("Google doesn't receive actual email addresses. Google's system transforms the contact information we have for Google accounts, like email addresses and phone numbers, into hashed codes using the secure hashing algorithm SHA256, a one-way hashing mechanism that is not unencrypted by Google.").

311. *See* James Ioannidis, *New Ways to Support Your Measurement with First-Party Data*, GOOGLE: GOOGLE ADS HELP (Apr. 15, 2021), https://support.google.com/google-ads/answer/10591309?hl=en [https://perma.cc/MJ5J-Y4PS].

312. *See id.* ("Consistent with how existing first-party cookies work, the new cookie will be unique and limited to users on your site only. Starting in May 2021, this first-party cookie will enable more accurate attribution of conversions, including instances where a user might engage with more than one of your ads before converting.").

313. *See Send Data to Server-Side Tag Manager*, GOOGLE: TAGS, https://developers.google.com/tag-platform/tag-manager/server-side/send-data?option=gtag [https://perma.cc/R6UQ-ERHR] (Nov. 12, 2024) ("[S]et up the Google Analytics 4 client in your server container to parse the additional parameters and create event data out of them.").

showing the sheer number of publishers sharing first-party data with Google.[314]

To improve first-party data collection and sharing, Google is making it significantly easier for publishers to share first-party data.[315] A significant development in this direction was the incorporation of Seller Defined Audiences (SDA) into Google Ad Manager in 2022.[316] The SDA feature simplifies the process for publishers to offer first-party data to potential buyers.[317] SDA enables the structuring of user data collected by publishers into a widely recognized format, facilitating seamless transactions with programmatic buyers throughout the advertising sector.[318] By standardizing this format, publishers can transfer data to Google more efficiently without incurring the additional overhead associated with proprietary data formats.[319]

Furthermore, towards the end of 2022, Google launched the Publisher Advertiser Identity Reconciliation (PAIR) service.[320] Whereas SDA facilitates the structured sharing of user data, PAIR enables the merging of first-party data between distinct publishers and advertisers.[321] This collaboration yields access to more nuanced user profiles, encompassing preferences, dislikes, demographics, and more.[322] Despite being promoted as an inclusive platform open to various industry players, PAIR is perceived by some as a Google-centric solution that primarily benefits the Google ecosystem.[323]

---

314. Munir et al., *supra* note 258, at 3502 tbl.5.

315. *See* Peentoo Patel, *New Ways for Publishers to Manage First-Party Data*, GOOGLE: AD MANAGER (Sept. 13, 2022), https://blog.google/products/admanager/new-ways-for-publishers-to-manage-first-party-data/ [https://perma.cc/C6K2-UT7Q].

316. *Id.* ("As a first step, we are integrating the IAB Tech Lab's Seller Defined Audiences into this solution. Publishers can use the IAB's Audience Taxonomy and Content Taxonomy to share signals with Google Ads and Display & Video 360 as part of our beta testing."); *see Curated Audiences*, IAB TECH LAB, https://iabtechlab.com/sda [https://perma.cc/TSW7-DG68] (Dec. 16, 2024) ("Curated Audiences (formerly Seller Defined Audiences or SDA) is an addressability specification incubated within Project Rearc. It allows publishers, DMPs and data providers to scale first-party data responsibly and reliably without data leakage or reliance on deprecated IDs and/or new, untested technologies.").

317. Patel, *supra* note 315 ("These signals make it easier for programmatic buyers to find and purchase audiences based on things like demographics, content interests or purchase intent across multiple sites and apps without tracking people's activity in apps or across the web.").

318. *See id.* ("Seller Defined Audiences was created by the advertising industry via Project Rearc, to help advertisers and publishers responsibly and reliably share first-party data at scale without the need for user identifiers. It's a great step forward to have Google Ad Manager adopt these new standards for their publisher partners, and help advance a new system for addressability and accountability that meets user privacy expectations." (quoting Anthony Katsur, CEO, IAB Tech Lab)).

319. *See id.* ("We're introducing publisher provided signals to help you categorize your first-party data into consistent audience or contextual segments and then share these signals with programmatic buyers.").

320. Dan Taylor, *Engage Your First-Party Audience in Display & Video 360*, GOOGLE: MKTG. PLATFORM (Oct. 11, 2022), https://blog.google/products/marketingplatform/360/engage-your-first-party-audience-in-display-video-360/ [https://perma.cc/3UWS-KD9Q] ("Publisher Advertiser Identity Reconciliation, or PAIR, is a new solution that gives publishers and advertisers the option to securely and privately reconcile their first-party data for audiences who have visited both an advertiser's and a publisher's site.").

The SDA and PAIR developments demonstrate Google's push towards collection of first-party data and its continued use in its existing services. First-party data is critical in ensuring that businesses are not impacted by any restrictions on third-party cookies by other browsers.[324] This collection of first-party data gives Google an unfair advantage over other competitors, who do not have access to such a vast amount of first-party data.[325]

---

321. *Id.* ("Publisher Advertiser Identity Reconciliation, or PAIR, is a new solution that gives publishers and advertisers the option to securely and privately reconcile their first-party data for audiences who have visited both an advertiser's and a publisher's site. Advertisers and publishers will be able to activate encrypted first-party information that is unique to their sites via aggregation.").

322. *See id.* ("PAIR gives advertisers the ability to more closely connect with their known audiences, while avoiding tracking individuals across the web. As a result, advertisers can show relevant ads to some of their highest-intent audiences, helping to increase advertising performance and hit marketing objectives, while respecting people's privacy expectations.").

323. Trey Titone, *What Is Google PAIR? A New First-Party Data Solution*, AD TECH EXPLAINED (Oct. 17, 2022), https://adtechexplained.com/google-pair-a-not-so-universal-identity-solution/ [https://perma.cc/HTM7-PQBQ] ("First, the most obvious difference is that PAIR is only a solution for Google's DSP, Display & Video 360 (DV360). While other solutions are DSP-agnostic, PAIR is a solution by Google for Google.").

324. *See* Brian Quinn, *How Marketers Can Thrive amid Signal Loss: Data Strategies in a Privacy-First World*, VENTURE BEAT (July 30, 2024, 7:20 AM), https://venturebeat.com/ai/how-marketers-can-thrive-amid-signal-loss-data-strategies-in-a-privacy-first-world/ [https://perma.cc/8RG9-VTA4] ("As a result, advertisers must increasingly rely on first-party data, invest in privacy-first solutions and adopt flexible budgeting strategies. The dominance of walled gardens like Google and Facebook, which have vast amounts of first-party data, also increases."); David Chan & Kelly Leger, *With or Without Third-Party Cookies, First-Party Data Remains Paramount*, WALL ST. J. (Aug. 14, 2024, 2:00 PM), https://deloitte.wsj.com/cmo/with-or-without-third-party-cookies-first-party-data-remains-paramount-4772b8a6 [https://perma.cc/8A22-C3YB] ("The cookieless world is already here—and first-party data ownership remains critical. By prioritizing first-party data, companies may be better positioned to avoid the loss of customer signals associated with the decline of third-party cookies. At the same time, investment in first-party data solutions is likely critical to enhanced personalization and more connected experiences—strategies that are vital for marketers today.").

325. *See The Impact of the Google Topics API*, *supra* note 297.

#### ii. Dark Pattern Usage by Google to Encourage Collection of First-Party Data

Google employs dark patterns to prompt users into sharing more first-party data, which enhances its data collection strategies.

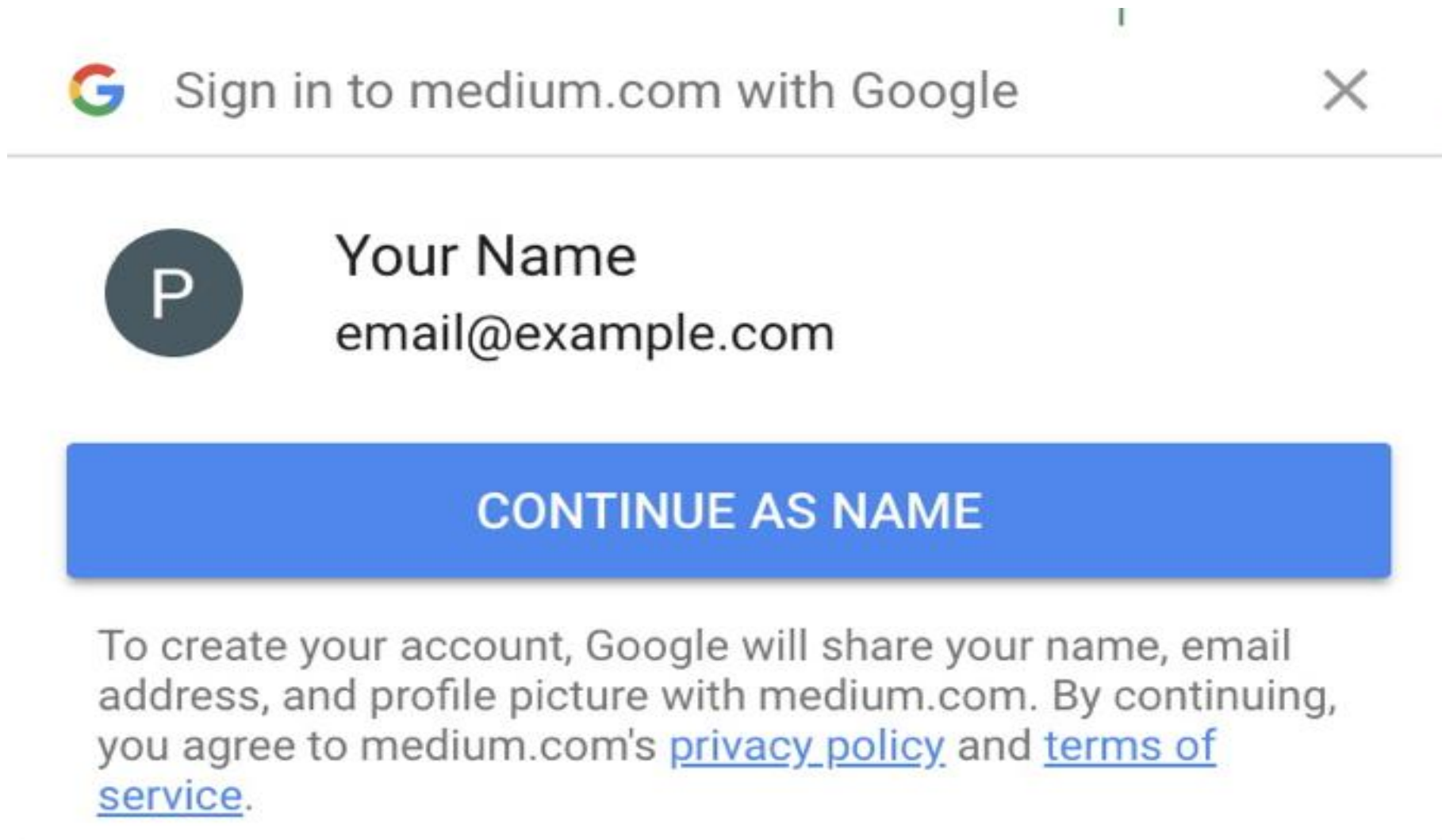

Figure 9. Example of Google SSO on medium.com

One common example of dark pattern practices is Google's integration of its single sign-on (SSO) service across various websites.[326] Users visiting sites that offer Google login options with a pop-up featuring their Google account details are encouraged to authenticate with Google's credentials.[327] This login method permits Google to monitor user interactions on third-party websites and logs the user into other Google services in the background, thereby broadening its first-party data acquisition scope.[328] One example of the additional data collection is through Google Ads' "DSID" cookie, which is deployed when

326. *See* Srivathsan G. Morkonda, Sonia Chiasson & Paul C. van Oorschot, *Empirical Analysis and Privacy Implications in OAuth-Based Single Sign-On Systems*, *in* WPES '21, PROCEEDINGS OF THE 20TH WORKSHOP ON PRIVACY IN THE ELECTRONIC SOCIETY 195, 202 (2021).

327. *See id.* at 206.

328. *See* Catalin Cimpanu, *Google Secretly Logs Users into Chrome Whenever They Log into a Google Site*, ZDNET (Sept. 23, 2018, 7:05 PM), https://www.zdnet.com/article/google-secretly-logs-users-into-chrome-whenever-they-log-into-a-google-site/ [https://perma.cc/Y5NT-VLNW] ("Now, with the revelations of this new auto-login mechanism, a large number of users are angry that this sneaky modification would allow Google to link that person's traffic to a specific browser and device with a higher degree of accuracy.").

a user logs in via Google on a third-party website.[329] This cookie assigns a unique ID to the user, which aggregates their advertisement preferences.[330] In response to this egregious use of pop-ups to coerce users into logging in, DuckDuckGo has developed an application and browser extension aimed at obstructing Google's ability to track users through this DSID cookie.[331]

Google has also restructured its services from distinct subdomains to centralized paths under its main domain. For example, Google transitioned from maps.google.com to google.com/maps.[332] This strategic move permits Google to apply the same data collection permissions across multiple services.[333] For instance, once a user consents to location sharing on google.com/maps, this permission extends seamlessly to other services like google.com/flights, without the user ever having given explicit permission to Google Flights for location sharing.[334] Such a unified permission model simplifies Google's data collection process across its various platforms.[335]

However, these deceptive patterns used by Google have resulted in legal troubles for the tech giant. One lawsuit arising in Arizona alleged the company's data collection practices to be opaque.[336] The lawsuit resulted in a settlement of $85 million due to Google's deceptive

329. *See How Google Uses Cookies*, GOOGLE: PRIV. & TERMS, https://policies.google.com/technologies/cookies?hl=en-US [https://perma.cc/S4LH-W3C6].

330. *See id.* ("[T]he 'DSID' cookie is used to identify a signed-in user on non-Google sites so that the user's ads personalization setting is respected accordingly. [It] lasts for 2 weeks.").

331. Hisan Kidwai, *DuckDuckGo Now Auto-Blocks Google Sign-in Pop-Ups on All Sites*, ANDROID HEADLINES (Dec. 26, 2022), https://www.androidheadlines.com/2022/12/duckduckgo-google-sign-in-auto-block-pop-ups.html [https://perma.cc/ECC6-89B8] ("[E]ven if you sign into sites not owned by Google, they can still track your behavior and collect your data on those sites.").

332. Ankit Ghosh, *Google Changed Google Maps URL: Your Location Data Is No Longer Safe*, SIMPLE ANALYTICS, https://www.simpleanalytics.com/blog/google-changed-google-maps-url-your-location-data-is-no-longer-safe [https://perma.cc/42Z5-GAVC] (Dec. 20, 2023).

333. Garrit Franke, *Smart Move, Google*, GARRIT'S NOTES BLOG (Nov. 23, 2022), https://garrit.xyz/posts/2022-11-24-smart-move-google [https://perma.cc/3GYH-538J] ("This implies that the permissions I give to Google Maps now apply to all of Googles services hosted under this domain." (emphasis omitted)).

334. *See id.* ("Congratulations, you now have permission to geo-track me across all of your services. Smart move, Google." (emphasis omitted)).

335. *See* Ghosh, *supra* note 332.

336. Press Release, Kris Mayes, Arizona Att'y Gen., Attorney General Mark Brnovich Achieves Historic $85 Million Settlement with Google (Oct. 4, 2022), https://www.azag.gov/press-release/attorney-general-mark-brnovich-achieves-historic-85-million-settlement-google [https://perma.cc/S6BB-2CPV] ("[A]n investigation of Google after a 2018 Associated Press article revealed that the company was misleading and deceiving consumers about the collection and use of their personal location data . . . .").

tracking of users' location data.[337] Moreover, internal communications revealed in discovery exposed criticisms by employees over the complexity of Google's privacy settings.[338] One employee noted that the user interface seemed "designed to make things possible, yet difficult enough that people won't figure it out . . . ."[339] Similarly, in a separate lawsuit arising in California, Google agreed to a $93 million settlement over allegations of unauthorized data collection.[340] The Arizona Superior Court found that Google was deceiving users by employing coercive methods to gather data without providing a clear and accessible opt-out option.[341]

These new tools, such as SDA integration in Google Ads Platform and PAIR services, and dark patterns employed by Google are designed to allow the collection and utilization of substantial volumes of first-party data.[342] The integration of first-party data with Google's proposed Topics API and other emerging standards significantly strengthens Google's competitive position.[343] This advantage becomes even more pronounced without access to third-party cookies because the data collection capabilities of Google's rivals become constrained.[344] In contrast, competitors like Criteo have observed performance disparities as high as fivefold.[345] Considering these changes are enacted on behest of Google's interests, they reveal a critical pattern of Google using its market share in Chrome to perpetuate its own advertising solutions and monopolize the market.

---

337. *Id.* ("The $85 million settlement directs the bulk of the money to the general fund and it will require legislative appropriation before it can be spent. Additionally, $5 million is specifically directed for attorney general education programs.").

338. Elizabeth Lopatto, *Even Google Engineers Are Confused About Google's Privacy Settings*, VERGE (Aug. 26, 2020, 5:30 PM), https://www.theverge.com/2020/8/26/21403202/google-engineers-privacy-settings-lawsuit-arizona-doubleclick [https://perma.cc/J679-4PDF] ("Google's privacy settings don't just confuse its users — they confuse its employees too, according to internal documents unsealed in a lawsuit over Google's data collection.").

339. *Id.*

340. Press Release, Rob Bonta, California Att'y Gen., Attorney General Bonta Announces $93 Million Settlement Regarding Google's Location-Privacy Practices (Sept. 14, 2023), https://oag.ca.gov/news/press-releases/attorney-general-bonta-announces-93-million-settlement-regarding-google's [https://perma.cc/M4QZ-ATAB].

341. *See* Expert Report of Colin M. Gray, Ph.D. at 35, Arizona *ex. rel.* Brnovich v. Google LLC, No. CV2020-006219 (Ariz. Super. Ct. May 4, 2022) ("In my opinion, these 'loopholes' that enabled Google to collect and use location data even when the user explicitly disabled location tracking through various combinations of settings is an example of the dark pattern strategies sneaking and forced action." (emphasis omitted)).

342. *See* Trey Titone, *Google Embraces Seller Defined Audiences*, AD TECH EXPLAINED (Sept. 20, 2022), https://www.adtechexplained.com/p/google-embraces-seller-defined-audiences [https://perma.cc/8EVZ-GRF4]; *About PAIR (Publisher Advertiser Identity Reconciliation) (Beta)*, GOOGLE: GOOGLE AD MANAGER HELP, https://support.google.com/admanager/answer/15067908?hl=en#:~:text=PAIR%20(Publisher%20Advertiser%20Identity%20Reconciliation)%20is%20a%20secure%20and%20privacy,reliance%20on%20third%2Dparty%20cookies [https://perma.cc/AY94-4KNS] (last visited Jan. 31, 2025).

343. *See* Anu Adegbola, *New Google Tools to Boost First-Party Data Strategies*, SEARCH ENGINE LAND (Aug. 28, 2024, 3:57 PM), https://searchengineland.com/new-google-tools-first-party-data-strategies-446107 [https://perma.cc/82M5-Q2UA].

*e. Sabotaging Web Standards to Maintain Market Dominance*

Google has also endeavored to solidify its advertising market dominance by shaping web standards in ways that predominantly advantage its own advertising services, consequently skewing the competitive landscape to its benefit.

In 2018, Google introduced Manifest v3 to Chrome, citing goals to enhance extension performance, privacy, and security.[346] This initiative, proposed with minimal engagement from other industry participants, has notably restricted the capabilities of adblocking, anti-tracking, and web security extensions such as AdBlock Plus, DuckDuckGo, and Norton Safe Web—tools that once significantly impacted Google's advertising revenues.[347] Manifest v3's shift from the webRequest API, which permitted extensions to scrutinize and block network requests, to the more restrictive DeclarativeNetRequest API, which restricts extensions to use prewritten, static rules to block tracking requests, has been a point of contention.[348] This shift from the webRequest API to the DeclarativeNetRequest API also includes limitations on the syntax and quantity of blocking rules that can be imposed by these extensions.[349] Despite considerable pushback from the adblocking community,[350] Google implemented Manifest v3 in Chrome v88 with plans to phase out the previous version, Manifest v2, in 2023.[351] In contrast, browsers like Brave and Firefox have opted to maintain support for Manifest v2, thus preserving the functionality of these third-party extensions.[352] Independent studies have challenged Google's justifications for Manifest v3, suggesting that it does not necessarily enhance browser performance and may degrade it by permitting more advertising and tracking scripts to load.[353] This episode around transition from Manifest v2 to v3 illustrates how Chrome has caused consumer harm by curbing the efficacy of third-party adblocking, anti-tracking web security extensions that users rely on for safety on the web.[354] Indeed, the Manifest v3 standard seems to primarily serve Google's interests in advertising and tracking.[355]

---

344. *See* Liptak, *supra* note 239.

345. *See* Selman, *supra* note 299.

346. James Wagner, *Trustworthy Chrome Extensions, by Default*, GOOGLE: CHROMIUM BLOG (Oct. 1, 2018), https://blog.chromium.org/2018/10/trustworthy-chrome-extensions-by-default.html [https://perma.cc/PF7P-U3GV] ("In 2019 we will introduce the next extensions manifest version. Manifest v3 will entail additional platform changes that aim to create stronger security, privacy, and performance guarantees.").

347. *See* Michael Crider, *Google Is Killing One of Chrome's Biggest Ad Blockers*, PCWORLD (Aug. 9, 2024 8:24 AM), https://www.pcworld.com/article/2423294/google-is-killing-one-of-chromes-biggest-ad-blockers.html [https://perma.cc/A4MD-PMCN] ("Most people on the internet use the Chrome browser, and most Americans are using ad blockers. And even though advertising puts food on my table, I can't blame them — because I block ads, too. Google, as the world's biggest advertising company, doesn't appreciate that.").

348. *See declarativeNetRequest*, MDN WEB DOCS, https://developer.mozilla.org/en-US/docs/Mozilla/Add-ons/WebExtensions/API/declarativeNetRequest [https://perma.cc/74DT-VX38] (last visited Jan. 31, 2025) ("The webRequest API is more flexible than the declarativeNetRequest API because it allows extensions to evaluate a request programmatically.").

---

349. See *id.* ("The number of rules in enabled static rulesets for all extensions must not exceed the global limit. Extensions shouldn't depend on the global limit having a specific value; instead, they should use getAvailableStaticRuleCount to find the number of additional rules they can enable.").

350. *See* Daly Barnett, *Chrome Users Beware: Manifest V3 Is Deceitful and Threatening*, ELEC. FRONTIER FOUND. (Dec. 9, 2021), https://www.eff.org/deeplinks/2021/12/chrome-users-beware-manifest-v3-deceitful-and-threatening [https://perma.cc/K4CA-JPEN] ("Like FLoC and Privacy Sandbox before it, Manifest V3 is another example of the inherent conflict of interest that comes from Google controlling both the dominant web browser and one of the largest internet advertising networks.").

351. Pete LePage, *New in Chrome 88*, GOOGLE: CHROME FOR DEVS. BLOG, https://developer.chrome.com/blog/new-in-chrome-88/ [https://perma.cc/3ELZ-4DLY] (Jan. 19, 2021) ("You can now upload extensions using manifest V3 to the Chrome Web Store."); David Li, *More Details on the Transition to Manifest V3*, GOOGLE: CHROME FOR DEVS. BLOG, https://developer.chrome.com/blog/more-mv2-transition [https://perma.cc/8WM3-4NTW] (Sept. 28, 2022) ("In January 2024, following the expiration of the Manifest V2 enterprise policy, the Chrome Web Store will remove all remaining Manifest V2 items from the store.").

352. *See* Ashwin, *Brave Confirms It Will Support Manifest V2 Extensions like uBlock Origin Even After Chrome Drops Them*, GHACKS, https://www.ghacks.net/2022/09/29/brave-browser-manifest-v2-extensions-after-v3-update/ [https://perma.cc/8M8S-R3T8] (Sept. 30, 2022) ("Brave has done the same to reassure its users that it too will support ad blocking after the dreaded update, and Manifest V2 extensions like uBlock Origin."); Rob Wu, *Manifest v3 Update*, MOZILLA (May 27, 2021), https://blog.mozilla.org/addons/2021/05/27/manifest-v3-update/ [https://perma.cc/ZJE7-LX5Y] ("We have not yet set a deprecation date for Manifest v2 but expect it to be supported for at least one year after Manifest v3 becomes stable in the release channel."); *What Manifest V3 Means for Brave Shields and the Use of Extensions in the Brave Browser*, BRAVE (June 27, 2024), https://brave.com/blog/brave-shields-manifest-v3/#:~:text=protection%20than%20extensions.-,Will%20MV2%20extensions%20still%20work%20in%20Brave%3F,your%20favorite%20extensions%20without%20interruption [https://perma.cc/2L45-PFLS] ("Yes, for now. We recognize the importance of supporting existing Manifest V2 extensions. We have force-enabled Manifest V2 support in the Brave browser, ensuring that you can continue to use your favorite extensions without interruption.").

353. *See* Kevin Borgolte & Nick Feamster, *Understanding the Performance Costs and Benefits of Privacy-Focused Browser Extensions*, *in* WWW '20: PROCEEDINGS OF THE WEB CONFERENCE 2020 2275, 2275, 2284–85 (2020), ("Our results highlight that privacy-focused extensions not only improve users' privacy, but can also increase users' browsing experience."); *see also* @gorhill, GITHUB (Jan. 21, 2019), https://github.com/uBlockOrigin/uBlock-

Google's introduction of the SameParty cookie attribute, which evolved into the First-Party Sets (FPS) standard and later was renamed to "Related Website Sets (RWS),"[356] was ostensibly aimed at softening the blow of enhanced privacy measures like third-party cookie blocking.[357] This standard would enable a data-collecting entity to

issues/issues/338#issuecomment-456134855 [https://perma.cc/LLH6-PPHC] ("The fact that they are planning to remove a proper blocking webRequest API with no word of an equivalent replacement is a sign of intent, that is, reducing the level of user agency in their user agent (aka Google Chrome).").

354. *See About Google Chrome's "This Extension May Soon No Longer Be Supported"*, GITHUB: UBLOCKORIGIN, https://github.com/uBlockOrigin/uBlock-issues/wiki/About-Google-Chrome's-%22This-extension-may-soon-no-longer-be-supported%22 [https://perma.cc/K6UK-4VJX] (Nov. 1, 2024) ("However the focus on reliability and efficiency in a Manifest v3 environment meant having to sacrifice many features beyond those not possible within a Manifest v3 framework."); Lena Cohen, *Google Breaks Promise to Block Third-Party Cookies*, ELEC. FRONTIER FOUND. (Aug. 2, 2024) [hereinafter Cohen, *Google Breaks Promise*], https://www.eff.org/deeplinks/2024/08/google-breaks-promise-block-third-party-cookies [https://perma.cc/WH9X-XKHD].

355. *See* Barnett, *supra* note 350 ("Manifest V3, or Mv3 for short, is outright harmful to privacy efforts. It will restrict the capabilities of web extensions—especially those that are designed to monitor, modify, and compute alongside the conversation your browser has with the websites you visit. Under the new specifications, extensions like these– like some privacy-protective tracker blockers– will have greatly reduced capabilities. Google's efforts to limit that access is concerning, especially considering that Google has trackers installed on 75% of the top one million websites.").

356. *Related Website Sets - The New Name for First-Party Sets in Chrome 117*, GOOGLE: PRIV. SANDBOX, https://developers.google.com/privacy-sandbox/blog/related-website-sets [https://perma.cc/M8SH-BHAB] (Aug. 31, 2023) ("In this post, we introduce Related Website Sets (RWS)—our new name for FPS that better reflects its purpose—and provide a refresher on key use cases along with an update on the associated subset domain limit.").

357. *See* Lily Chen, Kaustubha Govind, David Benjamin & Chris Fredrickson, *SameParty Cookie Attribute Explainer*, GITHUB: SAMEPARTY [Hereinafter Chen et al., *SameParty Cookie*], https://github.com/cfredric/sameparty [https://perma.cc/V2BL-8H3A] (last visited Jan. 20, 2025) ("The SameParty cookie attribute provides web developers a means to annotate cookies that are allowed to be set or sent in same-party, cross-site contexts."); *Related Website Sets (Formerly Known As: First-Party Sets)*, GITHUB: WEB INCUBATOR CG, https://github.com/WICG/first-party-sets [https://perma.cc/TP7R-QE5N] (last visited Jan. 31, 2025) ("[Related Website sets] will allow browsers to ensure continued operation of existing functionality that would otherwise be broken by blocking cross-domain cookies ('third-party cookies') . . . ."); Ronan Shields & Tim Peterson, *WTF Are Related Website Sets in Google's Privacy Sandbox?*, DIGIDAY (Feb. 12, 2024), https://digiday.com/media/wtf-are-related-website-sets-in-googles-privacy-sandbox/ [https://perma.cc/H6LM-NNWM] ("For example, a publisher with a primary domain, such as declare domains under its control as "first parties" to one another, which allows for shared cookie access.[358] For instance, a cookie set by google.com could be recognized by youtube.com, thus enabling cross-site tracking within Google's ecosystem.[359]

This framework was evaluated by the World Wide Web Consortium's (W3C) Privacy Community Group (PrivacyCG).[360] Despite Google's endorsement, several key browsers such as Apple and Brave voiced objections within PrivacyCG, leading to the abandonment of the RWS proposal due to a lack of broad multi-implementer interest.[361]

Nevertheless, Google independently integrated FPS into Chrome with version 115.[362] Google's unilateral move is poised to

---

'www.digiday.com,' may have several related country domains, such as 'www.digiday.co.uk' or other related properties. Until now, such publishers have used third-party cookies to track users across their various web properties, meaning they don't have to log in to each separate domain. The proposed end benefit of RWS is that website visitors still won't have to do so even after the demise of third-party cookies in Chrome.").

358. *See* Chen et al., *SameParty Cookie*, *supra* note 357.

359. *See* Google's Related Website Set, https://google.com/.well-known/related-website-set.json [https://perma.cc/9L32-VM2N] (last visited Feb. 15, 2025). Google has added YouTube, along with other sites, to its related sites, thereby allowing all first-party cookies and other first-party data to be easily shared across its different websites. *See id.*

360. *See Privacy Community Group*, GITHUB, https://github.com/privacycg/ [https://perma.cc/ZBQ6-53CF] (last visited Jan. 13, 2025).

361. *See* Posting of John Wilander, wilander@apple.com, to public-privacycg@w3.org (May 24, 2022), https://lists.w3.org/Archives/Public/public-privacycg/2022May/0006.html [https://perma.cc/ELJ2-62YP] ("Setting browser policy based on joint domain ownership will very likely go against the user's interest in many cases . . . ."); Peter Snyder, *First-Party Sets: Tearing Down Privacy Defenses Just as They're Being Built*, BRAVE: WEBSTANDARDS@BRAVE (May 19, 2022), https://brave.com/web-standards-at-brave/8-first-party-sets/ [https://perma.cc/XCR8-DHDY] ("The privacy harm from [first-party sets] is obvious: it would allow companies to automatically track you across sites, and without proper notification or consent."); Posting of Theresa O'Connor, hober@apple.com, to public-privacycg@w3.org (June 2, 2022), https://lists.w3.org/Archives/Public/public-privacycg/2022Jun/0003.html [https://perma.cc/337C-MBDP] ("As chairs of the W3C Privacy Community Group, we have decided to drop First-Party Sets as a Work Item in the group."). The W3C Technical Architecture Group (TAG) also reviewed and arrived at similar conclusions. *See* Daniel Appelquist & Martin Thomson, *TAG Review Feedback on Related Website Sets (Originally Named "First Party Sets")*, GITHUB: W3C TAG, https://github.com/w3ctag/design-reviews/blob/main/reviews/first_party_sets_feedback.md [https://perma.cc/5BH9-CT9G] (Dec. 2, 2024) ("[W]e consider the First Party Sets proposal harmful to the web in its current form.").

362. *See* Google Groups Message from Johann Hofmann, to blink-dev@chromium.org, cfredric@chromium.org, shuuran@chromium.org & kaustubhag@chromium.org (Mar. 20, 2023, 4:39 PM), https://groups.google.com/a/chromium.org/g/blink-dev/c/7_6JDIfE1as?pli=1 [https://perma.cc/AA9E-W8GN] ("First-Party Sets ('FPS') provides a framework for developers to declare relationships among sites, to enable limited cross-site cookie access for specific, user-facing purposes. This is facilitated through the use of the Storage Access API and requestStorageAccessFor API.").

diminish the effects of any restrictions on third-party cookies on Google's operations.[363] Conversely, the move advantages Google more as compared to its competitors because it owns and operates a wider array of domains.[364] Yet again, Google further entrenches its market dominance.

First-party sets are not the only way through which Chrome tries to gain an unfair advantage by not following web standards. Chrome dispatches a special HTTP header,[365] X-Client-Data, exclusively to the domains it owns.[366] Google asserts this header is instrumental for testing experimental functionalities, yet it has the potential for user tracking.[367] The header itself carries low entropy,[368]

---

363. *See Chrome Is Entrenching Third-Party Cookies for Some Sites in a Way That Will Predictably, Inevitably Mislead Users*, BRAVE (Aug. 26, 2024), https://brave.com/blog/related-website-sets/ [https://perma.cc/HC6U-AZZ9] ("Although Related Website Sets is being presented as a general Web proposal, the truth is that most of the Web has already considered and rejected it. Most browsers, including Brave, Firefox, and Safari, have publicly stated that they believe Related Website Sets (previously called First-Party Sets) is bad for users, and bad for the Web. The proposal has been removed from the W3C Privacy Community Group and is no longer being considered by any privacy-focused group in the W3C.").

364. *See* Andrew Allemann, *How Google Became a Strong Competitor for Domain Names*, DOMAIN NAME WIRE (Dec. 10, 2020), https://domainnamewire.com/2020/12/10/how-google-became-a-strong-competitor-for-domain-names/ [https://perma.cc/922M-5RHW] ("Google's domains under management have steadily climbed. Less than four years after it entered the market, Google was a top 10 registrar for .com with 1.4 million domains registered.").

365. HTTP headers are pieces of information sent with HTTP requests and responses that help the server and browser communicate. Reut Abolnik, *What Are HTTP Headers?*, NIMBLE (Jan. 4, 2024), https://www.nimbleway.com/blog/http-headers [https://perma.cc/X3WC-TLPM]. They include key details like the type of data being sent (Content-Type), who is making the request (User-Agent), or instructions for caching (Cache-Control). *See id.* These headers ensure that requests and responses are handled correctly and securely. *See id.* For example, the Authorization header is used to send login details, while the Content-Type header tells the server what kind of data it is receiving, like text or JSON.

366. *See* Thomas Claburn, *Google: You Know We Said That Chrome Tracker Contained No Personally Identifiable Info? Yeah, About That. . .*, THE REG. (Mar. 11, 2020, 8:03 AM), https://www.theregister.com/2020/03/11/google_personally_identifiable_info/ [https://perma.cc/BCM7-XVGU] ("In February, Arnaud Granal, a software developer who works on a Chromium-based browser called Kiwi, claimed the X-client-data header, which Chrome sends to Google when a Google webpage has been requested, represents a unique identifier that can be used to track people across the web.").

367. *See* Kyle Bradshaw, *Google Denies Chrome Tracking Allegation, Explains Use of 'X-Client-Data'*, 9TO5GOOGLE (Feb. 6, 2020, 8:45 AM), https://9to5google.com/2020/02/06/google-chrome-x-client-data-tracking/ [https://perma.cc/MQQ2-9VZF] ("Meanwhile, Arnaud Granal, the developer of Kiwi Browser, a Chromium-based alternative browser for Android—and thus someone who has a deep understanding of Chrome and Chromium—has pointed out that Chrome creates its own special bit of data called 'X-Client-Data.' Granal claims this could be used by Google to bypass any fingerprinting restrictions that Google Chrome would add.").

368. Entropy, in terms of information theory, measures how much information is contained in a dataset or system. Neha Seth, *Entropy in Machine Learning: Definition, Examples and Uses*, suggesting limited capability for individual user tracking.[369] However, when this data is combined with an IP address—which accompanies every network request—and other HTTP headers (or the first-party data Google collects), it can be used for user identification.[370]

The HTTP header emerged around the same time as the industry's move to limit the granularity of User-Agent strings, a standard HTTP header, as a measure to bolster privacy as higher amounts of information sent as part of the User-Agent string can be used to uniquely identify users.[371] Google's deployment of the X-Client-Data header thus stands as a strategic pivot; it preserves tracking capabilities for Google's domains amidst privacy advancements that impede similar tracking by other third parties.[372] As a result, while the utility of the User-Agent diminishes for tracking purposes, Google domains may continue to track users via this header.[373]

These strategies illustrate that Google's maneuvers extend beyond leveraging its dominant position for competitive advantage. The company appears to be preempting restrictions on current web

---

ANALYTICS VIDHYA, https://www.analyticsvidhya.com/blog/2020/11/entropy-a-key-concept-for-all-data-science-beginners/ [https://perma.cc/D6DK-LUGU] (Dec. 13, 2024). It reflects the degree of unpredictability or variability in the data. *Id.* In the context of user identification, higher entropy means more unique and diverse information, allowing for more precise identification of individual users. *See id.* Lower entropy, on the other hand, indicates less information and more predictability, which makes it harder to distinguish between users. *See id.*

369. *See* Claburn, *supra* note 366 ("X-client-data header, which comes in two variations, a low-entropy (13-bit) version that ranges from 0-7999 and a high-entropy version, which is what most Chrome users will send if they have not disabled usage statistic reporting.").

370. *See id.* ("As a user, in the current state, it's important to understand that no matter if you use a proxy, a VPN, or even Tor (with Google Chrome), Google (including DoubleClick) may be able to identify you using this X-Client-Data. Do you want Google to be able to recognize you even if you are not logged-in to your account or behind a proxy?" (quoting Arnaud Granal, software developer at Kiwi)).

371. Kyle Bradshaw, *Google Is Seeking to Deprecate Chrome's User Agent String, and That's a Win for Privacy*, 9TO5GOOGLE (Jan. 14, 2020, 11:02 AM), https://9to5google.com/2020/01/14/google-deprecate-chrome-user-agent-string-privacy/ [https://perma.cc/6SW3-UM6F] ("Google's Chromium team has submitted a new proposal that includes deprecating the User Agent string starting in Chrome 81."); *see* Roy Fielding, Mark Nottingham & Julian Reschke, *RFC 9110 HTTP Semantics*, INTERNET ENG'G TASK FORCE (June, 2020), https://www.rfc-editor.org/rfc/rfc9110#name-user-agent [https://perma.cc/2DXZ-7MYL] ("The 'User-Agent' header field contains information about the user agent originating the request, which is often used by servers to help identify the scope of reported interoperability problems, to work around or tailor responses to avoid particular user agent limitations, and for analytics regarding browser or operating system use.").

372. Bradshaw, *supra* note 367.

373. *Id.*

standards while simultaneously advocating for new standards that may consolidate its dominance, typically at the expense of user privacy and equitable market competition.[374] Standard-setting processes typically thrive on multi-stakeholder collaboration.[375] However, Google's actions are a deviation from this norm. Instead, Google leverages its browser market share to unilaterally craft standards that primarily serve its interests.[376]

*f. Implications for Consumer Welfare*

Google's anticompetitive practices not only cause financial loss to its market rivals but also bear adverse implications for consumer welfare. The disproportionate impact of tracking limitations enforced by browsers, coupled with the introduction of Google-favored standards, have hampered competitive dynamics within the online advertising industry.[377] Indeed, competitors face reduced options and elevated costs for advertisers seeking online visibility.[378] Constrained choice often translates into higher advertising rates, burdening businesses with increased marketing expenses that may, in turn, inflate consumer prices.[379]

374. Barnett, *supra* note 350; Taylor, *supra* note 320.

375. Parsons, *supra* note 301.

376. *Id.*

377. *Id.*

378. *See* IAB TECH. LAB, PRIVACY SANDBOX FIT GAP ANALYSIS FOR DIGITAL ADVERTISING 71, 74 (2024), https://iabtechlab.com/wp-content/uploads/2024/06/Privacy-Sandbox-Fit-Gap-Analysis-FINAL.pdf [https://perma.cc/QKJ3-L89S] ("Advertisers will need to allocate significantly more financial and operational resources to compensate for the operational cost imposed by Interest Group creation, increasing their overall marketing costs. Operationally, it might result in less efficient campaigns, as advertisers struggle to reach their desired audience. The impact on small to mid-sized brands can be even more pronounced.").

379. Adi Robertson, *Google Quietly Raised Ad Prices to Boost Search Revenue, Says Executive*, VERGE (Sept. 19, 2023, 9:01 AM), https://www.theverge.com/2023/9/19/23880275/google-search-ads-competition-auction-prices-doj-trial-antitrust [https://perma.cc/MK3M-78HB] ("Google's dominance lets it raise prices for advertisers with few repercussions — a claim backed up by Google ads executive Jerry Dischler on the stand."); Leah Nylen, *Google Changed Ad Auctions, Raising Prices 15%, Witness Says*, BLOOMBERG (Oct. 6, 2023, 2:13 PM), https://www.bloomberg.com/news/articles/2023-10-06/google-changed-ad-auctions-raising-prices-15-witness-says?embedded-checkout=true [https://perma.cc/KZX8-G8RV] ("Alphabet Inc.'s Google changed its advertising auction formula in 2017, raising prices by 15% and likely making the company billions of dollars in additional revenue, according to an economist testifying for the US Justice Department in the antitrust case against the search giant."); Kristen McCormick, *Google Ads Cost per Lead Has Increased for 91% of Industries YoY*, WORDSTREAM, https://www.wordstream.com/blog/ws/2022/11/10/search-advertising-benchmarks [https://perma.cc/YZL7-YZCL] (Jan. 23, 2024) ("[A] latest benchmark study showing that cost per lead has increased for 91% of industries year over year . . . .").

Beyond the advertisers, publishers also encounter financial pressures as the lack of competition allows Google to demand a larger share of advertising revenues.[380] This economic strain is evidenced by numerous online news outlets resorting to paywalls to offset diminished ad revenue.[381]

Consumers are harmed not only in financial terms but also from a privacy standpoint. While third-party cookies known for their tracking capabilities can be managed by privacy-aware users, the increasing shift towards the collection of first-party data presents more drastic challenges.[382] First-party data is often harvested and processed outside the purview of a user's browser, thereby escaping easy detection and removal.[383] Moreover, the nature of the accumulated data is more personal and, consequently, potentially more intrusive.[384] Yet, consumers are afforded minimal oversight or control over this collection and its subsequent use.[385] The result is a higher privacy cost for

380. Press Release, U.S. Dep't of Just., *supra* note 13.

381. *See* Susan Athey, Emilio Calvano & Joshua Gans, *The Impact of the Internet on Advertising Markets for News Media* 28 (Nat'l Bureau of Econ. Rsch., Working Paper No. 19419, 2013).

382. Munir et al., *supra* note 258, at 3491 ("Our analysis reveals that these trackers store identifiers in first-party cookies based on probabilistic and deterministic information. Unlike third-party cookies, blocking all first-party cookies is not practical, as some of these cookies might be required for legitimate website functionality.").

383. *See* Simon Waters, *Blocking Advertising Cookies Only Makes Facebook Stronger*, SILKTIDE, https://silktide.com/blog/blocking-advertising-cookies-only-makes-facebook-stronger/ [https://perma.cc/3HF7-2PCB] (last visited Jan. 31, 2025) ("When you fill in a form on a website or buy something through a checkout, you're giving your data over to the website owner. That's fair because without it the website couldn't process your transaction or inquiry, submit your comment, or whatever. Third parties don't get access to this personal information (I'm talking about names, email addresses, form values, etc.). It's not transferred in cookies. What Facebook is doing is encouraging business owners to give them this data at the point of transaction, so they can use it to match to a Facebook user. . . . The problem is that, for the consumer, there is no choice. No way to block or opt out of this behavior, and no way to know it's happening (without reading through the privacy policy of every website and assuming that businesses disclose all this information transfer as they should do).").

384. *Id.* ("Now, the Conversions API bypasses the need for third-party cookies entirely and collects far more PID than cookies ever could.").

385. Florian Eisenmenger, *Shifting to First-Party Data: Privacy Pitfalls Around Consent and Transparency*, IAPP (Mar. 28, 2023), https://iapp.org/news/a/shifting-to-first-party-data-privacy-pitfalls-around-consent-and-transparency/ [https://perma.cc/5LYN-YNFJ] ("Companies are increasingly pursuing first-party data approaches to move away from third parties that collect and process personal data on their behalf. Instead, they rely on personal data collected themselves, in particular to pursue personalized marketing activities. Naturally, this comes with a number of privacy challenges—most importantly, obtaining valid consent that meets transparency requirements.").

consumers stemming directly from the anticompetitive strategies implemented by Google.[386]

2. From Browser to Publishing Market

Google provides a host of first-party services, including Google Search, YouTube, and Gmail.[387] Google uses its dominance in the browser market to assert, perpetuate, and leverage its dominant position within these key markets with the help of Chrome.

*a. Self-Preferencing Google Search in Chrome*

Google Search is the most widely used search engine in the world.[388] Globally, Google Search's market share among online search engines exceeds 90%.[389] Its closest competitor, Bing, only has a market share of about 3%, as noted by Similarweb.[390] While Google has diversified its revenue across other business divisions, the financial health of Alphabet, Google's parent company, continues to hinge on the advertising revenue generated by Google Search.[391] In 2022, advertising revenue generated from Google Search constituted most of Alphabet's revenue.[392] Thus, maintaining the dominant position in search engines is Google's key interest.

Google uses Chrome's dominant position in the market to boost its search business. Google Search has been the default search engine on Chrome since its inception.[393] Relying on the "power of defaults" and

---

386. *Id.*; Waters, *supra* note 383.

387. *Data Use in Personalized Ads on Google Search, Gmail, and YouTube*, GOOGLE: ADVERT. POLICIES HELP, https://support.google.com/adspolicy/answer/6242605?hl=en [https://perma.cc/Z4DX-49HC] (last visited Feb. 16, 2025).

388. *Search Engines Market Share*, SIMILARWEB, https://www.similarweb.com/engines/ [PERMA] (last visited Jan. 22, 2025).

389. *Search Engine Market Share Worldwide, Dec 2023 - Dec 2024*, *supra* note 23.

390. *See Search Engines Market Share*, *supra* note 387.

391. Alphabet Inc., Annual Report, *supra* note 2.

392. Gennaro Cuofano, *Google Revenue Breakdown*, FOURWEEKMBA (Feb. 4, 2024), https://fourweekmba.com/google-revenue-breakdown/ [https://perma.cc/2DW3-8T8S] ("In 2023, Alphabet generated over $175B from Google search, $31.51B billion from the Network members (Adsense and AdMob), $31.31B billion from YouTube Ads, $33B from Google Cloud, and $34.69B billion from other sources (Google Play, Hardware devices, and other services)").

393. *See* Ryan Paul, *Hands-on with Chrome: Google's Browser Shines (Mostly)*, ARS TECHNICA (Sept. 2, 2008, 10:39 PM), https://arstechnica.com/information-technology/2008/09/hands-on-with-chrome-googles-browser-shines-mostly/ [https://perma.cc/8U53-BT6D] ("One aspect of the Chrome user interface that particularly impressed me is the autocompletion mechanism in the URL textbox, which Google refers to as the Omnibar. This feature is evidently inspired by the Firefox 3 AwesomeBar, but it embellishes on exploiting individuals' known psychological vulnerabilities (i.e., those studied in behavioral economics), Google ensures that its dominant market share among browsers is also translated into search engines.[394] Most users never change default options, which Google uses to its advantage on mobile and desktop apps.[395] This exploitation of user behavior by way of making Google the default search engine was penalized by the European Commission in 2018, which conducted an antitrust investigation into Google's behavior on Android and found the company guilty of unfair practices.[396] Google was fined €4.34 billion and is now required to present users in the European Union options to choose between different search engines during the initial setup.[397] Similar antitrust actions in India have resulted in Google loosening control of search engine defaults on Androids in other regions of the world.[398] As a result of these actions, Android has long had a default

---

the concept and adds some really nice additional features. In addition to providing completion suggestions based on the user's history and bookmarks, it will also provide domain recommendations and allow users to perform searches.").

394. Fowler, *supra* note 224.

395. See Jared Spool, *Do Users Change Their Settings?*, UIE (Sept. 14, 2011), https://archive.uie.com/brainsparks/2011/09/14/do-users-change-their-settings/ [https://perma.cc/9ZBY-Z648] ("What we found was really interesting. Less than 5% of the users we surveyed had changed any settings at all. More than 95% had kept the settings in the exact configuration that the program installed in." (emphasis omitted)).

396. *See* European Commission Press Release IP/18/4581, Antitrust: Commission fines Google €4.34 billion for illegal practices regarding Android mobile devices to strengthen dominance of Google's search engine (July 18, 2018) ("Our case is about three types of restrictions that Google has imposed on Android device manufacturers and network operators to ensure that traffic on Android devices goes to the Google search engine. In this way, Google has used Android as a vehicle to cement the dominance of its search engine. . . . Google has used Android as a vehicle to cement the dominance of its search engine. These practices have denied rivals the chance to innovate and compete on the merits. They have denied European consumers the benefits of effective competition in the important mobile sphere. This is illegal under EU antitrust rules.").

397. *Id.*; Sam Byford, *Google Will Give Android Users a Choice of Browser and Search Engine in Europe*, VERGE (Mar. 20, 2019, 1:55 AM), https://www.theverge.com/2019/3/20/18273888/google-eu-browser-search-choice [https://perma.cc/RLR2-F87Z] ("Google has announced that it will start asking European Android users which browser and search engine they would prefer to use on their devices, following regulatory action against the company for the way it bundles software in its mobile operating system.").

398. *See Google Says Will Allow Users in India to Choose Default Search Engine on Android Phones*, TIMES INDIA (Jan. 25, 2023, 8:22 PM), https://timesofindia.indiatimes.com/google-says-will-allow-users-in-india-to-choose-default-search-engine-on-android-phones/articleshow/97322310.cms [https://perma.cc/LG4A-4MLF] ("After failing to get a court order to block an antitrust ruling, Google on Wednesday said it will allow users in India to choose [sic] default search engine on Android-based smartphones.").

search engine selection screen.[399] However, the default search engine of Chrome desktop browser remains Google.[400] Due to self-preferencing their own search engine on a browser that commands a monopolizing share of the market, Google's actions on desktop deserve the same level of antitrust scrutiny.

### *b. Implications for Consumer Welfare*

Google's dominant position in the search engine market, which is further bolstered by its dominance in the browser market, has profound implications for consumer welfare. First, it stifles innovation and competition in the search engine market.[401] Google's dominance enables it to amass a vast amount of potentially sensitive consumer data, which increases the competitive imbalance in the market.[402] Competitors, lacking access to similar levels of consumer data, find it challenging to refine and improve their search algorithms to the same degree.[403] This lack of competition, in turn, leads to a poor consumer experience. Consumers are not only left with fewer choices but are also subject to a search engine ecosystem that evolves more slowly due to reduced competitive pressure.[404] Despite significant privacy concerns attached to Google's data collection practices, with few viable alternatives, consumers are often compelled to use a service that continually gathers personal information and potentially infringes on their privacy rights.[405]

---

399. *About the Choice Screen*, ANDROID, https://www.android.com/choicescreen/dma/ [https://perma.cc/W7MR-92A3] (Mar. 29, 2024).

400. *See* Günter Born, *EU: Google Chrome Will Display a Search Engine Choice for Desktops from 2024*, BORN'S TECH & WINDOWS WORLD BLOG (Dec. 23, 2023), https://borncity.com/win/2023/12/23/eu-google-chrome-will-display-a-search-engine-choice-for-desktops-from-2024/ [https://perma.cc/RHP4-2DNB] ("These EU rules also influence how Google Chrome has to react to certain situations. The browser is not pre-installed on the Windows and macOS desktop (Android has long had a selection screen for the browser and search engine). However, the default search engine in the Chrome browser is Google." (citation omitted)).

401. *See* Sumit Sharma, *Google Is a Monopoly, but Its Search Engine Does Not Have to Be*, TECH POL'Y PRESS (Apr. 23, 2024), https://www.techpolicy.press/google-is-a-monopoly-but-its-search-engine-does-not-have-to-be/ [https://perma.cc/B2UC-CCYQ] ("An uncompetitive search engine market stifles innovation and quality improvements as Google controls access to users.").

402. *Id.* ("The problem is twofold. The first is that Google, with over 85% market share in the US and over 90% market share globally, operates at a scale that is qualitatively different from any of its competitors. Google's ability to collect data on what users are searching for and get feedback on which results users find useful is unparalleled.").

403. *Id.* ("Google's ability to collect data on what users are searching for and get feedback on which results users find useful is unparalleled.").

404. *Id.*

405. *See id.*

Google's actions regarding the use of its Chrome browser have resulted in significant ramifications across markets, creating a self-reinforcing cycle where increased usage of Chrome leads to more Google Search users.[406] More Search users provide Google with more data, and more data further entrenches Google's market dominance.[407]

### *D. Flow of Dominance from Advertiser to Publisher Market*

Google engages in various anticompetitive behaviors by using the revenue generated by its advertising business to augment its position in the publisher market.[408] Google dominates advertisement business through Google Ads and Google Analytics, with each registering a presence of almost 50 percent of all websites.[409] In fiscal year 2023, out of $282 billion in revenue generated by Alphabet, 80 percent was generated by its advertising business.[410] Most of the Google's publishing business, especially Google Search and YouTube, rely on advertising to generate revenue.[411]

Therefore, Google's advertising business provides it with significant financial leverage. Google uses this leverage to further perpetuate its dominance in other market segments through engaging in pay-to-play behavior or outright competitor buy-outs.[412]

406. *See id.*

407. *Id.*

408. *See* Cory Doctorow, *Forcing Google to Spin Off Chrome (And Android?)*, MEDIUM (Nov. 19, 2024), https://doctorow.medium.com/https-pluralistic-net-2024-11-19-breaking-up-is-hard-to-do-shiny-and-chrome-5b6eaf08bb5a [https://perma.cc/KLM4-A7QZ].

409. *Google Ads*, *supra* note 268; *Google Analytics*, *supra* note 268.

410. Alphabet Inc., Annual Report (Form 10-K), *supra* note 2, at 9, 30.

411. *See How Does YouTube Make Money?*, YOUTUBE, https://www.youtube.com/howyoutubeworks/our-commitments/sharing-revenue/ [https://perma.cc/JP2G-BTKT] (last visited Jan. 31, 2025) ("YouTube's main source of revenue is advertising."); *Advertising Revenue of Google from 2001 to 2023*, STATISTA, https://www.statista.com/statistics/266249/advertising-revenue-of-google/#:~:text=Advertising%20accounts%20for%20the%20majority,billion%20U.S.%20dollars%20in%202023 [https://perma.cc/YJV9-58KP] (last visited Jan. 31, 2025).

412. *See* Ricky Sutton, *DOJ Alleges Google Used "Killer Acquisition" Strategy to Shutdown Ad Tech Rival AdMeld Post US$400m Buyout*, MI3 (Sept. 18, 2024, 9:46 AM), https://www.mi-3.com.au/18-09-2024/doj-alleges-google-used-killer-acquisition-strategy-shutdown-ad-tech-rival-admeld-post [https://perma.cc/9E6A-99ZX].

### 1. Engaging in Pay-to-Play Behavior

Google frequently pays off its competitors to ensure its publisher businesses such as Google Search, YouTube, and others remain dominant in their respective markets.[413]

To preserve its search engine dominance, Google has reportedly paid competitors to guarantee that its search tool is the default choice across major platforms.[414] In a striking demonstration of this market strategy, estimates suggest Google compensated Apple, a competitor in the smartphone, browser, and desktop OS markets, approximately $20 billion in 2022 to retain Google Search as the default in Safari on both Apple's desktop and mobile platforms.[415] The payments, escalating annually and surpassing inflation rates,[416] reflect not only Safari's increasing market share—from about 8 percent in 2012 to roughly 20 percent in 2023[417]—but also a strategic effort to deter Apple from entering the search engine market[418] or partnering with Google's

413. *See* Lauren Feiner, *Google Paid $26 Billion in 2021 to Become the Default Search Engine on Browsers and Phones*, CNBC [hereinafter Feiner, *Google Paid*], https://www.cnbc.com/2023/10/27/google-paid-26-billion-in-2021-to-become-a-default-search-engine.html [https://perma.cc/D9VP-2UJW] (Oct. 27, 2023, 3:38 PM).

414. *Id.*

415. Paul Kunert, *Google Pays Apple $18B to $20B a Year to Keep Its Search in iPhone*, REG. (Oct. 10, 2023, 5:31 PM), https://www.theregister.com/2023/10/10/google_pays_apple_18_20_claims_bernstein/ [https://perma.cc/LFM4-Q9YA] ("We estimate that the ISA is worth $18B-20B in annual payments from Google to Apple, accounting for 14-16 percent of Apple's annual operating profits." (quoting Bernstein report)).

416. Jeremy Bowman, *20 Billion Reasons Alphabet's Moat Isn't as Big as It Seems*, MOTLEY FOOL (Feb. 21, 2023, 5:00 AM), https://www.fool.com/investing/2023/02/21/20-billion-reasons-alphabets-moat-isnt-as-big-as-i/ [https://perma.cc/6X84-KZM5] ("In 2022, Alphabet was estimated to pay Apple as much as $20 billion to be the search engine of choice on Safari. The payment is not publicly disclosed, but it has ramped up significantly over the last decade as court filings in 2014 showed that Alphabet paid Apple just $1 billion."); Paresh Dave, *Would You Still Use Google if It Didn't Pay Apple $20 Billion to Get on Your iPhone?*, WIRED (May 2, 2024, 9:23 PM), https://www.wired.com/story/google-pay-apple-20-billion-to-get-on-your-iphone/ [https://perma.cc/H6E4-5QJN].

417. *See Browser Market Share Worldwide*, *supra* note 35.

418. Bowman, *supra* note 415. This is not a theoretical threat since Apple has been developing its own search engine for a while. Mark Gurman, *Apple Has What It Needs to Launch Its Own Google Replacement*, BLOOMBERG (Oct. 1, 2023, 8:00 AM), https://www.bloomberg.com/news/newsletters/2023-10-01/could-apple-replace-google-with-own-search-engine-it-s-possible-but-unlikely-ln7gywed [https://perma.cc/9SV4-KWAX]. This search engine comes, however, not with an online interface or ads, but is rather integrated into the "Spotlight" search on macOS and iOS. *Id.* Whenever individuals make a search through this in-built search functionality of their Apple devices, Apple's own search engine—rather than Google—gets invoked nowadays. *Search for Anything with Spotlight on Mac*, APPLE, https://support.apple.com/guide/mac-help/search-with-spotlight-mchlp1008/mac [https://perma.cc/HXU2-PJ9M] (last visited Jan. 31, 2025). As Apple, too, is trying to diversify its rivals.[419] Recent disclosures have shed light on a revenue-sharing component within the Google-Apple agreement—a remarkable feature for such deals.[420] But this revenue model does not just have practical implications for Google, it also has practical implications for consumer choice. For instance, iPhone users initially had no search engine options during device setup and, until the release of iOS 17 in September 2023,

---

revenue and move away from selling hardware to selling service, it might well choose to enter the market for search advertising in Spotlight—like how it already increasingly does in the Apple App Store. B Robson, *Navigating Apple's Diversified 2024 Business Model: Opportunities and Risks*, WDD MALAY. (Nov. 27, 2023, 8:01 AM), https://wdd.my/blog/navigation-apples-diversified-business-model-opportunities-and-risks/ [https://perma.cc/8WV2-CLLD]. Apple does compete against Google in the market for browsers but does not currently generate any direct income from its Safari browser. *BrandWagon Examiner: How Does Safari Makes Money?*, BRANDWAGON (Aug. 6, 2024, 11:42 AM), https://www.financialexpress.com/business/brandwagon-brandwagon-explainer-how-does-safari-makes-money-3574896/ [https://perma.cc/25TG-RXLU]. Again, as Apple moves ever more into services, this might change, given that the browser represents a rich source for data that might be valuable for Apple's own advertising business—like how Google does it already. Nicole Bogart, *Apple Clarifies Stance on Data Collection After Spotlight Search Backlash*, GLOB. NEWS (Oct. 21, 2014, 1:03 PM), https://globalnews.ca/news/1626493/apple-clarifies-stance-on-data-collection-after-spotlight-search-backlash/ [https://perma.cc/B5E7-EA2J].

419. Paul Wiseman, *Apple Leverages Idea of Switching to Bing to Pry More Money Out of Google, Microsoft Exec Say*, AP (Sept. 27, 2023, 5:07 PM), https://apnews.com/article/google-antitrust-microsoft-bing-search-engine-eee462713c9ab59f6f3e886940c11a88 [https://perma.cc/QD75-Q6EA]. In the FTC's antitrust proceedings against Google, recent testimonies shed light on the intricate dynamics between major industry players. *Id.* Microsoft revealed that Apple was effectively dissuaded from acquiring the Bing search engine, attributing this to the lucrative financial arrangement between Google and Apple. Nicola Agius, *Microsoft Blames Google for Apple Rejecting Offer to Buy Bing*, SEARCH ENGINE LAND (Sept. 29, 2023, 5:05 PM), https://searchengineland.com/microsoft-blames-google-apple-rejecting-bing-432689#:~:text=The%20tech%20giant%27s%20CEO%20of,the%20company%27s%20deal%20with%20Google [https://perma.cc/S9PP-PQRL]. Microsoft contended that "Apple is making more money on Bing existing than Bing does," implying that Google's substantial annual payments serve not only as a revenue stream for Apple but also as a strategic deterrent against Apple's potential entry into the search market, safeguarding Google's most profitable venture. Wiseman, *supra* note 418 ("Apple was never serious about replacing Google with Microsoft's Bing as the default search engine in Macs and iPhones, but kept the possibility open as a 'bargaining chip' to extract bigger payments from Google, a Microsoft executive testified Wednesday in the biggest U.S. antitrust trial in a quarter century.").

420. *See* Daniel Howley, *Google's Antitrust Loss Could Put Billions at Risk for Apple*, YAHOO FIN. (Aug. 5, 2024), https://finance.yahoo.com/news/googles-antitrust-loss-could-put-billions-at-risk-for-apple-215401052.html [https://perma.cc/JA84-65ZL] ("The gist of the deal is that Apple uses Google as its default search engine for its Safari browser, Spotlight Search, and Siri. In return, Google pays Apple 36% of the search revenue generated on Apple devices using Google's services.").

were restricted from altering the default search engine in private browsing mode.[421]

Google's pay-to-play strategy to maintain dominance is not limited to Safari. Firefox, a browser developed by Mozilla, was once a formidable player in the browser market, peaking at a 50 percent share before 2009.[422] As a successor to NetScape, Firefox is community-developed, and open-source.[423] However, the introduction of Chrome in 2008 precipitated a sharp decline in Firefox's market share as it struggled to keep pace with Google's and Microsoft's offerings.[424] A critical factor in this dynamic is Mozilla's lack of its own search engine and a robust advertising solution for revenue generation.[425] Consequently, Mozilla depends significantly on financial agreements with Google—approximately half a billion dollars in 2023—to maintain Google as the default search engine in Firefox.[426]

---

421. Ashley Capoot, *Apple Announces iOS 17 Release Date*, CNBC, https://www.cnbc.com/2023/09/13/apple-announces-ios-17-release-date-.html [https://perma.cc/S9LN-RDPG] (Sept. 13, 2023, 10:49 AM) ("iOS 17 will be available for users to download for free on Sept. 18."); William Gallagher, *See How to Set a Unique Search Engine for Private Browsing in iOS 17*, APPLEINSIDER (Sept. 21, 2023), https://appleinsider.com/inside/ios-17/tips/how-to-set-a-unique-search-engine-for-private-browsing-in-ios-17 [https://perma.cc/B77B-PHL6] ("Yet Apple seemingly doesn't think that's enough. So for iOS 17, it has beefed up private browsing by letting you choose a different search engine. Prior to iOS 17, Safari already had this option but it was a single control that applied to all searches. So whether you were in private browsing or not, you used the same search engine by default.").

422. Ken Kovash, *Is Firefox Approaching 50% Market Share*, MOZILLA: BLOG OF METRICS (Nov. 19, 2009), https://blog.mozilla.org/metrics/2009/11/19/is-firefox-approaching-50-market-share/ [https://perma.cc/3LXQ-MT9G] ("The chart . . . shows weekly browser market share data since the beginning of 2007 and it includes aggregated data from across nine countries—Czech Republic, Bulgaria, Estonia, Hungary, Lithuania, Latvia, Poland, Russia, and Ukraine.").

423. *See History of the Mozilla Project*, MOZILLA, https://www.mozilla.org/en-US/about/history/ [https://perma.cc/B6EF-PK7R] (last visited Jan. 31, 2025) ("[The Mozilla project] was intended to harness the creative power of thousands of programmers on the internet and fuel unprecedented levels of innovation in the browser market. Within the first year, new community members from around the world had already contributed new functionality, enhanced existing features and became engaged in the management and planning of the project itself. By creating an open community, the Mozilla project had become larger than any one company.").

424. *See* Steven Vaughan-Nichols, *The Fall of Firefox: Mozilla's Once-Popular Web Browser Slides into Irrelevance*, ZDNET (Jan. 5, 2024, 7:16 AM), https://www.zdnet.com/home-and-office/networking/the-fall-of-firefox-mozillas-once-popular-web-browser-slides-into-irrelevance/ [https://perma.cc/945V-QDAF] ("By the summer of 2010, Firefox reached its high point of 34.1% of the market. It's been all downhill since then.").

425. Gennaro Cuofano, *How Does Mozilla Make Money? Mozilla Business Model Analysis*, FOURWEEKMBA (June 17, 2024), https://fourweekmba.com/how-does-mozilla-make-money [https://perma.cc/TU4V-SB5L] ("The majority of Mozilla Corporation's revenue is from royalties earned through Firefox web browser search partnerships and distribution deals. Precisely about 88% of Mozilla revenues came through royalties received by search engines to be featured on its Mozilla Firefox browser.").

Firefox's payout underlines the asymmetry in market power and financial dependency among browser developers. Despite Firefox's market share plummeting by 90 percent over the past decade, Google's annual payments to Mozilla have remained relatively consistent.[427] Google's constant financial support has raised speculation that Google's rationale extends beyond the mere utility of being the default search engine.[428] Analysts suggest that Google's payments may be partly aimed at preserving the semblance of a competitive browser market.[429] Without Google's support, Firefox will falter as a viable option to users, which increases the risk of Chrome being perceived as a monopoly in the browser market.[430] Hence, while Google props up Firefox financially, the nonprofit and community-driven nature of Mozilla limits its ability to compete with Google on a financial front, as it is unlikely that Mozilla has the means or the drive to compete against Google in parallel markets such as search, advertising, and publishing.[431]

---

426. *Id.*; Joey Sneddon, *Mozilla Revenue Jumped In 2023, but Search Deal Cash Fell*, OMG UBUNTU, https://www.omgubuntu.co.uk/2024/12/mozilla-financial-report-2023-revenue-increase#:~:text=Mozilla%27s%20overall%20revenue%20saw%20a,~%24593%20million%20in%202022 [https://perma.cc/H76A-U3N9] (Dec. 19, 2024).

427. Cuofano, *supra* note 424.

428. Noam Cohen, *Why Has Google Spent a Half-Billion Dollars on Firefox?*, BLOOMBERG (May 5, 2023, 10:30 AM), https://www.bloomberg.com/news/newsletters/2023-05-05/why-google-keeps-paying-mozilla-s-firefox-even-as-chrome-dominates [https://perma.cc/D8HA-YQJX] ("'What a great foil for Google to then sponsor a nonprofit competitor that was never quite as good,' [Chris Messina, advocate for Firefox Broswer] says. Mozilla, Messina adds, has 'served its purpose and function as far as I'm concerned, and then stuck around with the spoils that came out of that success.'").

429. Noam Cohen, *Even $500 Million a Year From Google Isn't Enough to Save Firefox*, BLOOMBERG (May 5, 2023, 12:00 PM), https://www.bloomberg.com/news/articles/2023-05-05/google-chrome-dominance-has-mozilla-firefox-searching-for-answers?embedded-checkout=true [https://perma.cc/S47J-K3NN].

430. Steven Vaughan-Nichols, *Why Google's Legal Troubles Could Hasten Firefox's Slide into Irrelevance*, ZDNET (Dec. 10, 2024, 1:10 AM), https://www.zdnet.com/home-and-office/networking/why-googles-legal-troubles-could-hasten-firefoxs-slide-into-irrelevance/ [https://perma.cc/VJW9-36AR].

431. *See The Mozilla Manifesto*, MOZILLA, https://www.mozilla.org/en-US/about/manifesto/details/ [https://perma.cc/Y5XC-JY9D] (last visited Jan. 31, 2025) ("[The Mozilla Foundation will] build and enable open-source technologies and communities that support the Manifesto's principles; build and deliver great consumer products that support the Manifesto's principles; use the Mozilla assets (intellectual property such as copyrights and trademarks, infrastructure, funds, and reputation) to keep the internet an open platform; promote models for The financial entanglements of Google with key market players extend beyond browser partnerships. In a bid to mitigate the impact of ad filtering on its revenue streams, Google, along with tech giants Microsoft and Amazon, has paid substantial fees to Eyeo GmbH, the company behind AdBlock Plus, to have its advertisements whitelisted on various websites, including Google Search.[432] Eyeo's policy stipulates that large advertising entities like Google must remit 30 percent of ad revenue generated from users with AdBlock Plus to bypass the extension's filters.[433] Given Google's dominance in the browser market, it is the most significant contributor to—and beneficiary of—this model.[434] This revenue-sharing practice has attracted scrutiny, with calls from legislators for the FTC to investigate creating economic value for the public benefit; and promote the Mozilla Manifesto principles in public discourse and within the internet industry.").

432. Lara O'Reilly, *Google, Microsoft, and Amazon Are Paying Adblock Plus Huge Fees to Get Their Ads Unblocked*, BUS. INSIDER (Feb. 3, 2015, 5:57 AM), https://www.businessinsider.com/google-microsoft-amazon-taboola-pay-adblock-plus-to-stop-blocking-their-ads-2015-2 [https://perma.cc/DSF2-FBVZ] ("Google, Microsoft, Amazon, and Taboola are paying the owner of Adblock Plus to unblock ads on their websites at a fee of '30% of the additional ad revenues' they would have made were ads unblocked . . . ."); Darrell Etherington, *Google and Others Reportedly Pay Adblock Plus to Show You Ads Anyway*, TECHCRUNCH (July 6, 2013, 5:23 AM), https://techcrunch.com/2013/07/06/google-and-others-reportedly-pay-adblock-plus-to-show-you-ads-anyway/ [https://perma.cc/Z2CX-3YML] ("Adblock Plus [is] essentially acting as a gatekeeper meting out access to that sizeable chunk of consumers."); Alex Lekander, *Advertisers Are Paying Off Ad Blockers to Show 200 Million Users "Acceptable Ads"*, CYBERINSIDER (June 23, 2020), https://restoreprivacy.com/report-ad-blockers-allowing-acceptable-ads/ [https://perma.cc/TKM3-YC82] ("A handful of ad blocker companies are earning big bucks for not blocking ads from hundreds of advertisers, including big names like LinkedIn, Reddit, Amazon, and even Google."). Whitelisting, in the context of adblocking filter lists, refers to the practice of allowing certain domains, URLs, or scripts to bypass the adblocker and load on a webpage. James Bryant, *Ad Blocking Whitelist: How to Allow Websites You Trust*, POPER BLOCKER, https://poperblocker.com/ad-blocking-whitelist/#:~:text=Whitelisting%20is%20a%20process%20where,popups%20from%20sites%20you%20trust [https://perma.cc/GK8A-HUAH] (last visited Jan. 31, 2025). This is achieved by creating rules in the filter list that explicitly permit specific content, even if it matches general blocking criteria. James Bryant, *How Do Ad Blockers Work and Why You Need Them?*, POPER BLOCKER, https://poperblocker.com/how-ad-blockers-work/ [https://perma.cc/ZH8D-2C8D] (last visited Jan. 31, 2025).

433. Sapna Maheshwari, *Adblock Plus, Created to Protect Users From Ads, Instead Opens the Door*, N.Y. TIMES (Sept. 18, 2016), https://www.nytimes.com/2016/09/19/business/media/adblock-plus-created-to-protect-users-from-ads-opens-the-door.html [https://perma.cc/7NDC-WHVY].

434. *See Google's Attempts to Undermine Ad Blockers*, MAGIC LASSO, https://www.magiclasso.co/insights/google-undermines-adblock/ [https://perma.cc/864N-WSV4] (last visited Jan. 31, 2025).

the implications of such revenue sharing agreements as potentially unfair, deceptive, and anticompetitive.[435]

Google's strategic payments to competitors have effectively positioned Google Search as the default search engine across virtually all consumer platforms.[436] Notably, Firefox defaults to Google Search, as do iOS and MacOS through Safari, and Android via Google Chrome.[437] Furthermore, on personal computers running Microsoft Windows, Google Chrome holds a 65 percent market share.[438] This ubiquity of Google Search, entrenched by Google's payments to its competitors, ensures that alternative search engines face formidable barriers to entry, solidifying Google's search engine monopoly for the foreseeable future.

### 2. Google's Acquisitions of Competitors

Another strategy that Google has used to stifle competitors is strategic competitor acquisitions to bolster its services.[439] This strategy has resulted in Google maintaining a dominant position in several significant key markets. Some of these key acquisitions that have helped Google dominate significant publisher markets are listed below:

---

435. Letter from Ron Wyden, Sen., U.S., to Joseph J. Simons, Chairman, Fed. Trade Comm'n (Jan. 14, 2020), https://www.wyden.senate.gov/imo/media/doc/011420%20Wyden%20Ad%20Blocking%20Letter%20to%20FTC.pdf [https://perma.cc/Q4KX-F9GX] ("Accordingly, I urge the FTC to open an investigation into unfair, deceptive and anti-competitive practices in the ad blocking industry.").

436. *See* United States v. Google LLC, No. 20-cv-3010, 2024 WL 3647498, at *3–4 (D.D.C. Aug. 5, 2024).

437. Complaint at ¶ 4, United States v. Google LLC, No. 1:20-cv-03010 (D.D.C. Oct. 20, 2020) ("Google pays *billions* of dollars each year to distributors—including popular-device manufacturers such as Apple, LG, Motorola, and Samsung; major U.S. wireless carriers such as AT&T, T-Mobile, and Verizon; and browser developers such as Mozilla, Opera, and UCWeb . . . ."). Firefox is pre-installed on popular Linux distributions such as Ubuntu and Linux Mint. *Install Firefox on Linux*, MOZILLA: SUPPORT, https://support.mozilla.org/en-US/kb/install-firefox-linux#:~:text=Many%20Linux%20distributions%20come%20with,and%20install%20Firefox%20on%20Linux [https://perma.cc/99PQ-CQZ5] (Oct. 22, 2024).

438. *See* Dean, *supra* note 30.

439. Chris Walton, *A Deep Dive into Google's M&A Strategy: Key Factors for Acquisition Success*, ETON VENTURE SERVS. (Nov. 2, 2023), https://etonvs.com/ma/a-deep-dive-into-googles-ma-strategy-key-factors-for-acquisition-success/#:~:text=Companies%20with%20a%20track%20record,attention%20from%20Google%27s%20M&A%20team [https://perma.cc/YGM6-8EXT].

### *a. Search Engines*

- Outride (2001): Specialized in personalized search technology, which enhanced Google's ability to tailor search results to individual users.[440] This was a significant step in developing Google's targeted advertising capabilities.[441]
- Kaltix (2003): Focused on developing context-sensitive and personalized search, which contributed to the sophistication of Google Search in understanding user queries and preferences.[442]
- ITA Software (2010): A provider of flight information.[443] This acquisition played a key role in shaping Google's travel search functionalities.[444]

Because of these acquisitions, as of December 2024, Google Search has a market share of 90%, effectively capturing and monopolizing the search engine market.[445]

### *b. Social Media and Content Consumption*

- YouTube (2006): This acquisition marked Google's entry into the video streaming and content creation market, transforming YouTube into one of the world's leading social media platforms.[446] Despite the perception that Google is not a key player in social media following the shutdown of Google Plus in 2019, YouTube's acquisition negates this assumption.[447] YouTube has a market share of 98 percent in

---

440. *Google Acquires Technology Assets of Outride Inc.*, *supra* note 15.

441. *See* Amit Chowdhry, *A History of Google Acquisitions and Where They are Today*, PULSE 2.0 (Oct. 4, 2008), https://pulse2.com/a-history-of-google-acquisitions-and-where-they-are-today/ [https://perma.cc/SG6J-NAKZ].

442. *Google Acquires Kaltix Corp.*, *supra* note 15.

443. Amir Efrati & Gina Chon, *Google's Empire Expands to Travel*, WALL ST. J., https://www.wsj.com/articles/SB10001424052748703571704575341270531117614 [https://perma.cc/N7UG-PP6B] (July 2, 2010).

444. *Id.* ("The Internet search giant said the acquisition will make it easier for customers to comparison shop for flights and airfares and drive more potential customers to the $80 billion online travel market.").

445. *See Search Engine Market Share Worldwide, Dec 2023 - Dec 2024*, *supra* note 23.

446. Lucas Downey, *Google's Incredible YouTube Purchase 15 Years Later*, INVESTOPEDIA (Sept. 2, 2021, 3:54 PM), https://www.investopedia.com/google-s-incredible-youtube-purchase-15-years-later-5200225 [https://perma.cc/EHW7-TK7D]; *YouTube*, 6SENSE, https://6sense.com/tech/media-players-and-streaming-platforms/youtube-market-share [https://perma.cc/D5KG-B9BM] (last visited Jan. 31, 2025).

447. VOGELS ET AL., *supra* note 17; *Google Buys YouTube for $1.65 Billion*, *supra* note 17.

online streaming platforms.[448]

- Blogger (2003): One of Google's earliest acquisitions, Blogger helped Google gain a foothold in content creation and blog hosting service.[449]
- Songza (2014): Specializing in music curation and streaming, its features were integrated into Google Play Music and later YouTube Music.[450]

*c. Mapping and Location Services*

- Where 2, Keyhole, ZipDash (all in 2004): These companies laid the groundwork for Google Maps and Google Earth, revolutionizing how people navigate and interact with geographic information online.[451]
- Zagat (2011): A restaurant review and guide company, Zagat's content was integrated into Google Maps and Search, enriching local business information and reviews.[452]
- Waze (2013): By acquiring Waze, a popular community-based traffic and navigation app, Google not only eliminated a significant competitor but also integrated unique crowd-sourced traffic data into its mapping services.[453]
- Skybox Imaging (2014): A satellite imaging company, aiding Google Earth and Maps with real-time satellite pictures and

---

448. *YouTube*, *supra* note 446.

449. Neil McIntosh, *Google Buys Blogger Web Service*, GUARDIAN (Feb. 18, 2003, 12:13 PM), https://www.theguardian.com/business/2003/feb/18/digitalmedia.citynews [https://perma.cc/M6Z5-QS4N] ("Google, the world's most-used internet search engine, yesterday announced the acquisition of Blogger, a web service which has fueled the rapid rise of the web journals popularly known as weblogs.").

450. Jordan Crook, *Google Buys Songza*, TECHCRUNCH (July 1, 2014, 1:00 PM), https://techcrunch.com/2014/07/01/google-buys-songza/ [https://perma.cc/K7WX-4RRL] ("According to Google, Songza will remain intact for users and nothing will change about the service for now, though Songza's expertise will be applied to other products like Google Play Music and YouTube.").

451. Morris, *supra* note 19; Hines, *supra* note 19; Lunden, *supra* note 20.

452. Casey Johnston, *Google Dives Deep into Content-Generation Business with Zagat Purchase*, ARS TECHNICA (Sept. 8, 2011, 9:55 AM), https://arstechnica.com/information-technology/2011/09/google-acquires-entertainment-and-dining-review-company-zagat/ [https://perma.cc/T5AN-V3X2] ("Google plans to collaborate with Zagat to integrate its content with Google search results and Google Maps—likely, Zagat content will begin appearing on the Places pages of all the locations it has covered.").

453. Lunden, *supra* note 20.

data analysis.[454]

*d. Online Collaboration and Productivity Tools*

- Writely (2006): The technology behind Writely, an online word processing service, was integral in developing Google Docs.[455]
- DocVerse (acquired pre-2010): This acquisition was pivotal in creating Google Docs, allowing Google to venture into cloud-based productivity and collaborative working environments while challenging traditional office suite providers.[456]

As a result of these acquisitions, Google's office suite controls over 50 percent of the office productivity software market.[457]

*e. Artificial Intelligence (AI) and Machine Learning*

- DeepMind (2014): A leading AI research company, known for its work in deep learning and artificial neural networks.[458] DeepMind's technology has been instrumental in advancing Google's AI capabilities, particularly in areas like natural language processing and autonomous systems.[459]
- Dialogflow (2016, formerly known as Api.ai): A tool for building conversational interfaces, enhancing Google's capabilities in AI-driven chatbots and voice services.[460]

---

454. Thomas Claburn, *Google Buys Skybox Imaging*, INFO. WEEK (June 11, 2014), https://www.informationweek.com/machine-learning-ai/google-buys-skybox-imaging [https://perma.cc/L5F9-DP6F] ("Satellite company's technology will improve Google Maps and enhance Google's ability to provide business intelligence to organizations.").

455. Michael Arrington, *Writely Confirms Google Acquisition*, TECHCRUNCH (Mar. 9, 2006, 11:11 AM), https://techcrunch.com/2006/03/09/writely-confirms-google-acquisition/ [https://perma.cc/U85U-YHU3] ("This signals two things: a confirmation of Google's desire to hit Microsoft hard and attack their largest revenue product, and that they will do this at least partially through acquisition rather than building the office suite entirely in-house.")

456. Arrington, *Google Acquires Docverse*, *supra* note 22.

457. *See* Fabio Duarte, *Google Workspace User Stats (2024)*, EXPLODING TOPICS (Dec. 6, 2023), https://explodingtopics.com/blog/google-workspace-stats [https://perma.cc/8SD4-2BL5].

458. Catherine Shu, *Google Acquires Artificial Intelligence Startup DeepMind for More Than $500M*, TECHCRUNCH (Jan. 26, 2014, 5:20 PM), https://techcrunch.com/2014/01/26/google-deepmind/ [https://perma.cc/7KE8-3KLL].

459. *Id.* ("Google's hiring of DeepMind will help it compete against other major tech companies as they all try to gain business advantages by focusing on deep learning.").

460. Greg Kumparak, *Google Acquires API.AI, a Company Helping Developers Build Bots That Aren't Awful to Talk To*, TECHCRUNCH (Sept. 19, 2016, 3:32 PM), https://techcrunch.com/2016/09/19/google-acquires-api-ai-a-company-helping-developers-build-bots-that-arent-awful-to-

*f. Cloud Computing and Data Management*

- Looker (2020): A big data analytics company, enhancing Google Cloud's data visualization and business intelligence capabilities.[461]
- Apigee (2016): Specializing in API management, Apigee has bolstered Google Cloud's offerings in helping enterprises in digital transformation and developing API-driven ecosystems.[462]
- Firebase (2014): A platform for developing mobile and web applications, Firebase enhanced Google's offerings in cloud services and application development.[463]
- Mandiant (2022): A cybersecurity firm known for its expertise in incident response and threat intelligence, reinforcing Google Cloud's security offerings.[464]

As a result of these acquisitions, Google cloud has a market share of 11% as of 2024.[465]

---

talk-to/ [https://perma.cc/QXS8-NYAQ] ("Google has just disclosed that it has snatched up the team behind API.AI. API.AI provides tools to developers to help them build conversational, Siri-esque bots.").

461. Ron Miller, *Google Closes $2.6B Looker Acquisition*, TECHCRUNCH (Feb. 13, 2020, 8:35 AM), https://techcrunch.com/2020/02/13/google-closes-2-6b-looker-acquisition/ [https://perma.cc/C9P2-BU7H] ("Today, the company announced that deal has officially closed and Looker is part of the Google Cloud Platform.").

462. Ron Miller, *Google Will Acquire Apigee for $625 Million*, TECHCRUNCH (Sept. 8, 2016, 6:49 AM), https://techcrunch.com/2016/09/08/google-will-acquire-apigee-for-625-million/ [https://perma.cc/DJB8-YB43] ("The company, which helps customers build digital products with open APIs, has an impressive customer list including Walgreens, AT&T, Bechtel, Burberry, First Data and Live Nation.").

463. Frederic Lardinois, *Google Acquires Firebase to Help Developers Build Better Real-Time Apps*, TECHCRUNCH (Oct. 21, 2014, 10:30 AM), https://techcrunch.com/2014/10/21/google-acquires-firebase-to-help-developers-build-better-realtime-apps/ [https://perma.cc/CL8E-QAL9] ("Google today announced that it has acquired Firebase, a backend service that helps developers build realtime apps for iOS, Android and the web that can store and sync data instantly.").

464. Sam Shead, *Google to Acquire Cybersecurity Firm Mandiant for $5.4 Billion*, CNBC, https://www.cnbc.com/2022/03/08/google-plans-to-acquire-mandiant-for-5point4-billion.html [https://perma.cc/6XQB-WPPF] (Mar. 8, 2022, 10:06 AM) ("Mandiant will join Google's cloud computing division, which is yet to grow to the same size as Microsoft Azure or Amazon Web Services.").

465. *See* Felix Richter, *Amazon Maintains Cloud Lead as Microsoft Edges Closer*, STATISTA (Nov. 1, 2024), https://www.statista.com/chart/18819/worldwide-market-share-of-leading-cloud-infrastructure-service-providers/ [https://perma.cc/VMG7-CTU4] ("Amazon's market share in the worldwide cloud infrastructure market amounted to 31 percent in the third quarter of 2024, ahead of Microsoft's Azure platform at 20 percent and Google Cloud at 11 percent.").

#### *g. Photo Management and Editing*

- Picasa (2004): A leader in photo organization and editing software, Picasa's acquisition enabled Google to integrate advanced photo management tools into its suite of services, culminating in the development of Google Photos.[466]

Google Photos is among the top ten ranked applications in the United States in the photography category on both Google Play Store and Apple App Store.[467]

These numerous acquisitions have led to Google's dominance in different markets. Thanks to the flow of revenue from its advertising businesses, Google is monopolizing and consolidating power across different segments of the digital market.

### 3. Implications for Consumer Welfare

Google's extensive acquisitions and dominance across various digital markets, such as search engines, social media, and cloud computing have far-reaching implications for consumer welfare. Google's dominance has led to reduced consumer choice and a stifling of innovation.[468] In environments where a single entity like Google holds substantial market share across different platforms, diversity in consumer options tends to diminish. [469] This monopolization can lead to decreased incentives for innovation as competitive pressure to improve and evolve services lessens.[470]

Moreover, Google's extensive data collection practices, integral to its advertising business, pose significant privacy concerns. The company's capacity to collect and analyze vast amounts of user data across its platform challenges consumer privacy.[471] With limited alternatives in essential services such as search engines and email, consumers are often left with little choice but to use services that continuously harvest their personal data.[472] Forced consumer choice strengthens Google's market position by reinforcing its data monopoly while simultaneously creating barriers for competitors who lack similar data access.[473]

---

466. Morris, *supra* note 19.

467. *See Top Apps Ranking*, SIMILARWEB, https://www.similarweb.com/top-apps/google/photography/ [https://perma.cc/U9BR-2S5E] (Feb. 16, 2025).

468. Aron Solomon, Opinion, *If You Want to Know What a Monopoly Does, Google It*, NEWSWEEK, https://www.newsweek.com/if-you-want-know-what-monopoly-does-google-it-opinion-1934910. [https://perma.cc/TB9K-EM2Y] (Aug. 7, 2024, 11:55 AM).

469. *See* Borgolte & Feamster, *supra* note 353; Chen et al., *SameParty Cookie*, *supra* note 357; *Related Website Sets (Formerly Known As: First-Party Sets)*, *supra* note 357; Shields & Peterson, *supra* note 357.

470. Solomon, *supra* note 468.

471. Cohen, *Google Breaks Promise*, *supra* note 354.

472. Solomon, *supra* note 468.

Due to Google's vested interests in online advertising, which heavily relies on extensive data collection, any actions taken by the company to protect user privacy are inevitably scrutinized by regulators and competitors.[474] Google's recent indecision regarding the phaseout of third-party cookies exemplifies how its own interests can come at the expense of user privacy.[475] While other browsers like Firefox and Safari have restricted third-party cookies without major scrutiny of their intentions, Google's attempts have been met with protests and concerns by regulators and competitors.[476] This pressure has forced Google to abandon plans to phase out third-party cookies.[477] Although this decision may support market competitiveness, it raises important questions about Google's ability to protect users' online privacy while balancing its business interests.

Additionally, Google's market strategies, like making Google Search the default in Chrome and paying to maintain default status on other browsers, raise antitrust concerns.[478] Such practices not only consolidate Google's position in the search market but also create significant barriers to entry for new competitors, thereby undermining competitive market dynamics.[479] This monopolistic stance may impede the emergence of innovative competitors and alternative technologies, affecting consumer choice and the health of the digital market.

---

473. *About Google Ads*, *supra* note 264; *Set Up Analytics*, *supra* note 265.

474. Cyphers, *supra* note 79; Berke & Calacci, *supra* note 79.

475. *Supra* Section II.A; *see* Cohen, *Google Breaks Promise*, *supra* note 354 ("Google's decision to continue allowing third-party cookies, despite overwhelming evidence of their surveillance harms, is a direct consequence of their advertising-driven business model. Google makes most of its money from tracker-driven, behaviorally-targeted ads. If Google wanted, Chrome could do much more to protect your privacy. Other major browsers, like Safari and Firefox, provide significantly more protection against online tracking by default. Notably, Google is the internet's biggest tracker, and most of the websites you visit include Google trackers (including but not limited to third-party cookies). As Chrome leaves users vulnerable to tracking, Google continues to receive nearly 80% of their revenue from online advertising.").

476. Competition and Markets Authority, *Investigation*, *supra* note 151; *The Impact of the Google Topics API*, *supra* note 297; Selman, *supra* note 299.

477. Chavez, *New Path*, *supra* note 83.

478. *Supra* Section III.B.1.

479. *See* Sharma, *supra* note 401; Borgolte & Feamster, *supra* note 353; Chen et al., *SameParty Cookie*, *supra* note 357; *Related Website Sets (Formerly Known As: First-Party Sets)*, *supra* note 357; Shields & Peterson, *supra* note 357.

A particularly concerning aspect of this scenario is the creation of a vicious cycle where dominance in one market is leveraged to gain control in others. For instance, revenue and data obtained from Google's search business can be used to subsidize and promote other services, further entrenching its market position. This cycle leads to Google's consolidation of power, stifling competition across multiple sectors. As Google dominates each market and becomes a tool to further consolidate its position in other areas, it exacerbates the challenges for new entrants and innovators.

While the convenience and integration of Google's services offer benefits, the trade-offs in competition, innovation, and privacy are significant. This situation highlights the need for robust antitrust regulation and consumer protection. Ensuring a balanced digital market, where competition is encouraged, innovation is nurtured, and consumer data and privacy are protected, is crucial. Addressing these issues is essential to maintain a dynamic and fair digital ecosystem that serves the broad interests of consumers.

## IV. Recommendations

In this section, observations around different anticompetitive practices by Google inform this Article's recommended remedies that can ensure a fair and competitive online market landscape. The recommendations are divided into two categories: behavioral (concerning agreements, actions, and decisions of Google which affect competition) and structural (separating Chrome as a separate entity within Google, divestiture of Chrome).[480]

### *A. Behavioral Remedies*

In addressing Google Chrome's antitrust challenges, a potential strategy lies in implementing behavioral remedies. This Article's investigation has revealed a pattern in Google's integration of services across its browser and other platforms to lock users into its

480. It is important to note that at the time of writing this Article, the DOJ's case against Google's monopoly in search is still ongoing. Thus, the discussion of the DOJ's proposed remedies is limited and may not reflect the status quo of this ongoing discussion. Interestingly, following the publication of a pre-print of this Article, the DOJ picked up the idea of forcing Google to divest Chrome, an idea that was first discussed in detail in that pre-print. *See* Executive Summary of Plaintiffs' Proposed Final Judgement at 3–4, United States v. Google LLC, No. 1:20-cv-03010 (D.D.C. Nov. 20, 2024).

ecosystem.[481] This lock-in facilitates a more streamlined data collection process, thereby reinforcing Google's dominance in different markets.[482]

Central to these concerns is Google's practice of universal login across its services. Current protocols dictate that signing into one Google service inadvertently leads to automatic logins across others.[483] This system blurs the boundaries of user consent, especially since engagement with one service, like YouTube, does not prima facie equate to a blanket authorization for signing into other services such as Google Search, Google Docs, or even Google Chrome after version 70.[484] To remedy this, a separation between Google's various services is necessary. As a relatively noninvasive remedy, each service should require independent consent for user login coupled with a rigorous enforcement of the existing purpose limitation principle under EU data protection and privacy law.[485] Independent consent ensures that interactions with one platform do not result in unintended access to others, which also ensures that user data is not reused across different services in an anticompetitive manner. This approach not only supports user autonomy and privacy, but also restricts Google's ability to leverage its ecosystem to unfairly collect and monetize user data. Moreover, independent consent underscores the broader implications of informed consent in the digital age. It challenges the prevailing norms of data collection and usage by dominant market players like Google and advocates for a more user-centric approach to service integration.[486] This is not merely a technical adjustment but a fundamental shift toward respecting user choice and privacy in an increasingly interconnected digital landscape—a shift that is essential for restoring competitive balance and fairness in the digital market.

---

481. *See* Borgolte & Feamster, *supra* note 353; Chen et al., *SameParty Cookie*, *supra* note 357; *Related Website Sets (Formerly Known As: First-Party Sets)*, *supra* note 357; Shields & Peterson, *supra* note 357.

482. Cohen, *Google Breaks Promise*, *supra* note 354.

483. Cimpanu, *supra* note 328; *How Google Uses Cookies*, *supra* note 329; Green, *supra* note 223.

484. Green, *supra* note 223; Fowler, *supra* note 224.

485. *See* Commission Regulation 2016/679, 2016 O.J. (L 119) 35 (EU) ("[C]ollected for specified, explicit and legitimate purposes and not further processed in a manner that is incompatible with those purposes; further processing for archiving purposes in the public interest, scientific or historical research purposes or statistical purposes shall, in accordance with Article 89(1), not be considered to be incompatible with the initial purposes ('purpose limitation') . . . .").

486. *Id.*

Building on the need for greater user autonomy and privacy, it is crucial to consider regulatory frameworks that define and address corporate dominance in the digital space. A pertinent example is the European Union's 2022 DSA and DMA, which set out criteria for identifying dominant digital products and services and subjecting them to stringent obligations.[487] This approach to regulation acknowledges the substantial influence such services wield over digital markets and user experiences. It serves as a model for imposing accountability and mitigating the risks associated with excessive market power. By establishing dominance criteria, the DSA and DMA aim to ensure that large technology companies do not abuse their market position, while fostering a fairer, more competitive environment.[488] This regulatory measure is a step toward leveling the playing field by compelling dominant players like Google to adhere to higher standards of operation, particularly in aspects of consumer protection, privacy, and fair competition.

A critical aspect of addressing Google's antitrust practices also requires prohibiting Google from advertising its own services on platforms within its network. For example, Google's promotion of the Chrome browser within its search engine.[489] Because both Chrome and Google Search are under the same corporate umbrella, this internal cross-promotion presents a conflict of interest.[490] Google, in effect, can easily outbid competitors for advertising space, essentially transferring funds within its divisions. This tactic effectively sidelines other browsers as Google's worst-case scenario is merely the loss of potential additional revenue from these competitors' ads.

Similarly, Google's advertising of its array of services on YouTube is another manifestation of this issue.[491] Such practices consolidate Google's market dominance not through superior service or competitive pricing, but through leveraging its existing control over multiple high-traffic platforms. This self-promotion strategy restricts free competition and limits consumer choice. It allows Google to maintain and expand its market dominance in various sectors, from browsers to online video platforms, by using its established platforms as self-reinforcing advertisement channels. To foster a more competitive digital market, remedial measures should enforce a clear separation between Google's advertising entities and its other service

487. Commission Regulation 2022/2065, 2022 O.J. (L 277) 1 (EU).

488. *Id.*

489. *See supra* Section III.B.1.

490. *See* Doctorow, *supra* note 408.

491. *See* United States v. Google LLC, No. 20-cv-3010, 2024 WL 3647498, at *36 (D.D.C. Aug. 5, 2024).

platforms. These measures would prevent Google from utilizing its dominance in one area (such as search or browser) to unfairly promote its services in another, thereby ensuring a level playing field for all market participants.

Another pivotal recommendation is preventing Google's practice of entering exclusive contracts, also proposed by the DOJ, and exemplified by Google's agreement with Apple where Google remains the default search engine in exchange for sharing 36% of the revenue generated from Apple users.[492] These agreements, albeit lucrative for the parties involved, contribute to the entrenchment and concentration of market power in the hands of already dominant players. These exclusivity contracts act as barriers to entry for potential competitors in the search engine market.[493] They not only reinforce Google's dominance but also limit consumer choice by preemptively deciding the default service for vast user bases.[494] This practice stifles competition and innovation in the market as emerging players find it increasingly challenging to gain a foothold against entrenched agreements.

### *B. Structural Remedies*

The structural remedies are divided into internal separation of Chrome within Google and divestiture of Chrome from Google.

#### 1. Internal Separation

To address the potential abuse of Chrome's dominant market position, Chrome should operate as a structurally separate entity within Google. This separation is crucial to mitigate conflicts of interests in the operation of Chrome and other Google services, which could lead to anticompetitive practices. By separating Chrome, Google's ability to use its browser dominance to unfairly influence other market segments would be significantly curtailed.

The California Public Utilities Act of 1912 brought natural gas, electric, telephone, and water companies, as well as railroads and marine transportation companies, under the purview of the California Public Utilities Commission.[495] Inspired by historical interventions like

492. Howley, *supra* note 420.

493. Sharma, *supra* note 401.

494. Feiner, *Google Paid*, *supra* note 413.

495. *See* Max Thelen, *The Public Utilities Act and Its Relation to the Public*, *in* PUBLIC UTILITIES ACT OF CALIFORNIA 17, 18–19 (1912).

this, there is a potential to enact utility-style regulation for browsers. This perspective is grounded in the recognition that browsers, much like utilities, are essential conduits to critical services—in this case, the internet.[496] Consequently, imposing utility-style regulation on browsers could ensure a level playing field, like the regulatory frameworks governing electricity or water services. For example, due to the high cost and impracticality of duplicating utility infrastructure, electricity and water services are considered natural monopolies and regulated accordingly.[497] These regulations ensure that a single company owning all the utility infrastructure in a locality is unable to extract unfair prices from consumers.[498] Similar regulation in the context of the internet would not only maintain Chrome's functional utility within Google's broader ecosystem but also establish safeguards to prevent its use as a tool for market manipulation.

There is precedence of similar regulation restricting behavior of firms designated as gatekeepers in the DMA enforced in the European Union. For example, Article 6(7) of the DMA requires gatekeepers to ensure interoperability for certain products of other business users, which contributes to the neutrality of the platform.[499] This provision mitigates, among other aspects, the fact that Apple used to allow only its own Apple Pay service to use the near-field communication (NFC) feature of iPhones for contactless payment.[500] Now, other payment providers can implement similar contactless payment solutions on iPhones.[501]

---

496. *See generally* Commission Regulation 2022/1925, 2022 O.J. (L 265) 34–36 (EU).

497. *See, e.g.*, Federal Power Act (FPA), 16 U.S.C §§ 791–825c; Public Utility Regulatory Policies Act of 1978, 16 U.S.C. §§ 2601–2645; Telecommunication Act of 1996, Pub. L. No. 104-104, 110 Stat. 56 (codified as amended in scattered sections of 47 U.S.C.).

498. *See generally* Federal Power Act (FPA), 16 U.S.C §§ 791–825c; Public Utility Regulatory Policies Act of 1978, 16 U.S.C. §§ 2601–2645; Telecommunication Act of 1996, Pub. L. No. 104-104, 110 Stat. 56 (codified as amended in scattered sections of 47 U.S.C.).

499. *See* Commission Regulation 2022/1925, 2022 O.J. (L 265) 30–32, 34–36 (EU) ("An undertaking shall be designated as a gatekeeper if: (a) it has a significant impact on the internal market; (b) it provides a core platform service which is an important gateway for business users to reach end users; and (c) it enjoys an entrenched and durable position, in its operations, or it is foreseeable that it will enjoy such a position in the near future.").

500. *See* Margrethe Vestager, Exec. Vice President, Eur. Comm'n, Remarks by Executive Vice-President Vestager on the Decision to Make Binding Commitments Offered by Apple (July 10, 2024) ("Our preliminary finding was therefore that Apple abused its dominant position by refusing to supply the NFC technology to competing mobile wallet developers. . . . First, Apple commits to give access to NFC functionality to third-party mobile wallets. This access will be free of charge.").

501. *Id.* ("It will take place in what is called 'Host Card Emulation mode'. This is a software solution that allows rival wallets to make secure NFC payments.").

Similarly, Chrome is the de facto operating system for the web and has a strong economic position and impact on the digital advertising market.[502] Therefore, leveraging and potentially expanding the DMA could open the door to obligations which mitigate issues discussed previously in this Article.[503] Firstly, it could prevent Chrome from self-preferencing, acting as an effective measure against dark patterns. Secondly, it could allow other browsers to access and prevent Google from using excessive data collected with the help of Chrome.[504] Lastly, it could ensure interoperability between features provided by Chrome and other browsers. Ensuring interoperability could allow Chrome to follow standards more closely and remove hurdles for users looking to switch browsers but unable to do so because of cross-device sync and other features currently limited to Chrome.[505]

This structural approach aligns with the broader objective of antitrust law—to foster competitive markets and protect consumer welfare.[506] Redefining browsers as utilities and structurally separating them within conglomerates like Google, the market dynamics of the digital age could be recalibrated to ensure fairness and prevent the concentration of power in the hands of a few dominant players.

---

502. *See Search Engine Market Share Worldwide, Dec 2023 - Dec 2024*, *supra* note 23.

503. *See* Commission Regulation 2022/1925, 2022 O.J. (L 265) 33–34 (EU).

504. *See id.* at 33–36 ("The gatekeeper shall not do any of the following . . . combine personal data from the relevant core platform service with personal data from any further core platform services or from any other services provided by the gatekeeper or with personal data from third-party services . . . . The gatekeeper shall not treat more favourably, in ranking and related indexing and crawling, services and products offered by the gatekeeper itself than similar services or products of a third party. The gatekeeper shall apply transparent, fair and non-discriminatory conditions to such ranking.").

505. *See id.* at 36 ("The gatekeeper shall allow providers of services and providers of hardware, free of charge, effective interoperability with, and access for the purposes of interoperability to, the same hardware and software features accessed or controlled via the operating system or virtual assistant listed in the designation decision pursuant to Article 3(9) as are available to services or hardware provided by the gatekeeper. Furthermore, the gatekeeper shall allow business users and alternative providers of services provided together with, or in support of, core platform services, free of charge, effective interoperability with, and access for the purposes of interoperability to, the same operating system, hardware or software features, regardless of whether those features are part of the operating system, as are available to, or used by, that gatekeeper when providing such services."). While cross-device sync works between Firefox Desktop and Firefox Mobile, it does not work between Firefox Desktop and Chrome Mobile (which comes as default browser on all Android phones). *Sync Firefox Data*, MOZILLA: SUPPORT, https://support.mozilla.org/en-US/kb/sync [https://perma.cc/JCD9-RQSU] (Mar. 15, 2024). This incompatibility forces users to switch to the same browser on both their mobile and desktop devices. *See id.*

506. *See The Antitrust Laws*, ANTITRUST DIV., https://www.justice.gov/atr/antitrust-laws-and-you#:~:text=This%20law%20aims%20to%20promote,mergers%20that%20could%20lessen%20competition [https://perma.cc/5QH8-KTJG] (last visited Jan. 31, 2025).

### 2. Divestiture

The divestiture of Chrome from Google is another remedy that can be considered if other remedies prove insufficient in mitigating Chrome's market dominance. The ruling of the US District Court of the District of Columbia finding that Google is in violation of Section 2 of the Sherman Act also motivates this solution.[507] The DOJ filed a proposed final judgment which suggests divesting parts of Google as a remedy to this antitrust ruling.[508] In this scenario, Chrome would be spun off into an independent entity or sold to another party, free from Google's influence, which might prompt concerns regarding the financial viability of the newly independent Chrome. As an example, 83 percent of all revenue generated by Firefox's parent company, Mozilla, comes from payments by Google to maintain its search engine exclusivity.[509] An independent Chrome may also be similarly reliant. The precedent set by Mozilla's financial dependence on Google highlights potential challenges for Chrome in maintaining its operations without similar support. These concerns were also raised by Google in response to the antitrust judgment by the US District Court of the District of Columbia and the corresponding remedies suggested by DOJ.[510]

---

507. *See* United States v. Google LLC, No. 20-cv-3010, 2024 WL 3647498, at *3 (D.D.C. Aug. 5, 2024) ("After having carefully considered and weighed the witness testimony and evidence, the court reaches the following conclusion: Google is a monopolist, and it has acted as one to maintain its monopoly. It has violated Section 2 of the Sherman Act.").

508. *See* Executive Summary of Plaintiffs' Proposed Final Judgement at 3–4, United States v. Google LLC, No. 1:20-cv-03010 (D.D.C. Nov. 20, 2024) ("Google's ownership and control of Chrome and Android—key methods for the distribution of search engines to consumers—poses a significant challenge to effectuate a remedy that aims to 'unfetter [these] market[s] from anticompetitive conduct' and 'ensure that there remain no practices likely to result in monopolization in the future.' To address these challenges, Google must divest Chrome, which has 'fortified [Google's] dominance'. . . . (quoting United States v. Microsoft Corp., 253 F.3d 34, 103 (D.C. Cir. 2001))).

509. Cohen, *supra* note 429.

510. *See* Lee-Anne Mulholland, *DOJ's Radical and Sweeping Proposals Risk Hurting Consumers, Businesses, and Developers*, GOOGLE: KEYWORD (Oct. 9, 2024), https://blog.google/outreach-initiatives/public-policy/doj-search-remedies-framework/ [https://perma.cc/M92T-24E8] ("We've invested billions of dollars in Chrome and Android. Chrome is a secure, fast, and free browser and its open-source code provides the backbone for numerous competing browsers. Android is a secure, innovative, and free open-source operating system that has enabled vast choice in the smartphone market, helping to keep the cost of phones low for billions of people. Because both Chrome and Android help people access the web and use our While there is merit to the concerns surrounding the financial viability of Chrome without Google's support, the existence of these concerns cannot be considered a strong case against Chrome's divestiture from Google. Google's investment of billions of dollars on Chrome makes little financial sense without expectation of a reciprocal return on investment (ROI).[511] Google's continued investment in Chrome signals that Chrome is meeting its ROI expectations, specifically, Chrome helps Google attain and retain users of its other services.[512] In the scenario where Chrome is divested, Google's continued financial and technical investments in Chrome would likely yield similar returns, provided that Chrome remains open and free for all (as per Google's current claims regarding Chrome), including Google. While Google should not see any negative impact due to this divestment, Chrome and the online ecosystem would benefit from this. As an independent entity, Chrome could attract investment from other parties who are currently unable to partake (technically and financially) in its development due to Google's control on the browser.[513] To ensure this interest, it is essential to couple the divestiture with the implementation of the previously discussed behavioral remedies. These measures would ensure that, while Google and other interested parties may have the ability to contribute to Chrome, they would not be able to forge exclusive contracts that could yield an unfair competitive advantage, similar to those currently employed by Google.

Additionally, reclassifying browsers as utilities could play a crucial role in this context. Under such a classification, Chrome's agreements and decisions would be subject to heightened scrutiny, particularly to safeguard consumer welfare and prevent exclusionary

---

products, we offer them (and their underlying code) for free. Few companies would have the ability or incentive to keep them open source, or to invest in them at the same level we do.").

511. *See* Jason Feernando, *What Is Return on Investment (ROI) and How to Calculate It*, INVESTOPEDIA, https://www.investopedia.com/terms/r/returnoninvestment.asp [https://perma.cc/CG6F-WDR2] (Nov. 18, 2024).

512. *See* Mulholland, *supra* note 510.

513. *See Google's Potential Chrome Sell-Off: A Game-Changer Decision Makers Can't Ignore*, MARKEDIUM: TECH DESK [hereinafter *Google's Potential Chrome Sell-Off*], https://markedium.com/googles-potential-chrome-sell-off-a-game-changer-decision-makers-cant-ignore/#:~:text=The%20DOJ's%20Antitrust%20Rationale,advertising%20market%20and%20foster%20competition [https://perma.cc/3TFK-LP83] (Nov. 23, 2024).

practices.[514] For example, if Pacific Gas and Electric (PG&E) seeks to increase rates, it must go through a regulatory process which is overseen by California Public Utilities Commission.[515] Similarly, agreements that might impact Chrome's freedom and fairness would need to go through a regulatory process as well.

In addition to divestiture of Chrome, it is critical to re-examine Google's past mergers and analyze whether those mergers resulted in less market competition. The Google-DoubleClick merger serves as a prime example, where initial regulatory approval failed to anticipate potential long term anticompetitive effects.[516] By examining the FTC's 2007 approval of the Google-DoubleClick merger, the limitations of forward-looking assessments that failed to foresee the enhanced market power and anticompetitive behavior arising from the combination of Google's dominance in other markets with DoubleClick's advertising technology are evident.[517] Yet, despite concerns from within the FTC, the merger was approved.[518] While the FTC claimed to keep a close watch on the market and promised to "act quickly" should there be

514. *See* Hugh Langley & Lara O'Reilly, *The DOJ Wants Google to Sell Its Chrome Browser. Here Are the Winners and Losers if That Happens*, BUS. INSIDER AFR. (Nov. 21, 2024, 4:47 AM), https://africa.businessinsider.com/news/the-doj-wants-google-to-sell-its-chrome-browser-here-are-the-winners-and-losers-if/6z78gvs [https://perma.cc/P7PP-GG99].

515. *See 2023 General Rate Case*, PG&E, https://www.pge.com/en/regulation/general-rate-case.html [https://perma.cc/EWK3-B5TM] (last visited Jan. 31, 2025) ("The CPUC requires PG&E and other regulated utilities to submit a GRC proposal every four years. The proposals determine fair energy rates. The rates are based on the cost of operating, maintaining and improving the safety and reliability of our electric and natural gas systems. The GRC is a thorough and public regulatory proceeding. It includes a series of filings, hearings and negotiations. It also includes feedback and input from customers, customer advocates and stakeholders. The process ensures energy rates are based on the actual costs.").

516. *See* Press Release, Fed. Trade Comm'n, Federal Trade Commission Closes Google/DoubleClick Investigation (Dec. 20, 2007), https://www.ftc.gov/news-events/news/press-releases/2007/12/federal-trade-commission-closes-googledoubleclick-investigation [https://perma.cc/S8T2-42W9] ("The evidence also showed that it was unlikely that Google could manipulate DoubleClick's third-party ad serving products in a way that would competitively disadvantage Google's competitors in the ad intermediation market. Further, the evidence demonstrated that any aggregation of consumer and competitive data resulting from the acquisition is unlikely to harm competition in the ad intermediation market.").

517. *See* Binns & Bietti, *Dissolving Privacy*, *supra* note 132.

518. *See* Fed. Trade Comm'n, Dissenting Statement of Commissioner Pamela Jones Harbour, F.T.C. File No. 071-0170, at 1 (Dec. 20, 2007), https://www.ftc.gov/sites/default/files/documents/public_statements/statement-matter-google/doubleclick/071220harbour_0.pdf [https://perma.cc/CA26-UVNF] ("I dissent because I make alternate predictions about where this market is heading, and the transformative role the combined Google/DoubleClick will play if the proposed acquisition is consummated. If the Commission closes its investigation at this time, without imposing any conditions on the merger, neither the competition nor the privacy interests of consumers will have been adequately addressed.").

anticompetitive behavior, the present market situation illustrates the shortcomings of this approach.[519] Therefore, it is essential that a mechanism aimed at revisiting these decisions is established by retrospective divestiture approaches on previous mergers. Additionally, it is also important to apply lessons learned from re-examination of these acquisitions to future acquisitions by Google and other large technology firms.

This dual strategy of divestiture and regulatory oversight would create a more equitable browser market. It would ensure that Chrome operates independently, both structurally and financially, while remaining subject to regulatory frameworks that prevent anti-competitive contracts and actions detrimental to consumer interests and a competitive market.

## V. CONCLUSIONS

Google's market dominance in the realms of web browsing, publishing, and advertising is a critical barrier to fair competition in the digital age.[520] The company's strategic acquisitions, coupled with tactics that subtly coerce users and disadvantage competitors, illuminate a complex web of dominance that extends far beyond traditional market boundaries.[521]

The exploration of potential remedies—both behavioral and structural—is imperative in addressing the multifaceted nature of Google's market power. Behavioral remedies, while targeting specific anticompetitive practices, may fall short in dismantling the entrenched dominance Google holds.[522] The imposition of structural remedies, calling for an internal reorganization to disentangle Google's advertising and browser divisions, could serve as a more robust

519. *See* Fed. Trade Comm'n, Statement of Federal Trade Commission Concerning Google/DoubleClick, F.T.C. File No. 071-0170, at 13 (Dec. 20, 2007), https://www.ftc.gov/system/files/documents/public_statements/418081/071220googledc-commstmt.pdf [https://perma.cc/8QCT-SHFH].

520. *See* Montoya, *supra* note 3.

521. *See id.*

522. *See* Makan Delrahim, Remarks at the American Bar Association Antitrust Section Fall Forum (Nov. 16, 2017), https://www.justice.gov/opa/speech/file/1012086/dl [https://perma.cc/Q4WB-UMLS].

approach to ensuring fair play.[523] However, even this might not be enough to fully address the overarching issue of market dominance. Drawing parallels with historical regulatory actions, the divestiture of Google Chrome into an independent organization, potentially as a public utility, emerges as a potentially transformative solution. This approach would not only curb Google's ability to exploit Chrome for monopolistic gain but also pave the way for a more equitable digital marketplace.[524] Such a transition would mark a significant shift towards competition and innovation in the digital economy and transform Chrome from a tool aimed at market capture and monopolization into an entity bound to promote public interest.

In the ever-evolving digital landscape, the need for vigilant regulatory oversight and proactive measures is more urgent than ever. As Google demonstrates, unchecked dominance in one market can lead to cross-market abuses and the distortion of competition and innovation.[525] It is essential for regulatory bodies and policymakers to adapt to these new challenges and ensure that the digital domain remains a competitive, diverse, and vibrant space. Managing to hire top tech talent at these authorities, rather than exclusively at the leading tech companies, will be an important and necessary step towards to goal.

## VI. ACKNOWLEDGEMENTS

Shaoor Munir and Zubair Shafiq are supported in part by the National Science Foundation under grant numbers 2103439 and 2138139.

Anastasia Shuba was employed by DuckDuckGo during the project, but this work was completed independently and does not reflect DuckDuckGo's position.

Konrad Kollnig is supported by the RegTech4AI AiNed Fellowship Grant, which is funded by Dutch National Growth Fund (NGF) under file number NGF.1607.22.028.

523. *See Google Could Be Forced to Sell Parts of Its Business if the DOJ Gets Its Way*, FAST CO. (Oct. 9, 2024), https://www.fastcompany.com/91206215/google-could-forced-sell-parts-business-doj-gets-way [https://perma.cc/E7UF-VHCD].

524. *See Google's Potential Chrome Sell-Off*, *supra* note 513.

525. *See* Press Release, U.S. Dep't of Just., *supra* note 13.